\documentclass{article}
\usepackage{authblk}%
\usepackage{amsmath}
\usepackage{array, amsfonts}
\usepackage{amsthm}
\usepackage{amssymb}
\usepackage{subcaption}
\usepackage{enumerate}
\usepackage{lineno}
\usepackage{comment}
\usepackage{url}
\usepackage{fontenc}
\usepackage{xcolor}
\makeatletter
\renewcommand{\textcolor}[2]{#2}
\renewcommand{\color}[1]{}
\makeatother
\usepackage{graphicx}
\usepackage{amstext}
\usepackage{graphics}
\usepackage{titlesec} 
\usepackage{float}
\usepackage{booktabs}
\usepackage{caption} 
\usepackage{subcaption} 
\usepackage{graphicx}
\usepackage{longtable}
\usepackage[all]{nowidow}
\usepackage{tikz}
\usetikzlibrary{er,positioning,bayesnet}
\usepackage{multicol}
\usepackage{algpseudocode,algorithm,algorithmicx}
\usepackage[draft]{minted}
\usepackage{xcolor} 

\makeatletter
\def\PYGdefault@reset{\let\PYGdefault@it=\relax \let\PYGdefault@bf=\relax%
    \let\PYGdefault@ul=\relax \let\PYGdefault@tc=\relax%
    \let\PYGdefault@bc=\relax \let\PYGdefault@ff=\relax}
\def\PYGdefault@tok#1{\csname PYGdefault@tok@#1\endcsname}
\def\PYGdefault@toks#1+{\ifx\relax#1\empty\else%
    \PYGdefault@tok{#1}\expandafter\PYGdefault@toks\fi}
\def\PYGdefault@do#1{\PYGdefault@bc{\PYGdefault@tc{\PYGdefault@ul{%
    \PYGdefault@it{\PYGdefault@bf{\PYGdefault@ff{#1}}}}}}}
\def\PYGdefault#1#2{\PYGdefault@reset\PYGdefault@toks#1+\relax+\PYGdefault@do{#2}}

\expandafter\def\csname PYGdefault@tok@gd\endcsname{\def\PYGdefault@tc##1{\textcolor[rgb]{0.63,0.00,0.00}{##1}}}
\expandafter\def\csname PYGdefault@tok@gu\endcsname{\let\PYGdefault@bf=\textbf\def\PYGdefault@tc##1{\textcolor[rgb]{0.50,0.00,0.50}{##1}}}
\expandafter\def\csname PYGdefault@tok@gt\endcsname{\def\PYGdefault@tc##1{\textcolor[rgb]{0.00,0.27,0.87}{##1}}}
\expandafter\def\csname PYGdefault@tok@gs\endcsname{\let\PYGdefault@bf=\textbf}
\expandafter\def\csname PYGdefault@tok@gr\endcsname{\def\PYGdefault@tc##1{\textcolor[rgb]{1.00,0.00,0.00}{##1}}}
\expandafter\def\csname PYGdefault@tok@cm\endcsname{\let\PYGdefault@it=\textit\def\PYGdefault@tc##1{\textcolor[rgb]{0.25,0.50,0.50}{##1}}}
\expandafter\def\csname PYGdefault@tok@vg\endcsname{\def\PYGdefault@tc##1{\textcolor[rgb]{0.10,0.09,0.49}{##1}}}
\expandafter\def\csname PYGdefault@tok@m\endcsname{\def\PYGdefault@tc##1{\textcolor[rgb]{0.40,0.40,0.40}{##1}}}
\expandafter\def\csname PYGdefault@tok@mh\endcsname{\def\PYGdefault@tc##1{\textcolor[rgb]{0.40,0.40,0.40}{##1}}}
\expandafter\def\csname PYGdefault@tok@go\endcsname{\def\PYGdefault@tc##1{\textcolor[rgb]{0.53,0.53,0.53}{##1}}}
\expandafter\def\csname PYGdefault@tok@ge\endcsname{\let\PYGdefault@it=\textit}
\expandafter\def\csname PYGdefault@tok@vc\endcsname{\def\PYGdefault@tc##1{\textcolor[rgb]{0.10,0.09,0.49}{##1}}}
\expandafter\def\csname PYGdefault@tok@il\endcsname{\def\PYGdefault@tc##1{\textcolor[rgb]{0.40,0.40,0.40}{##1}}}
\expandafter\def\csname PYGdefault@tok@cs\endcsname{\let\PYGdefault@it=\textit\def\PYGdefault@tc##1{\textcolor[rgb]{0.25,0.50,0.50}{##1}}}
\expandafter\def\csname PYGdefault@tok@cp\endcsname{\def\PYGdefault@tc##1{\textcolor[rgb]{0.74,0.48,0.00}{##1}}}
\expandafter\def\csname PYGdefault@tok@gi\endcsname{\def\PYGdefault@tc##1{\textcolor[rgb]{0.00,0.63,0.00}{##1}}}
\expandafter\def\csname PYGdefault@tok@gh\endcsname{\let\PYGdefault@bf=\textbf\def\PYGdefault@tc##1{\textcolor[rgb]{0.00,0.00,0.50}{##1}}}
\expandafter\def\csname PYGdefault@tok@ni\endcsname{\let\PYGdefault@bf=\textbf\def\PYGdefault@tc##1{\textcolor[rgb]{0.60,0.60,0.60}{##1}}}
\expandafter\def\csname PYGdefault@tok@nl\endcsname{\def\PYGdefault@tc##1{\textcolor[rgb]{0.63,0.63,0.00}{##1}}}
\expandafter\def\csname PYGdefault@tok@nn\endcsname{\let\PYGdefault@bf=\textbf\def\PYGdefault@tc##1{\textcolor[rgb]{0.00,0.00,1.00}{##1}}}
\expandafter\def\csname PYGdefault@tok@no\endcsname{\def\PYGdefault@tc##1{\textcolor[rgb]{0.53,0.00,0.00}{##1}}}
\expandafter\def\csname PYGdefault@tok@na\endcsname{\def\PYGdefault@tc##1{\textcolor[rgb]{0.49,0.56,0.16}{##1}}}
\expandafter\def\csname PYGdefault@tok@nb\endcsname{\def\PYGdefault@tc##1{\textcolor[rgb]{0.00,0.50,0.00}{##1}}}
\expandafter\def\csname PYGdefault@tok@nc\endcsname{\let\PYGdefault@bf=\textbf\def\PYGdefault@tc##1{\textcolor[rgb]{0.00,0.00,1.00}{##1}}}
\expandafter\def\csname PYGdefault@tok@nd\endcsname{\def\PYGdefault@tc##1{\textcolor[rgb]{0.67,0.13,1.00}{##1}}}
\expandafter\def\csname PYGdefault@tok@ne\endcsname{\let\PYGdefault@bf=\textbf\def\PYGdefault@tc##1{\textcolor[rgb]{0.82,0.25,0.23}{##1}}}
\expandafter\def\csname PYGdefault@tok@nf\endcsname{\def\PYGdefault@tc##1{\textcolor[rgb]{0.00,0.00,1.00}{##1}}}
\expandafter\def\csname PYGdefault@tok@si\endcsname{\let\PYGdefault@bf=\textbf\def\PYGdefault@tc##1{\textcolor[rgb]{0.73,0.40,0.53}{##1}}}
\expandafter\def\csname PYGdefault@tok@s2\endcsname{\def\PYGdefault@tc##1{\textcolor[rgb]{0.73,0.13,0.13}{##1}}}
\expandafter\def\csname PYGdefault@tok@vi\endcsname{\def\PYGdefault@tc##1{\textcolor[rgb]{0.10,0.09,0.49}{##1}}}
\expandafter\def\csname PYGdefault@tok@nt\endcsname{\let\PYGdefault@bf=\textbf\def\PYGdefault@tc##1{\textcolor[rgb]{0.00,0.50,0.00}{##1}}}
\expandafter\def\csname PYGdefault@tok@nv\endcsname{\def\PYGdefault@tc##1{\textcolor[rgb]{0.10,0.09,0.49}{##1}}}
\expandafter\def\csname PYGdefault@tok@s1\endcsname{\def\PYGdefault@tc##1{\textcolor[rgb]{0.73,0.13,0.13}{##1}}}
\expandafter\def\csname PYGdefault@tok@sh\endcsname{\def\PYGdefault@tc##1{\textcolor[rgb]{0.73,0.13,0.13}{##1}}}
\expandafter\def\csname PYGdefault@tok@sc\endcsname{\def\PYGdefault@tc##1{\textcolor[rgb]{0.73,0.13,0.13}{##1}}}
\expandafter\def\csname PYGdefault@tok@sx\endcsname{\def\PYGdefault@tc##1{\textcolor[rgb]{0.00,0.50,0.00}{##1}}}
\expandafter\def\csname PYGdefault@tok@bp\endcsname{\def\PYGdefault@tc##1{\textcolor[rgb]{0.00,0.50,0.00}{##1}}}
\expandafter\def\csname PYGdefault@tok@c1\endcsname{\let\PYGdefault@it=\textit\def\PYGdefault@tc##1{\textcolor[rgb]{0.25,0.50,0.50}{##1}}}
\expandafter\def\csname PYGdefault@tok@kc\endcsname{\let\PYGdefault@bf=\textbf\def\PYGdefault@tc##1{\textcolor[rgb]{0.00,0.50,0.00}{##1}}}
\expandafter\def\csname PYGdefault@tok@c\endcsname{\let\PYGdefault@it=\textit\def\PYGdefault@tc##1{\textcolor[rgb]{0.25,0.50,0.50}{##1}}}
\expandafter\def\csname PYGdefault@tok@mf\endcsname{\def\PYGdefault@tc##1{\textcolor[rgb]{0.40,0.40,0.40}{##1}}}
\expandafter\def\csname PYGdefault@tok@err\endcsname{\def\PYGdefault@bc##1{\setlength{\fboxsep}{0pt}\fcolorbox[rgb]{1.00,0.00,0.00}{1,1,1}{\strut ##1}}}
\expandafter\def\csname PYGdefault@tok@kd\endcsname{\let\PYGdefault@bf=\textbf\def\PYGdefault@tc##1{\textcolor[rgb]{0.00,0.50,0.00}{##1}}}
\expandafter\def\csname PYGdefault@tok@ss\endcsname{\def\PYGdefault@tc##1{\textcolor[rgb]{0.10,0.09,0.49}{##1}}}
\expandafter\def\csname PYGdefault@tok@sr\endcsname{\def\PYGdefault@tc##1{\textcolor[rgb]{0.73,0.40,0.53}{##1}}}
\expandafter\def\csname PYGdefault@tok@mo\endcsname{\def\PYGdefault@tc##1{\textcolor[rgb]{0.40,0.40,0.40}{##1}}}
\expandafter\def\csname PYGdefault@tok@kn\endcsname{\let\PYGdefault@bf=\textbf\def\PYGdefault@tc##1{\textcolor[rgb]{0.00,0.50,0.00}{##1}}}
\expandafter\def\csname PYGdefault@tok@mi\endcsname{\def\PYGdefault@tc##1{\textcolor[rgb]{0.40,0.40,0.40}{##1}}}
\expandafter\def\csname PYGdefault@tok@gp\endcsname{\let\PYGdefault@bf=\textbf\def\PYGdefault@tc##1{\textcolor[rgb]{0.00,0.00,0.50}{##1}}}
\expandafter\def\csname PYGdefault@tok@o\endcsname{\def\PYGdefault@tc##1{\textcolor[rgb]{0.40,0.40,0.40}{##1}}}
\expandafter\def\csname PYGdefault@tok@kr\endcsname{\let\PYGdefault@bf=\textbf\def\PYGdefault@tc##1{\textcolor[rgb]{0.00,0.50,0.00}{##1}}}
\expandafter\def\csname PYGdefault@tok@s\endcsname{\def\PYGdefault@tc##1{\textcolor[rgb]{0.73,0.13,0.13}{##1}}}
\expandafter\def\csname PYGdefault@tok@kp\endcsname{\def\PYGdefault@tc##1{\textcolor[rgb]{0.00,0.50,0.00}{##1}}}
\expandafter\def\csname PYGdefault@tok@w\endcsname{\def\PYGdefault@tc##1{\textcolor[rgb]{0.73,0.73,0.73}{##1}}}
\expandafter\def\csname PYGdefault@tok@kt\endcsname{\def\PYGdefault@tc##1{\textcolor[rgb]{0.69,0.00,0.25}{##1}}}
\expandafter\def\csname PYGdefault@tok@ow\endcsname{\let\PYGdefault@bf=\textbf\def\PYGdefault@tc##1{\textcolor[rgb]{0.67,0.13,1.00}{##1}}}
\expandafter\def\csname PYGdefault@tok@sb\endcsname{\def\PYGdefault@tc##1{\textcolor[rgb]{0.73,0.13,0.13}{##1}}}
\expandafter\def\csname PYGdefault@tok@k\endcsname{\let\PYGdefault@bf=\textbf\def\PYGdefault@tc##1{\textcolor[rgb]{0.00,0.50,0.00}{##1}}}
\expandafter\def\csname PYGdefault@tok@se\endcsname{\let\PYGdefault@bf=\textbf\def\PYGdefault@tc##1{\textcolor[rgb]{0.73,0.40,0.13}{##1}}}
\expandafter\def\csname PYGdefault@tok@sd\endcsname{\let\PYGdefault@it=\textit\def\PYGdefault@tc##1{\textcolor[rgb]{0.73,0.13,0.13}{##1}}}

\makeatother

\makeatletter
\def\PYG@reset{\let\PYG@it=\relax \let\PYG@bf=\relax%
    \let\PYG@ul=\relax \let\PYG@tc=\relax%
    \let\PYG@bc=\relax \let\PYG@ff=\relax}
\def\PYG@tok#1{\csname PYG@tok@#1\endcsname}
\def\PYG@toks#1+{\ifx\relax#1\empty\else%
    \PYG@tok{#1}\expandafter\PYG@toks\fi}
\def\PYG@do#1{\PYG@bc{\PYG@tc{\PYG@ul{%
    \PYG@it{\PYG@bf{\PYG@ff{#1}}}}}}}
\def\PYG#1#2{\PYG@reset\PYG@toks#1+\relax+\PYG@do{#2}}

\expandafter\def\csname PYG@tok@gd\endcsname{\def\PYG@tc##1{\textcolor[rgb]{0.63,0.00,0.00}{##1}}}
\expandafter\def\csname PYG@tok@gu\endcsname{\let\PYG@bf=\textbf\def\PYG@tc##1{\textcolor[rgb]{0.50,0.00,0.50}{##1}}}
\expandafter\def\csname PYG@tok@gt\endcsname{\def\PYG@tc##1{\textcolor[rgb]{0.00,0.27,0.87}{##1}}}
\expandafter\def\csname PYG@tok@gs\endcsname{\let\PYG@bf=\textbf}
\expandafter\def\csname PYG@tok@gr\endcsname{\def\PYG@tc##1{\textcolor[rgb]{1.00,0.00,0.00}{##1}}}
\expandafter\def\csname PYG@tok@cm\endcsname{\let\PYG@it=\textit\def\PYG@tc##1{\textcolor[rgb]{0.25,0.50,0.50}{##1}}}
\expandafter\def\csname PYG@tok@vg\endcsname{\def\PYG@tc##1{\textcolor[rgb]{0.10,0.09,0.49}{##1}}}
\expandafter\def\csname PYG@tok@m\endcsname{\def\PYG@tc##1{\textcolor[rgb]{0.40,0.40,0.40}{##1}}}
\expandafter\def\csname PYG@tok@mh\endcsname{\def\PYG@tc##1{\textcolor[rgb]{0.40,0.40,0.40}{##1}}}
\expandafter\def\csname PYG@tok@go\endcsname{\def\PYG@tc##1{\textcolor[rgb]{0.53,0.53,0.53}{##1}}}
\expandafter\def\csname PYG@tok@ge\endcsname{\let\PYG@it=\textit}
\expandafter\def\csname PYG@tok@vc\endcsname{\def\PYG@tc##1{\textcolor[rgb]{0.10,0.09,0.49}{##1}}}
\expandafter\def\csname PYG@tok@il\endcsname{\def\PYG@tc##1{\textcolor[rgb]{0.40,0.40,0.40}{##1}}}
\expandafter\def\csname PYG@tok@cs\endcsname{\let\PYG@it=\textit\def\PYG@tc##1{\textcolor[rgb]{0.25,0.50,0.50}{##1}}}
\expandafter\def\csname PYG@tok@cp\endcsname{\def\PYG@tc##1{\textcolor[rgb]{0.74,0.48,0.00}{##1}}}
\expandafter\def\csname PYG@tok@gi\endcsname{\def\PYG@tc##1{\textcolor[rgb]{0.00,0.63,0.00}{##1}}}
\expandafter\def\csname PYG@tok@gh\endcsname{\let\PYG@bf=\textbf\def\PYG@tc##1{\textcolor[rgb]{0.00,0.00,0.50}{##1}}}
\expandafter\def\csname PYG@tok@ni\endcsname{\let\PYG@bf=\textbf\def\PYG@tc##1{\textcolor[rgb]{0.60,0.60,0.60}{##1}}}
\expandafter\def\csname PYG@tok@nl\endcsname{\def\PYG@tc##1{\textcolor[rgb]{0.63,0.63,0.00}{##1}}}
\expandafter\def\csname PYG@tok@nn\endcsname{\let\PYG@bf=\textbf\def\PYG@tc##1{\textcolor[rgb]{0.00,0.00,1.00}{##1}}}
\expandafter\def\csname PYG@tok@no\endcsname{\def\PYG@tc##1{\textcolor[rgb]{0.53,0.00,0.00}{##1}}}
\expandafter\def\csname PYG@tok@na\endcsname{\def\PYG@tc##1{\textcolor[rgb]{0.49,0.56,0.16}{##1}}}
\expandafter\def\csname PYG@tok@nb\endcsname{\def\PYG@tc##1{\textcolor[rgb]{0.00,0.50,0.00}{##1}}}
\expandafter\def\csname PYG@tok@nc\endcsname{\let\PYG@bf=\textbf\def\PYG@tc##1{\textcolor[rgb]{0.00,0.00,1.00}{##1}}}
\expandafter\def\csname PYG@tok@nd\endcsname{\def\PYG@tc##1{\textcolor[rgb]{0.67,0.13,1.00}{##1}}}
\expandafter\def\csname PYG@tok@ne\endcsname{\let\PYG@bf=\textbf\def\PYG@tc##1{\textcolor[rgb]{0.82,0.25,0.23}{##1}}}
\expandafter\def\csname PYG@tok@nf\endcsname{\def\PYG@tc##1{\textcolor[rgb]{0.00,0.00,1.00}{##1}}}
\expandafter\def\csname PYG@tok@si\endcsname{\let\PYG@bf=\textbf\def\PYG@tc##1{\textcolor[rgb]{0.73,0.40,0.53}{##1}}}
\expandafter\def\csname PYG@tok@s2\endcsname{\def\PYG@tc##1{\textcolor[rgb]{0.73,0.13,0.13}{##1}}}
\expandafter\def\csname PYG@tok@vi\endcsname{\def\PYG@tc##1{\textcolor[rgb]{0.10,0.09,0.49}{##1}}}
\expandafter\def\csname PYG@tok@nt\endcsname{\let\PYG@bf=\textbf\def\PYG@tc##1{\textcolor[rgb]{0.00,0.50,0.00}{##1}}}
\expandafter\def\csname PYG@tok@nv\endcsname{\def\PYG@tc##1{\textcolor[rgb]{0.10,0.09,0.49}{##1}}}
\expandafter\def\csname PYG@tok@s1\endcsname{\def\PYG@tc##1{\textcolor[rgb]{0.73,0.13,0.13}{##1}}}
\expandafter\def\csname PYG@tok@sh\endcsname{\def\PYG@tc##1{\textcolor[rgb]{0.73,0.13,0.13}{##1}}}
\expandafter\def\csname PYG@tok@sc\endcsname{\def\PYG@tc##1{\textcolor[rgb]{0.73,0.13,0.13}{##1}}}
\expandafter\def\csname PYG@tok@sx\endcsname{\def\PYG@tc##1{\textcolor[rgb]{0.00,0.50,0.00}{##1}}}
\expandafter\def\csname PYG@tok@bp\endcsname{\def\PYG@tc##1{\textcolor[rgb]{0.00,0.50,0.00}{##1}}}
\expandafter\def\csname PYG@tok@c1\endcsname{\let\PYG@it=\textit\def\PYG@tc##1{\textcolor[rgb]{0.25,0.50,0.50}{##1}}}
\expandafter\def\csname PYG@tok@kc\endcsname{\let\PYG@bf=\textbf\def\PYG@tc##1{\textcolor[rgb]{0.00,0.50,0.00}{##1}}}
\expandafter\def\csname PYG@tok@c\endcsname{\let\PYG@it=\textit\def\PYG@tc##1{\textcolor[rgb]{0.25,0.50,0.50}{##1}}}
\expandafter\def\csname PYG@tok@mf\endcsname{\def\PYG@tc##1{\textcolor[rgb]{0.40,0.40,0.40}{##1}}}
\expandafter\def\csname PYG@tok@err\endcsname{\def\PYG@bc##1{\setlength{\fboxsep}{0pt}\fcolorbox[rgb]{1.00,0.00,0.00}{1,1,1}{\strut ##1}}}
\expandafter\def\csname PYG@tok@kd\endcsname{\let\PYG@bf=\textbf\def\PYG@tc##1{\textcolor[rgb]{0.00,0.50,0.00}{##1}}}
\expandafter\def\csname PYG@tok@ss\endcsname{\def\PYG@tc##1{\textcolor[rgb]{0.10,0.09,0.49}{##1}}}
\expandafter\def\csname PYG@tok@sr\endcsname{\def\PYG@tc##1{\textcolor[rgb]{0.73,0.40,0.53}{##1}}}
\expandafter\def\csname PYG@tok@mo\endcsname{\def\PYG@tc##1{\textcolor[rgb]{0.40,0.40,0.40}{##1}}}
\expandafter\def\csname PYG@tok@kn\endcsname{\let\PYG@bf=\textbf\def\PYG@tc##1{\textcolor[rgb]{0.00,0.50,0.00}{##1}}}
\expandafter\def\csname PYG@tok@mi\endcsname{\def\PYG@tc##1{\textcolor[rgb]{0.40,0.40,0.40}{##1}}}
\expandafter\def\csname PYG@tok@gp\endcsname{\let\PYG@bf=\textbf\def\PYG@tc##1{\textcolor[rgb]{0.00,0.00,0.50}{##1}}}
\expandafter\def\csname PYG@tok@o\endcsname{\def\PYG@tc##1{\textcolor[rgb]{0.40,0.40,0.40}{##1}}}
\expandafter\def\csname PYG@tok@kr\endcsname{\let\PYG@bf=\textbf\def\PYG@tc##1{\textcolor[rgb]{0.00,0.50,0.00}{##1}}}
\expandafter\def\csname PYG@tok@s\endcsname{\def\PYG@tc##1{\textcolor[rgb]{0.73,0.13,0.13}{##1}}}
\expandafter\def\csname PYG@tok@kp\endcsname{\def\PYG@tc##1{\textcolor[rgb]{0.00,0.50,0.00}{##1}}}
\expandafter\def\csname PYG@tok@w\endcsname{\def\PYG@tc##1{\textcolor[rgb]{0.73,0.73,0.73}{##1}}}
\expandafter\def\csname PYG@tok@kt\endcsname{\def\PYG@tc##1{\textcolor[rgb]{0.69,0.00,0.25}{##1}}}
\expandafter\def\csname PYG@tok@ow\endcsname{\let\PYG@bf=\textbf\def\PYG@tc##1{\textcolor[rgb]{0.67,0.13,1.00}{##1}}}
\expandafter\def\csname PYG@tok@sb\endcsname{\def\PYG@tc##1{\textcolor[rgb]{0.73,0.13,0.13}{##1}}}
\expandafter\def\csname PYG@tok@k\endcsname{\let\PYG@bf=\textbf\def\PYG@tc##1{\textcolor[rgb]{0.00,0.50,0.00}{##1}}}
\expandafter\def\csname PYG@tok@se\endcsname{\let\PYG@bf=\textbf\def\PYG@tc##1{\textcolor[rgb]{0.73,0.40,0.13}{##1}}}
\expandafter\def\csname PYG@tok@sd\endcsname{\let\PYG@it=\textit\def\PYG@tc##1{\textcolor[rgb]{0.73,0.13,0.13}{##1}}}

\makeatother
\usepackage[bookmarks=true]{hyperref}
\usepackage[numbers]{natbib}
\usepackage{color}
\newcommand{\IF}{{\sf if}\:}
\newcommand{\THEN}{\:{\sf then}\:}
\newcommand{\ELSE}{\:{\sf else}\:}
\theoremstyle{definition}

\title{Integrative adaptive indexes from noisy routine haematological markers can predict and discriminate health status and biological age}
\author[1]{Santiago Hern\'andez-Orozco}
\author[1]{Abicumaran Uthamacumaran}
\author[1]{Francisco Hern\'andez Quiroz}
\author[1,2]{Kourosh Saeb-Parsy}
\author[1,3,4,5]{\\Hector Zenil\thanks{Corresponding author: hector.zenil@kcl.ac.uk}}
\affil[1]{\normalsize{ }Oxford Immune Algorithmics, Oxford University Innovation, Oxford, U.K.}
\affil[2]{\normalsize{ }Department of Surgery, University of Cambridge, Cambridge, U.K.}
\affil[3]{\normalsize{ }British Society for Research on Ageing, U.K.}
\affil[4]{\normalsize{ }King's Institute for Artificial Intelligence, King's College London, U.K.}
\affil[5]{\normalsize{ }Algorithmic Dynamics Lab, Research Departments of Biomedical Computing and Digital Twins, School of Biomedical Engineering and Imaging Sciences, King's College London, U.K.}

\date{}

\begin{document}
  \maketitle

\vspace{-1cm}
\begin{abstract}
For more than two decades, advances in personalised medicine and precision healthcare have largely been based on genomics and other omics data. These strategies aim to tailor interventions to individual patient profiles, promising greater treatment efficacy and more efficient allocation of healthcare resources. Here, we show that widely collected common haematological markers can reliably predict and discriminate individual chronological age and health status from even noisy sources. Our analysis includes synthetic and real retrospective patient data, including medically relevant and extreme cases. We combine fully explainable risk assessment scores with machine intelligence to focus on clinically significant patterns and characteristics without functioning as a ``black box" model and allowing interpretation and control tested on the US CDC NHANES database (100\,000 participants) and validated with the UK Biobank (500\,000 participants). Despite the noisy nature of these databases due to self-reporting and sparse data, the results remained relevant and statistically significant. Unlike current biological ageing indicators, this approach may offer rapid, and scalable implementations for personalised, precision and predictive approaches to healthcare and medicine without or before requiring other specialised, uncommon, or costly tests.\\


\end{abstract}

\maketitle 



\section{Introduction}

In a recent retrospective study, Complete Blood Count tests of 12,000 adults over 20 years showed that individuals are stable over time while being healthy and have unique characteristics that effectively act like health fingerprints, specific patterns or personalised signatures that when deviate from their personal baselines are indicative of health and even mortality~\cite{harvard}.

These personalised benchmarks can significantly improve early disease detection by providing a better and more precise picture of individual health as suggested in various immunological studies~\cite{brodin1,brodin2}. In contrast to previous studies~\cite{harvard}, the present work focusses on defining a minimal, interpretable index or risk assessment score grounded in routine haematological and circulatory biomarkers. Taking advantage of population-normalised Complete Blood Count (CBC) parameters, our approach achieves discriminatory performance not explored in~\cite{harvard} while remaining computationally efficient, reproducible and transparent for clinical purposes. The index therefore is the proactive natural consequence of what was reported in ~\cite{harvard} regarding personal setpoints and clinically relevant deviations even within normal population reference values.

We investigate whether these findings can be leveraged for risk assessment by identifying clinically relevant immune patterns. We further demonstrate that the same test allows robust detection of patterns predictive of age and health status, even in large, retrospective, and noisy datasets. In this paper~\cite{naturescore}, it was shown that 60 blood markers were predictive of biological age. While our paper precedes theirs as posted online, in this one we show that only a small subset of those markers are actually necessary for predictive purposes and that ad hoc artificial weighting of markers is not necessary.

This contribution has two main goals. First, we evaluate whether immune scores derived from routine blood tests can predict and triage (i.e., assessment of priority for healthcare services) health status using a translatable and explainable approach. Second, we assess whether immune-derived age diverges from chronological age in a clinically meaningful way. The results are validated across the US NHANES and UK Biobank cohorts to assess generalisability. 

Building on this need, we propose immune-based indexes derived from standard blood tests as quantitative predictors of health risk and immune ageing, offering valuable tools for preventive and precision medicine. Risk assessment tools are often needed to improve patient safety and patient outcomes.

Risk assessment scores used for initial patient classification or diagnosis and triaging are often qualitative and inter-subjective. Robust quantitative tools with clinical relevance are required to optimise health monitoring choices across all levels of healthcare systems. Digital medicine aims to improve the quality of patient care and outcome trajectories through clinical informatics, such as risk assessment metrics based on clinical data.

Risk scores are valuable predictive tools that, when implemented, can be used by clinicians with positive results, leading to better patient outcomes in precision medicine. Risk assessment measures for patient triage are often needed to optimise patient pathways and health monitoring across all levels of healthcare.

However, most scores are defined for very specific conditions, such as heart disease, cardiovascular disease (CVD) or venous thromboembolism (VTE), to name just a few. Others reflect factors such as risk of severe disease~\cite{1,2}, clinical acuity~\cite{3,4}, and long-term outcomes~\cite{5,6}.

For example, the Intermountain risk index~\cite{intermountain} provides a Complete Blood Count (CBC/FBC) risk index based on the size of red blood cells, shown to be associated with bleeding and in-hospital mortality. Originally used as an anaemia predictor~\cite{13}, it is often reported in FBC/CBC panels.

As precision healthcare and digital medicine continue to advance, these risk scores will likely integrate more advanced data-driven approaches to improve prediction accuracy and ensure relevance across a broad spectrum of clinical conditions.

\subsection{Biological Ageing}

Biological ageing clocks have been explored from various perspectives and approaches from skin to epigenetic~\cite{levine2018epigenetic} using blood to microbime~\cite{mitrabio2023} (skin and gut) to proteomic RNA~\cite{lehallier2019undulating}.

Epigenetic measurements are limited in scope due to limited range of molecules measures, while proteomics offers a comprehensive means for measuring ageing. However, whole-body proteomic signals are noisy due to different organ contribution to the full signal, and skin ageing for proteomics leads to tissue damage.

Gut microbiome and other age-related scores, require specialised tests and not routine tests. For example, immune age-related scores based on inflammation markers have  traditionally required highly specialised and therefore uncommon and expensive tests that are not suitable for mass adoption or mass deployment in current clinical pathways, as they are difficult to scale to cover large populations~\cite{alberro2021inflammaging, ferrucci2018inflammageing,
fulop2019biomarkers,
alpert2019immuneage, fulop2021immunology}. 

In contrast, the most popular medical test, a blood test, the CBC or Full Blood Count, has been found to often contain enough data to provide key clinically information~\cite{cbc,jupiter,harvard}, and even specific disease signatures~\cite{lord2024covid}. Although it is a popular low-cost laboratory test that is almost universally used, its risk predictive information content is underused~\cite{cbc2,harvard}, often in favour of more sophisticated omics tests that are also expensive and are not currently scalable.

Routine blood tests have been associated with the prediction of short-term mortality and with the ability to significantly improve emergency department triage~\cite{routine}.

Most risk scores depend on multiple external factors that are unique to a single test result in and of itself, and have proven useful when available, even in the context of the immune system~\cite{factorial}. Several of these scores are based on factors such as physiological data, medical and family history, and lifestyle choices, and reference a range of things, from age and sex to drinking and smoking habits.

Although numerical scores and colour-based scores are relatively common, no single general numerical and colour-based blood test risk assessment index exists that is based on a combination of distances from the population and personalised average values, in particular for CBC/FBC and immune-related data, used to guide Machine Learning approaches for discrimination of individual health predictors.

Here, we introduce a computational intelligence approach based on or guided by explainable risk stratification scores to quantify any number of blood markers relevant to health and immune age. The risk model introduced is not data expensive when run ab initio, and yet it is shown to be informative and predictive, and implementation remains interpretable even in combination with machine and deep learning, while allowing for clinical adaptation based on patient profiles. 

\textcolor{red}{
Further, from a mechanistic viewpoint, age-associated perturbations or changes in CBC analytes may reflect early detection markers of ageing-related disease progression and immunosenescence processes, including myeloid skewing, reduced lymphopoiesis, increased inflammatory tone (``inflammaging''), clonal hematopoiesis (or of intermediate potential) and epigenetic or genetic alterations in hematopoietic stem cell population dynamics, erythropoiesis, and platelet activation ~\cite{alberro2021inflammaging, ferrucci2018inflammageing,jaiswal_clonal_CHIP, abegunde_tet2_tnfa_2018, bruserud_hematopoiesis_inflammaging}.}
\textcolor{red}{
Chronic psychosocial and physiological stressors have also been linked to sustained immune activation and epigenetic remodeling of hematopoietic lineages, potentially accelerating immune-aging phenotypes detectable in peripheral blood indices~\cite{geiger_inflammation_epigenetic_2020, zhu_inflammation_epigenetics_aging_2021}. Emerging discussions in the brain-gut-immune axes are key converging points. CBC indices have been independently associated with systemic inflammatory burden, immune-microenvironmental remodelling, epigenetic alterations, and vascular ageing in prior literature.~\cite{barrett_stress_immune, oriordan_gut_immune_brain_2025}.}

\textcolor{red}{Thus, the composite index integrates distributed but biologically grounded ageing-related predictive signatures.}

\section{Methods}

A Complete Blood Count (CBC) or Full Blood Count (FBC) test is the most popular blood test with over 4 Billion of them performed every year world wide, 500 million only in the U.S.

It traditionally includes the total number of red blood cells (RBCs); the total number of white blood cells (WBCs) and platelets (PLT); haemoglobin (HGB); haematocrit (HCT); estimations related to concentration, shape and volume of red cells (MCV, RDW, MCH and MCHC) and counts of different WBC subtypes, including Lymphocytes (LYMP), Monocytes (MONO), Neutrophils (NEUT), Eosinophils (EOS) and Basophils (BAS).

A CBC is a routine blood test used to assess overall health and detect a wide range of disorders, including anaemia, infection and leukaemia. A CBC test is deeply related to the immune system and is a key diagnostic tool that can provide information on a variety of diseases and conditions, including infection, inflammation, cancer, anaemia, heart disease, and more. CDC NHANES data were used to estimate deficiencies and toxicities of specific nutrients in the population and subgroups in the US, to extract population reference data, and to estimate the contribution of diet, supplements, and other factors to whole blood levels of nutrients. This repository includes CBC data can be used for research purposes and is publicly available~\cite{hanes}.

A marker, analyte or test parameter is a potentially non-mutually exclusive property related to a blood test. Analytes or indices can include protein-based substances, antibodies, biochemical entities and any other product or sub-product of cellular function related to the blood or to the immune system. For example, in a Full (or Complete) Blood Count, there are usually about 15 indices, although the proposed index is not limited to any particular set or number of such analytes or features, as it can include cell size, cell and cell nuclei morphology, and cell maturity among a wide range of indices/analytes beyond cell counts.


The scores introduced take the form of quantitative measures that aggregate data from a set of numerical blood test results, combining and condensing them into single numerical values that can facilitate rapid classification, sorting, and assessment of immune age. The main purpose of the index is to learn from, monitor and provide a quick assessment of divergence from personalised health values or individual setpoints shown to hold high clinical value~\cite{harvard}. 

We will call this risk assessment index the ``immune index", given its connection to the immune system through blood cell subpopulation counts, specifically as registered in the most popular blood test, FBC or CBC. These scores will then be combined with more traditional Machine Learning (ML) and Deep Learning (DL) approaches, where the scores will help guide the search for specific features, in what constitutes a neuro-symbolic approach to health risk stratification and immune age characterisation in digital medicine. 

The immune index constitutes a dimensional reduction technique based on a single real-value number and a colour scheme that takes a multidimensional blood test space with dimension size equal to the number of analytes, where each value of a marker or analyte corresponds to the coordinate of that value in that dimension. The index itself can be seen as the norm of a suitable vector transformation that pinpoints the health status of the patient in that space, relative to the test and the set of markers or analytes. The full mathematical description is given in the Supplementary Information.

Thus, the numerical vector value integrates all the analyte's spatial dimensions. We call the associated vector space for blood cell analytes the \emph{multidimensional immune space}, or \emph{immune space} for short.


\subsection{Numerical index and Age Prediction}

\textcolor{black}{The following abbreviations are used throughout the paper: NIS (Numerical Immune Score), CCIS (Colour-Coded Immune Score), NCCIS (Normalised Colour-Coded Immune Score), and Immune Index (collective term encompassing all variants). In brief, the NIS represents the raw numerical formulation based solely on analyte deviations; the CCIS extends this by implementing a categorical colour-coding system for visually assisted interpretability; and the NCCIS normalises these values across population-level reference intervals, improving cross-cohort comparability. The Immune Index denotes the general framework comprehensive of all three forms.(See Table~\ref{tab:index_definitions})}

\textcolor{red}{
While CBC measurements may exhibit slight calibration differences across institutions, both NHANES and the UK Biobank inherently incorporate multi-site variability. Thereby, the data embeds real-world technical heterogeneity into the training distributions. Population-level normalization lowers systematic variations, and the adaptive personalised baseline further reduces sensitivity to fixed calibration bias. In prospects, if laboratory-specific calibration coefficients are available, the framework can incorporate them as buffers or corrections prior to index computation, preserving generalisability while maintaining clinical interpretability.
}

\begin{table}[H]
\centering
\begin{tabular}{p{3cm} p{8cm}}
\toprule
\textbf{Abbreviation} & \textbf{Description} \\ 
\midrule
\textbf{NIS}\\(Numerical Immune Score) & Quantitative index derived from haematological analyte deviations; represents raw numerical immune deviation without categorical scaling. \\ 
\textbf{CCIS}\\(Colour-Coded Immune Score) & Categorical extension of NIS using colour-coded bins to improve interpretability and visual comparison between individuals. \\ 
\textbf{NCCIS}\\(Normalised Colour-Coded Immune Score) & Normalised version of CCIS adjusted to population reference intervals (NHANES, NHS-UK), allowing generalisation and comparison across diverse datasets and laboratories. \\ 
\textbf{Immune Index} & Collective term referring to the overall framework encompassing NIS, CCIS, and NCCIS for generalisable immune health quantification. \\ 
\bottomrule
\end{tabular}
\caption{\textbf{Summary of immune-related indices and their definitions.}  
Each variant of the immune index represents a step toward increased interpretability and cross-dataset comparability, forming a hierarchical structure from raw numerical scoring (NIS) using the CBC analytes to population-normalised colour-coded indices (NCCIS).}
\label{tab:index_definitions}
\end{table}

This first formulation of the index takes the form of a linear function because it does not take into account the possible interactions between analytes. CBC blood cell counts can be affected by interactions that reflect immune system responses, but the NIS index will be defined by taking into consideration the equally weighted compounded distances of each analyte/index in a bag of analyte/index results of interest from upper and lower population reference values. In addition, the NCCIS displays a warning colour based on the number of standard deviations away from the reference values. An adaptive version is then applied to the learnt reference values over time, based on the patient persistence values over time dynamically. 

The sex groups were separated by health status determined by dividing immune scores and ages by the healthy range thresholds indicated in Table ~\ref{table:nhs} (see Sup. Inf.).

These thresholds are as follows: Red Blood Cell (RBC) count (between 3.8 to 5.8), and the following healthy ranges for the multivariate measure: normal Mean Corpuscular Haemoglobin (MCH) (27-33), Haematocrit (HCT) (37-47), lymphocytes (LYMP) (1.0-4.0), and neutrophils (NEUT) (1.8-7.5). Above or below these healthy count ranges, the individuals were categorised into unhealthy groups.

In addition, error rates analysis for age prediction between actual ages and immune index computed ages was performed using these binned subsets. Zero counts were removed from the NHANES data prior to all analysis. \textcolor{black}{Zero values were excluded only when they represented non-physiological or missing data artifacts (e.g., unrecorded analytes). True physiological zeros, where relevant, were retained.} However, for the quantitative analysis, sex was replaced by 1 for males and 2 to denote females. Table~\ref{hanes:table:means:sd} (see Sup. Inf.) highlights sex-specific differences in immune analyte values, for reference. \textcolor{black}{Because the NHANES and UK~Biobank datasets compile data from multiple laboratories, we standardized all analyte values using the corresponding national population reference intervals (NHANES and NHS-UK) prior to computing the indices. This population-level normalization mitigates inter-laboratory cut-off variability and supports cross-site comparability.}


The immune system plays an important role in protecting against infections and maintaining human health. Blood is the medium in which most immune cells circulate through the body.

Ageing is associated with complex changes and dysregulation of cellular processes.  Nine tentative hallmarks that represent common denominators of ageing in different organisms have been described~\cite{Ref1}. As lifespan increases in many countries, there is a concomitant increase in age-related morbidities, particularly cardiovascular disease, and increased susceptibility to seasonal infections, cancers, and neurodegenerative disorders. A unifying link between the higher rates of these disparate diseases observed in ageing populations is the progressive decline in immune function. 

It has been postulated that inflammation plays a critical role in the regulation of physiological ageing. Inflammatory components of the immune system are often chronically elevated in aged individuals, a phenomenon that has been termed ``inflammageing''~\cite{Ref2, Ref3, Ref4, Ref5, Ref6, alpert2019immuneage}. However, the dynamics of this process at the individual level have not been characterised, hindering quantification of an individual's `immune age'. In previous work~\cite{alpert2019immuneage}, various `omics' technologies are used to capture population- and individual-level changes in the human immune system in individuals of different ages sampled longitudinally over a nine-year period. They observed high inter-individual variability in the rates of change of cellular frequencies that was dictated by the individual's baseline values. This allowed identification of steady-state levels toward which a cell subset converged and the ordered convergence of multiple cell subsets toward older adult homeostasis.

In~\cite{Ref8}, an IMM-AGE index was described that captured an individual's immune-ageing process. The IMM-AGE index was correlated with age, yet it captured additional metrics such as cell-cytokine response better than chronological age.

Here, we present a method where a single blood test, as in a Complete Blood Count (CBC) of 13 blood analytes, provides insight into immune age compared to chronological age. \textcolor{black}{Chronological age simply indicates the age of the individual, determined by the length of time someone has lived or has lived since their birth. This term should not be confused with the context of other literature, such as age estimates predicted by DNA methylation patterns from the Horvath algorithm.}


The UK Biobank is a long-term prospective nationwide study that houses the deidentified biological samples and health-related data of half a million people. Between 2006 and 2010 volunteers aged 40 to 69 years were recruited from all over Great Britain and consented to share their health data and to be followed for at least 30 years thereafter; with the objective of allowing scientific discoveries to be made in the prevention, diagnosis, and treatment of disease~\cite{biobank1,biobank2}.

We tested the NIS index (which is the NCCIS index without the warning colouring component, see Supp. Inf.), to separate large disease categories in the UK Biobank.

The database does not contain a clean healthy control group, meaning that all participants had potentially some reported health problem, making our separation results more relevant given that we had to create a Rest group different from the large disease categories identified as a noisy control group.

\section{Results}


The results show that routine and common markers from CBC tests can be exploited for personalised healthcare and precision medicine. That CBC markers can be grouped into numerical scores that can be used for triaging purposes even from noisy self-reported sources, such as in the NHANES database (containing 100\,000 unfiltered surveys and tests). Filtering by normal and abnormal results improves the detection of health-related patterns and significantly improves the classification of health status, demonstrating that the approach performs better under less noisy conditions, as expected.

In contrast to the behaviour of the immune index, NCCISS is a non-linear function that displays a \textit{step-wise} behaviour by design~\ref{Score1and2:figure} (Sup Inf.), while remaining independent of the number of analytes.

To assess group differences in immune scores, we first conducted a graphical analysis of the residuals followed by Shapiro–Wilk tests to evaluate normality. Due to violations of normality assumptions and the ordinal nature of the NCCIS score, we employed the non-parametric Kruskal–Wallis test to assess overall group differences. All comparisons were statistically significant (p $<$ 0.0001). To evaluate differences between specific group pairs, we used Mann–Whitney U tests with Bonferroni correction for multiple comparisons, retaining significance across all contrasts.

\begin{table}[H]
\centering
\begin{tabular}{l c c}
\toprule
\textbf{Comparison} & \textbf{Sample Size (n)} & \textbf{p-value} \\ 
\midrule
Healthy vs Unhealthy & 7007 vs 6074 & $< 0.0001$ \\ 
Self-reported Healthy vs Unhealthy & 11976 vs 6074 & $< 0.0001$ \\ 
Healthy vs Anaemia & 7007 vs 980 & $< 0.0001$ \\ 
Healthy vs HIV & 7007 vs 62 & $< 0.0001$ \\ 
Healthy vs Hodgkin’s Disease & 7007 vs 37 & $< 0.0001$ \\ 
Healthy vs Leukaemia & 7007 vs 18 & $< 0.0001$ \\ 
Healthy vs Blood Cancer & 7007 vs 8 & $< 0.0001$ \\ 
\bottomrule
\end{tabular}
\caption{\textbf{Group-wise comparison of NIS mean values.}
Kruskal–Wallis tests revealed significant differences across all groups, followed by Mann–Whitney U tests with Bonferroni correction for pairwise comparisons.  Values illustrate the robustness of the NIS in differentiating between healthy and disease cohorts.}
\label{nis:anova}
\end{table}

\begin{table}[H]
\centering
\begin{tabular}{l c c}
\toprule
\textbf{Comparison} & \textbf{Sample Size (n)} & \textbf{p-value} \\ 
\midrule
Healthy vs Unhealthy & 7007 vs 6074 & $< 0.0001$ \\ 
Self-reported Healthy vs Unhealthy & 11976 vs 6074 & $< 0.0001$ \\ 
Healthy vs Anaemia & 7007 vs 980 & $< 0.0001$ \\ 
Healthy vs HIV & 7007 vs 62 & $< 0.0001$ \\ 
Healthy vs Hodgkin’s Disease & 7007 vs 37 & $< 0.0001$ \\ 
Healthy vs Leukaemia & 7007 vs 18 & $< 0.0001$ \\ 
Healthy vs Blood Cancer & 7007 vs 8 & $< 0.0001$ \\ 
\bottomrule
\end{tabular}
\caption{\textbf{Group-wise comparison of NCCIS mean values.}
Kruskal–Wallis tests revealed significant differences across all cohorts, 
followed by Bonferroni-corrected Mann–Whitney U tests for pairwise comparisons. 
These results confirm the NCCIS as a robust and generalizable metric for distinguishing immune and haematological states. 
Effect sizes are provided in Table~\ref{tab:effect_sizes}.}
\label{nccis:anova}
\end{table}

\begin{table}[htbp]
\centering
\begin{tabular}{l c c}
\toprule
\textbf{Parameter} & \textbf{Cohen’s d} & \textbf{p-value} \\ 
\midrule
Immune Age (by Health) & 2.28 & $< 0.0001$ \\ 
NIS Index (by Health) & 1.36 & $< 0.0001$ \\ 
Immune Age (by Sex) & 0.11 & $< 0.0001$ \\ 
\bottomrule
\end{tabular}
\caption{\textbf{Effect sizes of immune-related indices.}
Cohen’s \textit{d} values quantify the magnitude of group differences in immune parameters across health and sex categories. 
Results demonstrate strong statistical and practical significance of the immune indices, highlighting their discriminative capacity for biological age and health status.}
\label{tab:effect_sizes}
\end{table}

The Immune Age by Health Status shows a strong Cohen's d (2.28)~\ref{tab:effect_sizes}, indicating that it is a strong and robust measure to differentiate health status and predict health risks based on blood parameters with precision. In contrast, the weaker effect size by sex (0.11) suggests that sex is unlikely to be a confounding factor, reinforcing the precision of the Immune Age prediction based on health status.

\begin{figure}[p]
    \centering

    \scalebox{0.6}{\includegraphics{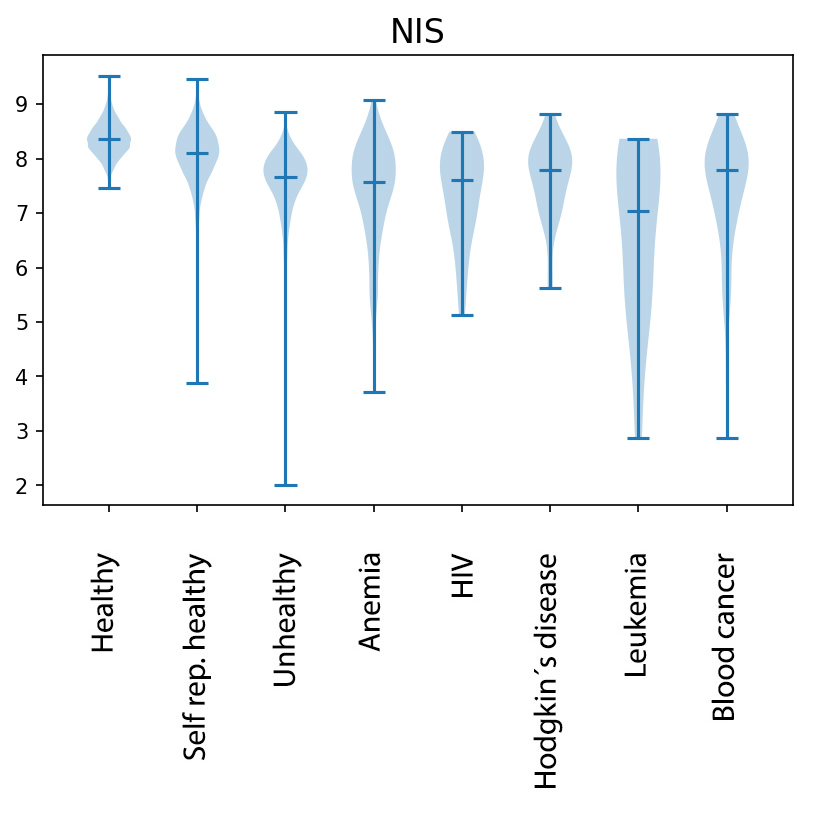}}

    \vspace{0.3cm}

    \scalebox{0.6}{\includegraphics{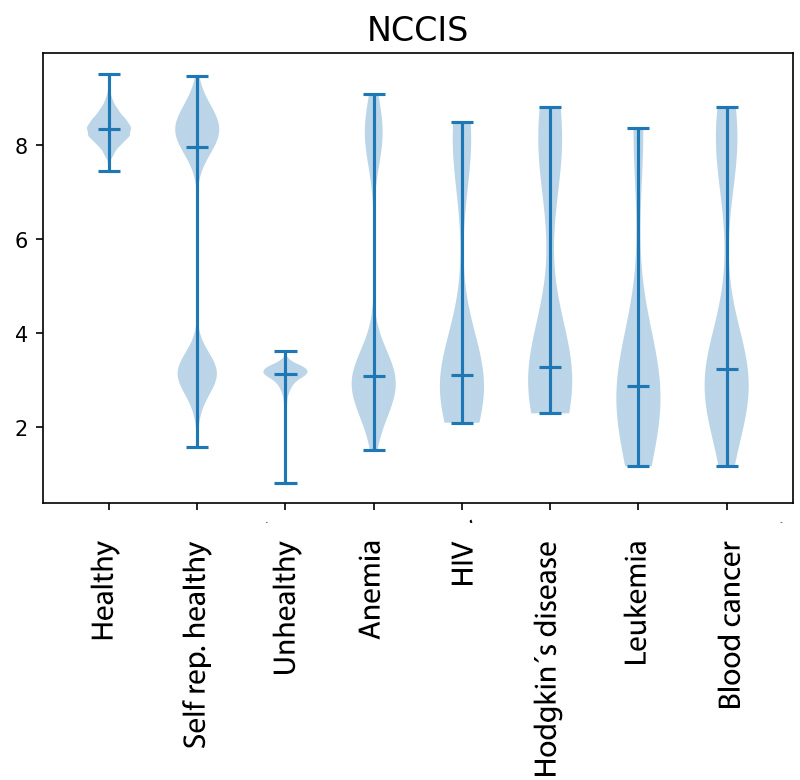}}

    \caption{Distribution of the Normalised Colour-Coded Immune index (NCCIS) and Numerical Immune Score (NIS).}
    \label{nccis:test}
\end{figure}

\begin{figure}[p]\ContinuedFloat
    \caption{\textcolor{black}{Distribution of the Normalised Colour-Coded Immune index (NCCIS) from 0 to 10 ($y$ axis) among adult individuals in the NHANES 2000-2018 database. Individuals are defined as \textit{Healthy} if all their analytes are within the normal ranges as defined by the NHS (\ref{table:nhs}) (see Sup. Inf.) and \textit{Unhealthy} otherwise. \textit{Self rep. healthy} comprises individuals who self-reported health of 3 (``Good'') or less. HIV is an example of chronic infection while all the diseases can fall in the category of immune and blood-related conditions. While separation between overtly healthy and disease groups may be expected, the novelty of our proposed metric lies in its ability to provide a generalizable, quantitative, and interpretable single numerical score that scales across heterogeneous conditions and noisy datasets, offering a standardized baseline for predictive stratification. The Numerical Immune Score (NIS) shows a clear separation between healthy and disease groups on the same standardized scale, with healthy individuals clustering at higher scores and more confined distributions. Disease states exhibit broader shapes and hence, greater variability and declining scores reflecting increased immune dysregulation. Overlap across conditions highlights NIS’s sensitivity to graded immune perturbations rather than binary health–disease discrimination.}}
\end{figure}

\clearpage

Table~\ref{table:cases:test} shows examples of CCIS and its corresponding NCCIS values for a sample of real-life and artificial cases. The second column of the table shows the NIS, for purposes of comparison.

It is worth noting that these examples were generated using the NHS values for normal healthy subjects and based on the provisional assumption that the distance between the upper and normal ranges is twice the standard deviation.

The examples were taken from \cite{upmc} as they were intended as very preliminary tests. We are aware that their source clearly states that they are meant only for teaching purposes. Here, they are used for purposes of illustration only, and the development of the NIS and NCCIS is not dependent on them.

In Table~\ref{tab:effect_sizes}, we provide Cohen's d values that suggest practical applicability of the index, making it clinically relevant for triaging purposes even at the level of individual classification and prediction.

One can observe how the NCCIS values meet the requirements we have set in advance of their definition. We tested NIS and NCCIS against real data from the 2003-2016 National Health and Nutrition Examination Survey (NHANES), provided in~\cite{hanes}. One measure of success was the ability to show how accurately the index discriminates between healthy subjects and patients suffering from various diagnosed diseases.

We calculated the distribution, mean values, and standard deviations for each of the analytes (see Fig.~\ref{Score1and2:figure} (top)). We used means and standard deviations from the data to define the metrics for both the NIS and NCCIS. Finally, we calculate NIS and NCCIS for thousands of cases in the NHANES database to find individual scores. The resulting values were grouped as belonging to healthy individuals or to a selected list of common diseases. Furthermore, Fig.~\ref{Score1and2:figure}(Bottom) (see Suppl. Info) shows how machine learning predictions can dynamically adjust the expected range of analyte values, detecting stable trends while identifying outliers, thus improving precision in monitoring patient health.

To maximise the discriminatory power of both scores, we tried different alternative combinations of means, normal ranges, and standard deviations, taking either these values from NHS values as of early 2022 or inferring them from the NHANES database ~\ref{Score1and2:figure}(top). We settled for (1)~NHS normal healthy reference values and means and standard deviations for the NIS; (2)~means and standard deviations calculated from the NHANES database, and normal ranges of NHS for NCCIS. NHANES means and standard deviations are shown in Table~\ref{colours} (see Sup. Inf.)


\textcolor{black}{The table \ref{nis:anova} presents the number of data points used, together with the respective Mann Whitney U test results to test pairwise statistical significance}. In principle, smaller sample sizes in one group could lead to unstable variability in p-values. However, the consistent statistical significance of all comparisons and the large effect sizes shown in Table~\ref{tab:effect_sizes} indicate that the results are robust. Furthermore, Cohen's d is based on the entire dataset, using group means, variances, and sample sizes to compute the pooled standard deviation, thereby accounting for sample sizes in the reported effect sizes. The large effect size of immune age indicates that it significantly distinguishes healthy from unhealthy individuals, reflecting its statistical and practical significance in discrimination and prediction of health status. It should be noted that the self-reported conditions were not independently confirmed. In addition, the survey participants did not distinguish between current or past diagnoses. This could be expected to introduce some noise, as some currently healthy people will be labelled as having a condition and some people with conditions will not have been diagnosed. We hope to improve accuracy by filtering the data or by adding other data sources in the future.

However, the NIS was able to discriminate between healthy and nonhealthy individuals, since most healthy individuals are clustered around very low NIS values. In contrast, different conditions produced higher NIS values on average.

The method is shown to be valid for predicting immune profiles/trends (variation over time) observed in diverse patient groups when considering socio-clinical metadata parameters such as age and sex, thereby suggesting their clinical applicability for purposes such as triaging and fast screening and extensions to deviant clinical cases in prospective studies. In Fig.~\ref{nccis:test}, it is shown that reported medical conditions produced a higher average NCCIS value, but also induced some anomalous clustering. Furthermore, the distribution of scores was even wider than that of the NIS. We therefore conclude that the NCCIS did enhance the insights already gleaned from the NIS and the CCIS.Separation of groups of self-reported data suggests that they could be separated from those that were not actually healthy at the time of testing.

\textcolor{black}{The self-reported health information may not always correspond to objectively verified clinical status. Thus, the observation possibly reflects statistical discrepancies between self-reported and haematological patterns rather than confirmed diagnostic misclassification. Accordingly, while individuals who self-reported not being healthy generally exhibited blood profiles consistent with non-healthy patterns, this is suggested as a potential interpretation rather than clinical distinction.} The number of individuals in each cohort is shown in Table~\ref{nccis:anova}. One possible explanation for the bimodality of some of these plots is the nature of the source data as self-reported, where respondents may or may not have been ill at the time of giving their answers, or from the application of the separating cutoff value from NIS.

\begin{table}[H]
\newcommand{\adjustline}{\vrule width0pt height1ex depth0.75ex\relax}
\caption{Different tested scores perform differently according to each index's definition.}

\begin{center}

\begin{tabular}{c|r|c|r}
\toprule
\textbf{Individual} & \textbf{NIS} & \textbf{CCIS} & \textbf{NCCIS} \\
\midrule
\parbox[c]{0.5\textwidth}{\noindent\flushleft
Adult male with all values at the mean\adjustline} & 10.00 & Green & 10.00 \\
\hline
\parbox[c]{0.5\textwidth}{\noindent\flushleft
Adult male values slightly removed from mean within normal healthy reference interval\adjustline} & 8.16 & Green & 8.16 \\
\hline
\parbox[c]{0.5\textwidth}{\noindent\flushleft
Adult male with all abnormal values\adjustline} & 0.00 & Red & 0.00 \\
\hline
\parbox[c]{0.5\textwidth}{\noindent\flushleft
Adult male with leukocytosis and diabetes\adjustline} & 5.85 & Red & 2.40 \\
\hline
\parbox[c]{0.5\textwidth}{\noindent\flushleft
Adult male with pancytopenia\adjustline} & 4.46 & Red & 1.82 \\
\hline
\parbox[c]{0.5\textwidth}{\noindent\flushleft
Adult male with mycosis\adjustline} & 5.31 & Red & 2.17 \\
\hline
\parbox[c]{0.5\textwidth}{\noindent\flushleft
Adult female with fatigue\adjustline} & 6.93 & Amber & 4.57 \\
\hline
\parbox[c]{0.5\textwidth}{\noindent\flushleft
Adult female with shortness of breath\adjustline} & 3.30 & Red & 1.32 \\
\hline
\parbox[c]{0.5\textwidth}{\noindent\flushleft
Adult female with thrombocytopenia\adjustline} & 4.50 & Red & 1.84 \\
\hline
\parbox[c]{0.5\textwidth}{\noindent\flushleft
Adult (female) with infection\adjustline} & 6.37 & Red & 2.61 \\
\bottomrule
\end{tabular}

\end{center}
\label{table:cases:test}
\end{table}


\begin{table}[h!]
\centering
\begin{tabular}{c|r|r}
\toprule
\textbf{Test} & \textbf{Statistic} & \textbf{p-value} \\
\midrule
\parbox[c]{0.5\textwidth}{\noindent\flushleft Kruskal-Wallis Test (Across all three scores)} & H-statistic = 10.49 & 0.0053 \\
\hline
\parbox[c]{0.5\textwidth}{\noindent\flushleft Pearson Correlation (NIS vs NCCIS)} & Coefficient = 0.89 & $<$ 0.0001 \\
\hline
\parbox[c]{0.5\textwidth}{\noindent\flushleft Wilcoxon Test (NIS vs NCCIS)} & N/A & 0.018 \\
\bottomrule
\end{tabular}
\caption{Statistical test results for the three scores: NIS, CCIS and NCCIS.}
\label{table:stat_results}
\end{table}

Statistical tests assessed the relationship among the three individualised scores, NIS, NCCIS, and CCIS, for various health conditions. Table~\ref{table:stat_results} presents significant statistical associations between these scores.

The Kruskal-Wallis test, a non-parametric alternative to ANOVA was performed due to the ordinal nature of the CCIS index. The significant p-value ($<$0.001) indicates that there are statistically significant differences in the distribution of the NIS, NCCIS, and CCIS scores in the conditions.

A Pearson correlation assessed the linear relationship between NIS and NCCIS. The strong positive correlation coefficient (0.89) with a highly significant p-value ($<$ 0.0001) determined from a two-tailed independent samples t-test suggests a robust linear association between these two scores.

Lastly, a Wilcoxon Signed-Rank test compared paired differences between NIS and NCCIS. The significant result (p = 0.018) indicates that there are significant differences between these paired scores and therefore each can capture different aspects of blood data.

\subsection{Adaptive Index Behaviour Over Time}\label{variation}

\begin{figure}[hbp]
\centering
\includegraphics[scale=0.6]{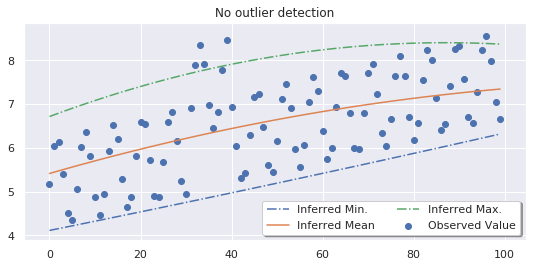}
\includegraphics[scale=0.6]{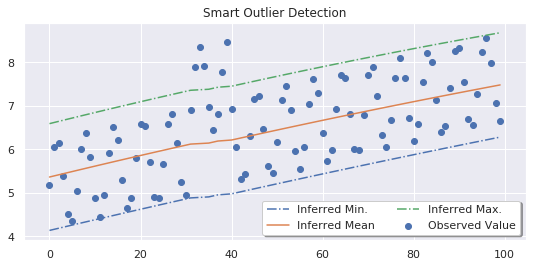}
\caption{Simulation of a process of learning personalised reference values based on NIS index behavious. The index is used to adapt the range of expected values for all analytes over time, showing stable trends while also detecting abnormal values from a personal previously established setpoint over time. Dashed lines are personalised adaptive reference values for two types of adaptation with and without outlier detection.}
\label{fig:smart}
\end{figure}

In~\ref{fig:smart}, we tested the behaviour of the NIS index over time by simulating patients evolving in different directions with outliers. We show that the index can be used to learn setpoints or baselines of patients over time based on the behaviour of the index flagging for changes even within the normal values of the population.

We also hypothesised that as the patient ages, the immune index would vary over time. In particular, we expected to see an upward trajectory in direct relation to age.

\begin{figure}[hbp]
\centering
\subfloat[Mean]{\includegraphics[width=9cm]{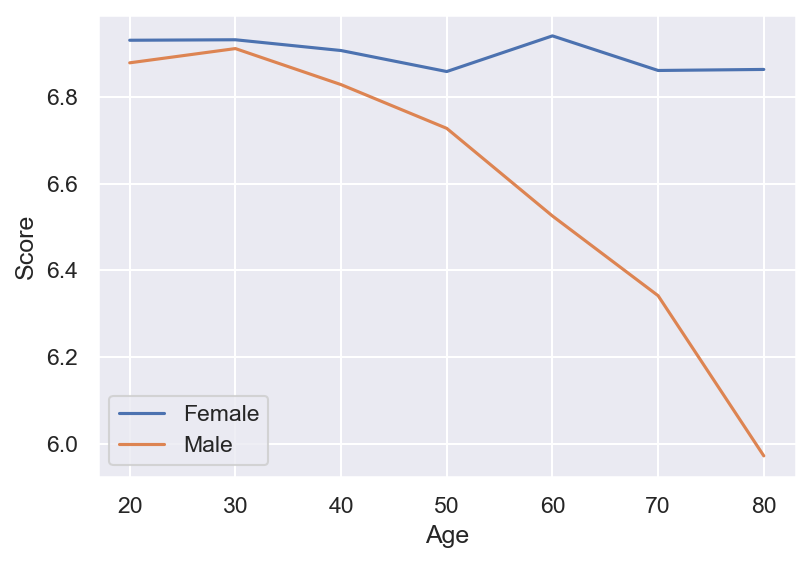}} \\
\subfloat[Distribution]{\includegraphics[width=8.3cm]{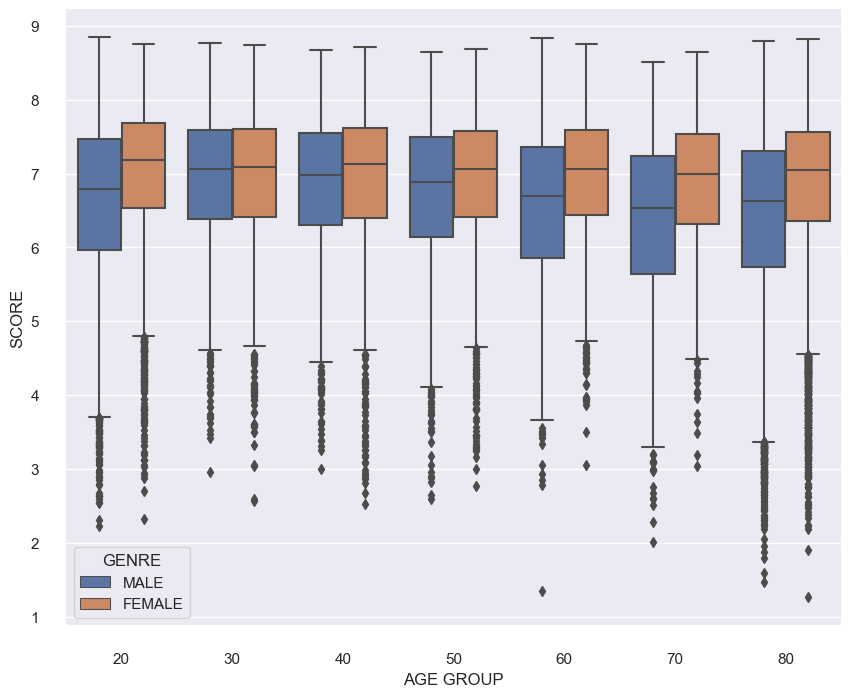}}

\caption{(a) The mean immune age as a function of age for NHANES data. The reference ranges were computed over the population aged 20 to \textcolor{black}{85} for males and females, respectively. Each of the data points represents the mean immune index for the cohort $[age-5,age+5]$; for example, 50 represents the immune index for the population aged 45 to 55 years. (b)The distribution of immune age as a function of age. \textcolor{black}{While a correlation trend is visible (especially for males), across sex the correlation is weak because healthy and non-healthy cohorts are confounded at the source given that both and the patient's actual current unhealthy states are in the same category.}}\label{ap}
\label{some example}
\end{figure}

To test the hypothesis, \textcolor{black}{we selected from the dataset two cohorts comprising all the data entries of individuals with ages ranging between 20 and 90 years, separating them by sex (male or female).} These cohorts were used to build two immune spaces by computing the respective mean and standard deviations for each of the 13 analytes. Afterward, for each sex, we selected seven cohorts of adults aged 20 to 25, 25 to 35, 35 to 45, 45 to 55, 55 to 65, 65 to 75, and 75 to 85 years of age, making a total of 14 cohorts. Finally, we computed the immune index for each entry. The mean immune age for each of the cohorts is shown in Fig. \ref{ap}.

The results show significantly different behaviour over time for male and female cohorts. For subjects over the age of 30, it is evident that the male population presents an upward trajectory with respect to age for the mean immune index, as well as an increase in the 50th percentile and variance. \textcolor{black}{The immune index decreases slightly for males between 20 and 25 year old}. However, the trend in women slows down (compared to men) between ages 40 and 50 and reverses between ages 50 and 60. 

\textcolor{red}{
This observed inflection in females is reported descriptively rather than causally interpreted. While menopause represents a temporally coincident physiological transition that may influence haematological parameters, the present dataset does not include menopause status, menopause-specific hormonal measurements, or neuro-immune-endocrine axes required for mechanistic validation.}

\textcolor{red}{Therefore, the association should be considered as suggestive systemic explanations or prospective study-generating hypotheses. Pregnancy and menopause have been associated with alterations in red blood cell counts, haematocrit levels, mean cell volume (MCV), and haemoglobin concentrations \cite{kovanen, nancy, cruickshank, nakada, aldrighi}, suggesting plausible mechanistic pathways.}

\textcolor{red}{However, given the multiscale physiological complexity underlying these processes, dedicated stratified longitudinal analyses incorporating reproductive and hormonal status would be necessary to formally validate this interpretation.
}
\subsection{Discounting Confounding Factors from Age}

We define immune age as the reverse function of the relation between the expected immune index and the age of an individual. Formally: 

For a given immune index $\alpha$, the \emph{immune age} $A:\alpha \mapsto a$ is defined as the function:
$$A(\alpha) = a.\overline{\{S(p):p\text{ age is }a\})}=\alpha,$$where $\overline{\{S(p):p\text{ age is }a\})}$ is the mean immune index for all the individuals of age $a$.

In other words, the immune age is defined as the age in the function of the expected immune index; the immune age for a given index is the age cohort for which such a index is expected.

Given the observed behaviour of the mean immune index over time, we assert that we can apply the given definition \emph{immune age} to men aged 20 or older. However, we have shown that the functional relation is much weaker, or nonexistent, in the female population.

\begin{figure}[H]
\centering
\includegraphics[width=8cm]{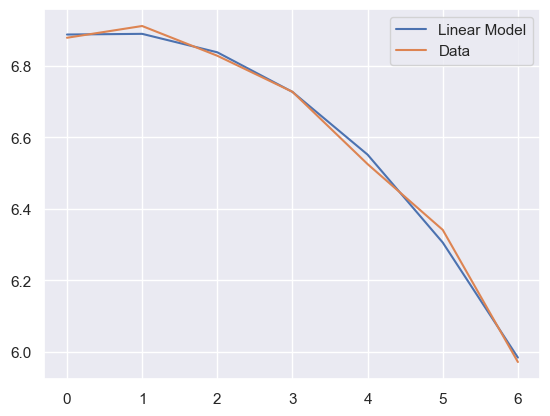}

\caption{A linear approximation of degree 3 to the mean index over seven age cohorts from the healthy groups only. The linear model smooths the curve and allows for extrapolation. }\label{lm}
\end{figure}

Similarly to that presented in Section \ref{variation}, we compiled the data into a list of seven values that represent the mean immune index for each of the seven age cohorts. We fitted a linear model of degree 3 to the data, obtaining a monotonously decreasing function that allows us to interpolate and extrapolate the expected immune index over specific ages. The resulting curve, which we call an \textit{immune age curve}, is shown in Fig. \ref{lm}. The corresponding ages are 20 to 25, 25 to 35, 35 to 45, 45 to 55, 55 to 65, 65 to 75 and 75 to 85.

\textcolor{black}{We will only consider integer ages in years. Therefore we can represent the \textit{immune age curve} by a list of 70 real values that contains the inferred immune index for each age between 30 and 85. Let us denote this curve by the list of pairs $\text{IAC} = \langle (20,6.87),(30, 6.91),\dots,(a,\alpha),\dots,(80, 5.97) \rangle$.}

We now can obtain the immune age $A$ by means of the following function:
\textcolor{black}{
\begin{equation*}
A(\alpha)=\begin{cases}
          20 \quad &\text{if} \, \alpha \geq 6.91  \\
          85 \quad &\text{if} \, \alpha \leq 5.624  \\
          a.(a, \alpha^*)\in \text{IAC} \quad & \text{elsewhere}
     \end{cases}
\end{equation*}
}
where $\alpha^*$ is the closest index in $IAC$ to $\alpha$. If there are two scores at the same distance, we choose the leftmost option. The resulting function is shown in Fig. ~\ref{lm}. The curve shows some minor bumps resulting from the discretisation used. We maintain that this distortion is not significant. 

\begin{figure}[ht!]
\centering
\includegraphics[width=8cm]{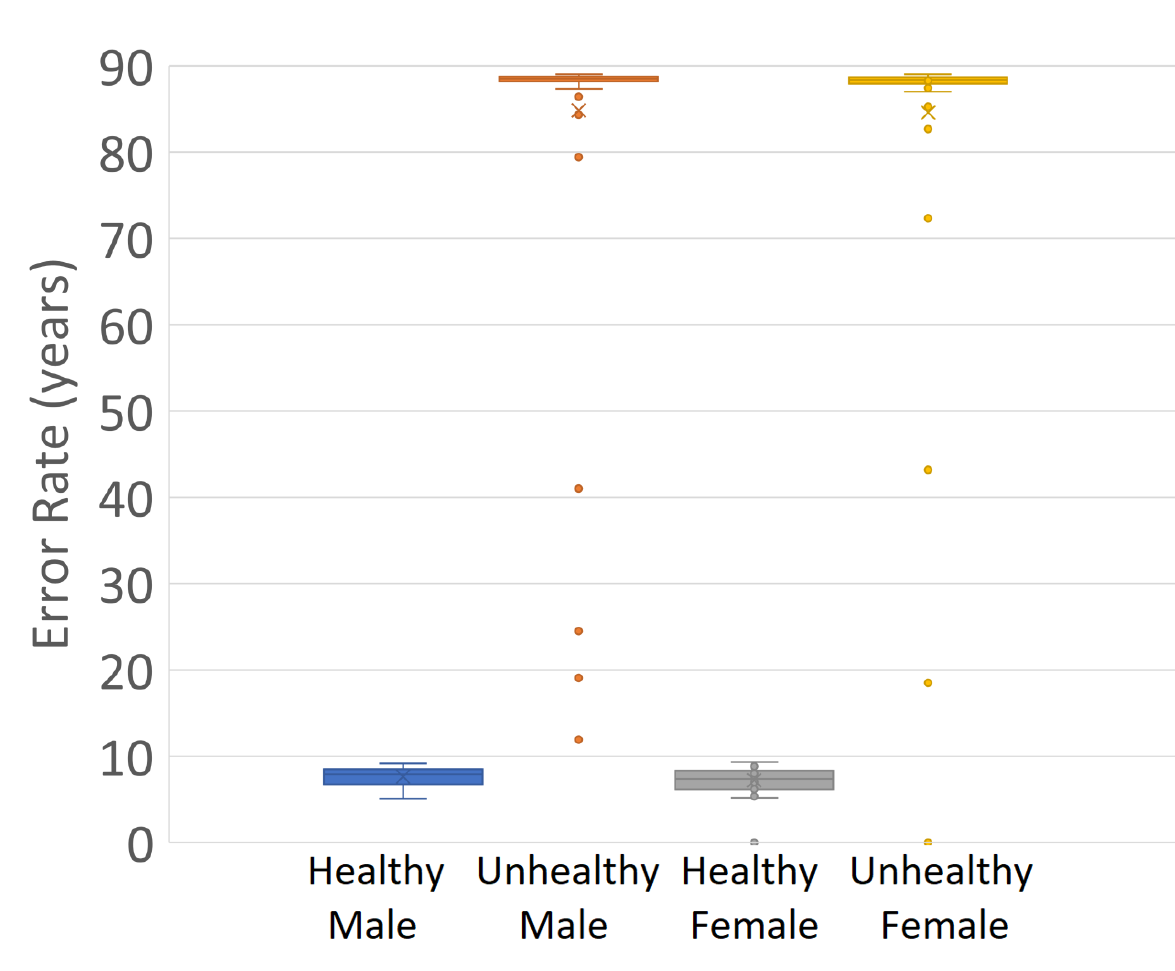}
\caption{Age Prediction Error Rates between actual patient (biological) age and Immune-index predicted age from NHANES dataset by CBC counts. The standard error analysis shown includes the prediction error rates in years (as box plots) across different health and sex groups, computed by assessing the variability in age prediction errors using interquartile ranges (IQR). \textcolor{black}{The correlation between predicted (immune) and chronological age, and the associated error distribution, are depicted in Fig.~5, which functionally serves as a scatter representation of these relationships.}The health status was determined by the best blood analyte correlate to separate healthy from unhealthy groups by sex.}
\label{ErrorRates}
\end{figure}

As a tool that returns an estimated age with respect to a index, we can analyse its predictive power. Using the NHANES data and focusing on males aged 20 to 90, we can measure the error of using the index to predict a person's age. As shown in Figure ~\ref{ErrorRates}, prediction error rates are much lower in healthy NHANES data, compared to unhealthy groups, for both sexes, and therefore can be used to estimate the chronological age of healthy patients. Prediction error rates for the immune index age were overestimated by approximately 7 years, for the healthy groups of both sexes, allowing a correction (mean error rate of 7.33 $\pm$ 1.21 for healthy females, and 7.62 $\pm$ 1.17 for healthy males). Furthermore, there is a statistically significant difference between the healthy and non-healthy groups (regardless of sex) in the error rates of age prediction, determined by two-tailed, two-sample t-tests. Pearson's correlation statistics on the age error rates predicted by the health status of the red blood cell count as discussed before, were an R square value of 0.293 and 0.168, for men (P$<$ 0.0001) and women (P$<$0.013) respectively. Age prediction error rates were high and not significant when self-reported health measures were used to separate health status groups, but were significant when taking abnormal CBC values only.

\textcolor{red}{
We emphasise that the immune age metric is not intended to function as a high-precision 'chronological' ageing clock comparable to existing methods, such as epigenetic or proteomic biological age estimators (e.g., methylation-based clocks).}

\textcolor{red}{Rather, its purpose is to detect divergence between routinely accessible physiological immune patterns and expected age-related norms derived from CBC blood counts.}

\textcolor{red}{The observed mean error largely reflects systematic offset rather than random dispersion. Post-correction bias is reduced and variance narrows substantially, particularly within healthy cohorts. Importantly, prediction error is lower in physiologically stable (healthy) groups, reinforcing its role in deviation detection rather than exact age estimation.}

\textcolor{red}{Accordingly, the index is positioned as a population-level screening and stratification tool, not a diagnostic clock, designed to support accessible, affordable, and equitable health assessment with translation towards personalised patient care.
}

\begin{table}[h]
\centering
\caption{\textbf{Statistics of each Complete Blood Count (CBC) analyte as a predictor or discriminant of health and age.}}
\begin{tabular}{l l l c c c}
\toprule
\textbf{Analyte} & \textbf{Full Name} & \textbf{Unit} & \textbf{Healthy} & \textbf{Unhealthy} & \textbf{Age} \\
\midrule
BASO & Basophils & $10^{-9}$/L & 0.01 & 0.02 & 0.00 \\
EOS & Eosinophils & $10^{-9}$/L & 0.00 & 0.00 & 0.00 \\
HCT & Haematocrit & L/L & 0.00 & 0.01 & 0.07 \\
HGB & Haemoglobin & g/L & 0.12 & 0.14 & 0.06 \\
LYMP & Lymphocytes & $10^{-9}$/L & 0.04 & 0.50 & 0.09 \\
MCH & Mean Cell Haemoglobin & pg & 0.08 & 0.00 & 0.15 \\
MCHC & MCH Concentration & g/L & 0.00 & 0.00 & 0.00 \\
MCV & Mean Cell Volume & fL & 0.27 & 0.34 & 0.22 \\
MONO & Monocytes & $10^{-9}$/L & 0.01 & 0.00 & 0.01 \\
NEUT & Neutrophils & $10^{-9}$/L & 0.01 & 0.01 & 0.02 \\
RBC & Red Blood Cells & $10^{-12}$/L & 0.03 & 0.00 & 0.00 \\
PLT & Platelets & $10^{-9}$/L & 0.09 & 0.33 & 0.10 \\
\bottomrule
\label{analytes}
\end{tabular}
\end{table}

Table ~\ref{analytes} shows the statistics of the predictors and separators of age for healthy and unhealthy populations. These plots show that common blood marker dynamics are predictive of age and health.

There are small but potentially meaningful differences between reference ranges. This could affect the classification of out-of-range counts when applying mismatched country-specific ranges. Using uniform, population-appropriate references is important for generalisable models, but our main contribution is to learn each individual's personal reference values, thereby making the reference discrepancies per country less relevant. The best single analyte features for predicting health status and age are summarized in Table~\ref{Bestpredictors}. Notably, MCV, LYMP, PLT, and HGB emerge as strong discriminators between healthy and unhealthy classes, each displaying distinct patterns across categories.

\begin{table}[htp!]
\centering
\caption{Best predictors for each category}
\begin{tabular}{ccc}
\toprule
\textbf{Healthy} & \textbf{Unhealthy} & \textbf{Age} \\
\midrule
MCV & LYMP & MCV \\
HGB & MCV & MCH \\
PLT & PLT & PLT \\
MCH & HGB & LYMP \\
LYMP & BASO & HCT \\
RBC & NEUT & HGB \\
MONO & HCT & NEUT \\
NEUT & MCHC & MONO \\
BASO & MONO & EOS \\
HCT & MCH & MCHC \\
MCHC & EOS & RBC \\
EOS & RBC & BASO \\
\bottomrule
\end{tabular}
\label{Bestpredictors}
\end{table}

Fig.~\ref{ErrorRates} highlights the fact that despite inter-group error rate variability, ML and computational methods can accurately differentiate between healthy and unhealthy groups, as evidenced by the well-separated box plots (medians and IQRs). The results show that the immune-index predicted age has the lowest prediction error rate among healthy groups, for both sexes, and overestimates age when the individuals are not healthy. Outliers beyond the 1.5 times IQR were not imputed to the nearest whisker point to showcase the variability and distribution of patient data. 

\clearpage

\subsection{Machine Learning Guiding index}

We explored how we could exploit the strong results from using the index as a risk assessment score and biological age predictor combined with traditional Machine Learning methods. We found that combining the symbolic strengths of a risk assessment index with the machine learning methods helped the algorithms to perform better while making the combined neurosymbolic result, a more interpretable approach given the explainability power of the index alone.

\begin{table}[ht]
\centering
\begin{tabular}{c|c}
\hline
\textbf{Feature} & \textbf{Importance index} \\
\hline
Hb & 0.1148 \\
Hct & 0.102 \\
Plt & 0.09722 \\
RBC & 0.0967 \\
MCV & 0.0927 \\
Lymphocytes & 0.08623 \\
MCH & 0.0818 \\
WBC & 0.08 \\
Neutrophils & 0.07409 \\
\hline
\end{tabular}
\caption{Feature importance metrics for different blood analyte features using RF model (with 10-fold stratified cross-validation to help prevent overfitting. The importance index is calculated as the average decrease in Gini entropy across all trees when a feature is used to split the data.)}
\label{tab:performance_metrics}
\end{table}

\begin{table}[ht]
\centering
\begin{tabular}{c|c|c|c|c}
\hline
\textbf{Model} & \textbf{Accuracy} & \textbf{MSE} & \textbf{Sensitivity} & \textbf{Specificity} \\
\hline
Random Forest & 0.94 & 0.06 & 0.40 & 0.99 \\
SVM & 0.92 & 0.08 & 0.01 & 1.00 \\
KNN & 0.91 & 0.09 & 0.13 & 0.98 \\
\hline
\end{tabular}
\caption{ML Performance metrics for different models using blood analytes. The models were evaluated using 10-fold stratified cross-validation.}
\label{tab:model_ML}
\end{table}

\begin{table}[ht]
\centering
\begin{tabular}{c|c|c|c|c}
\hline
\textbf{Model} & \textbf{Accuracy} & \textbf{MSE} & \textbf{Sensitivity} & \textbf{Specificity} \\
\hline
RandomForest & 0.94 & 0.06 & 0.36 & 0.99 \\
SVM & 0.92 & 0.08 & 0.01 & 0.99 \\
KNN & 0.91 & 0.09 & 0.06 & 0.99 \\
ANN & 0.99 & 0.01 & 0.04 & 1.00 \\
Transformer-ANN & 0.99 & 0.01 & 0.05 & 1.00 \\
\hline
\end{tabular}
\caption{ML performance metrics for NIS immune index using different models evaluated using 10-fold stratified cross-validation.
}
\label{tab:model_performance}
\end{table}

\begin{table}[ht]
\centering
\begin{tabular}{c|c|c}
\hline
\textbf{Feature} & \textbf{Model} & \textbf{AUC} \\
\hline
NIS & Transformer-ANN & 0.91 \\
NIS & ANN & 0.91 \\
NIS & KNN & 0.68 \\
NIS & RandomForest & 0.70 \\
NIS & SVM & 0.59 \\
\hline
\end{tabular}
\caption{AUC scores for each model under NIS index.}
\label{tab:auc_scores}
\end{table}

\textcolor{red}{
The interpretability of the hybrid neurosymbolic approach emerges from the fact that the neural networks function on a single scalar immune index (NIS) rather than on high-dimensional raw feature vectors.}

\textcolor{red}{Feature importance rankings derived from Random Forest models (mean Gini decrease; Table 9) were used to identify the most influential CBC analytes contributing to discrimination, providing an interpretable predictive ranking of variables.}

\textcolor{red}{The neural networks were then applied either to the full analyte set or directly to the scalar NIS score (Tables 10–12). Thus, when operating on the NIS, the neural component refines decision boundaries within a symbolically defined biological parameter space, rather than discovering latent representations de novo.}

\textcolor{red}{This preserves explainability of the contributing variables while allowing nonlinear pattern detection through computational intelligence.
}
\textcolor{red}{Future extensions of this neurosymbolic AI framework could incorporate architecture-dependent feature pruning, knowledge destination with hybrid combination of architectures, sparsity-constrained optimisation, or regularisation-driven layer compression to further align neural models' predictive performance with biologically meaningful features identified through neurosymbolic ranking. }

\begin{figure}[htp]
    \centering
    \includegraphics[width=\textwidth, keepaspectratio]{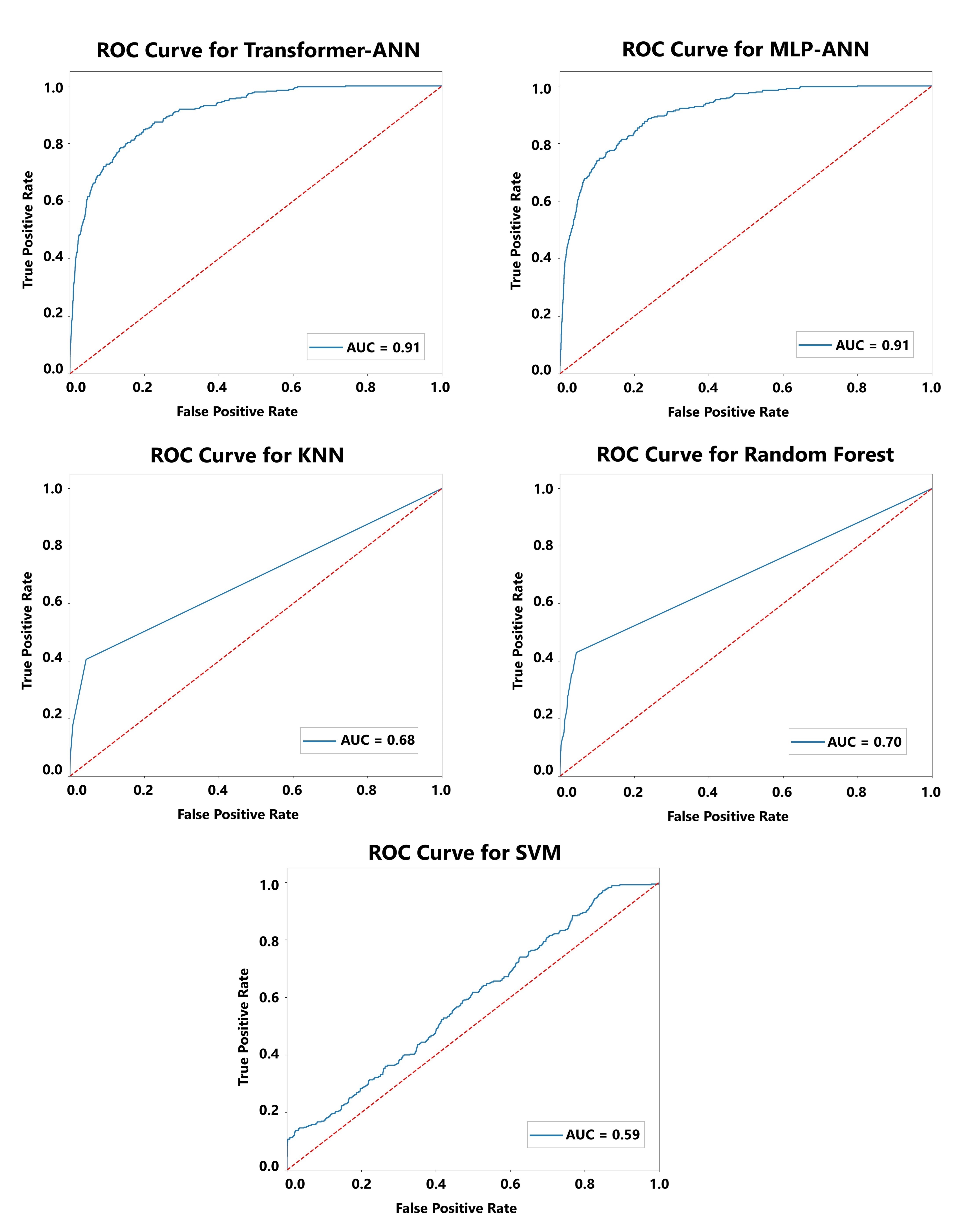}
    \caption{ROC Curves for NIS Immune index. Using more complex ML algorithms like a feed-forward neural network (MLP-ANN) and a transformer-based attentional neural network shows an AUC above 0.9, enhancing the predictive performance of NIS immune scores. The higher performance of ANN and transformer models coupled with NIS scores suggests the ability of ANN systems to capture complex, non-linear relationships in the NIS scores, leading to enhanced predictive accuracy with computational intelligence.}
    \label{fig:roc_index}
\end{figure}

As shown in Table ~\ref{tab:model_performance} for the different AI models, the NIS immune index, a single quantitative measure (scalar value), also showed significant classification performance, particularly when coupled with computational AI-driven predictive power using ANN networks. While other basic ML classifiers did not perform as well, the ANN paired with the NIS immune index achieved an optimal predictive accuracy of 0.987, a high specificity of 0.999, and an AUC index of 0.91, comparable to the individual analytes themselves. However, the sensitivity was poor compared to the analytes. A low sensitivity in AI classification implies it is not able to correctly classify true positives. In contrast, the basic machine learning classifiers, such as RandomForest (RF), SVM, and KNN, did not perform as well as the ANN models with the NIS immune index, achieving lower AUC scores with 10-fold cross-validation, despite their high classification accuracy. 
While the feedforward ANN (MLP) detects local features through its dense layers, the Transformer-ANN captures long-range dependencies using self-attention mechanisms.

Table~\ref{tab:performance_metrics} provides the highest features importance metrics (i.e., mean Gini entropy decrease) for different blood analytes using the RF algorithm. Table~\ref{tab:model_ML} shows the performance of all blood analytes collectively, and they perform suboptimally against ANN performance on NIS Immune scores shown in Table ~\ref{tab:model_performance}. The introduced index captures in a single numerical value the power of this and other analytes that may be of interest for other patient profiles, also giving the clinician the opportunity and option to add or remove analytes or assign different weights to different analytes. Table~\ref{tab:auc_scores} shows that both Transformer-ANN and ANN models demonstrate superior performance with an AUC of 0.91 to predict outcomes based on the NIS index, significantly outperforming traditional models such as KNN, RF and SVM. However, their sensitivity lacks robustness and requires further optimisation through threshold adjustments, enhanced feature selection, or targeted optimisation techniques. Future studies with larger validation datasets can include bootstrapping methods and dropout layers in the ANNs to help prevent overfitting. 

Fig.~\ref{fig:roc_index} illustrates these findings via ROC curves for the NIS immune index across different machine-learning models. The high AUC values achieved by the ANN models highlight their enhanced ability to capture complex, non-linear relationships in the NIS data. Therefore, they can significantly improve predictive modelling and accuracy, suggesting their potential to improve diagnostic precision using immune scores.

\subsection{UK Biobank Subset Validation}

We also analyzed a subset of the UK Biobank data as an independent validation cohort of the NIS immune index, which included three longitudinal timepoints for each CBC parameter. This allowed for more robust temporal averaging and normalisation prior to scoring.  While differences in cohort's demographic distribution, and data acquisition methods between the CDC NHANES and the UK Biobank limit direct comparability, our objective was to assess whether the immune index retained its ability to accurately distinguish health status in a broader population context with the same level of statistical predictability and hence, generalizability. As shown, significant separations between healthy and disease groups (e.g., cardiovascular, cancer, infection) were observed, reinforcing the discriminative power of the NIS immune index. These results provide preliminary evidence supporting the robustness and clinical translatability of the index, suggesting that it captures immunological patterns in complex blood datasets. 

\textcolor{red}{
Furthermore, the availability of repeated CBC measurements in the UK Biobank subset allows partial real-world assessment of temporal robustness related to the adaptive baseline introduced in Fig.~2. While the adaptive learning itself was illustrated through simulation, the averaging of three longitudinal timepoints prior to scoring demonstrates that the immune index retains discriminatory power under real-world temporal variability. The primary objective of this study is to demonstrate that the immune index discriminates health status and approximates biological age using routine CBC measurements, with multi-timepoint averaging in the UK Biobank serving as a robustness validation rather than the central methodological requirement. Thus, the index remains stable and predictive when applied to multi-timepoint clinical data.
}

\begin{table}[htbp]
\centering
\caption{Pairwise comparisons of Composite NIS between patient groups, including effect sizes and uncertainty estimates.}
\begin{tabular}{lccccc}
\hline
\textbf{Comparison} & \textbf{n} & \textbf{p-value} & \textbf{Cohen's} & \textbf{Cohen's} & \textbf{Cliff's} \\
&&&\textbf{d}&\textbf{d CI} & \textbf{$\delta$}\\
\hline
Rest v Cancer & 1741 v 438 & $<$0.00001 & 3.540 & [3.36, 3.7] & 0.99 \\
Rest v CVD & 1741 v 417 & $<$0.00001 & 3.49 & [3.33, 3.6] & 0.99 \\
Rest v Inf. & 1741 v 367 & $<$0.00001 & 3.424 & [3.2, 3.5] & 0.98 \\
Cancer v CVD & 438 v 417 & 0.702 & 0.029 & [-0.12, 0.16] & -0.015 \\
Cancer v Inf. & 438 v 367 & 0.0033 & -0.16 & [-0.31, -0.046] & -0.12 \\
CVD v Inf. & 417 v 367 & 0.1 & -0.19 & N/A & N/A \\
\hline
\end{tabular}
\label{tab:nis_comparisons}
\end{table}

\begin{figure}[!h]
    \centering
    \includegraphics[width=0.84\textwidth]{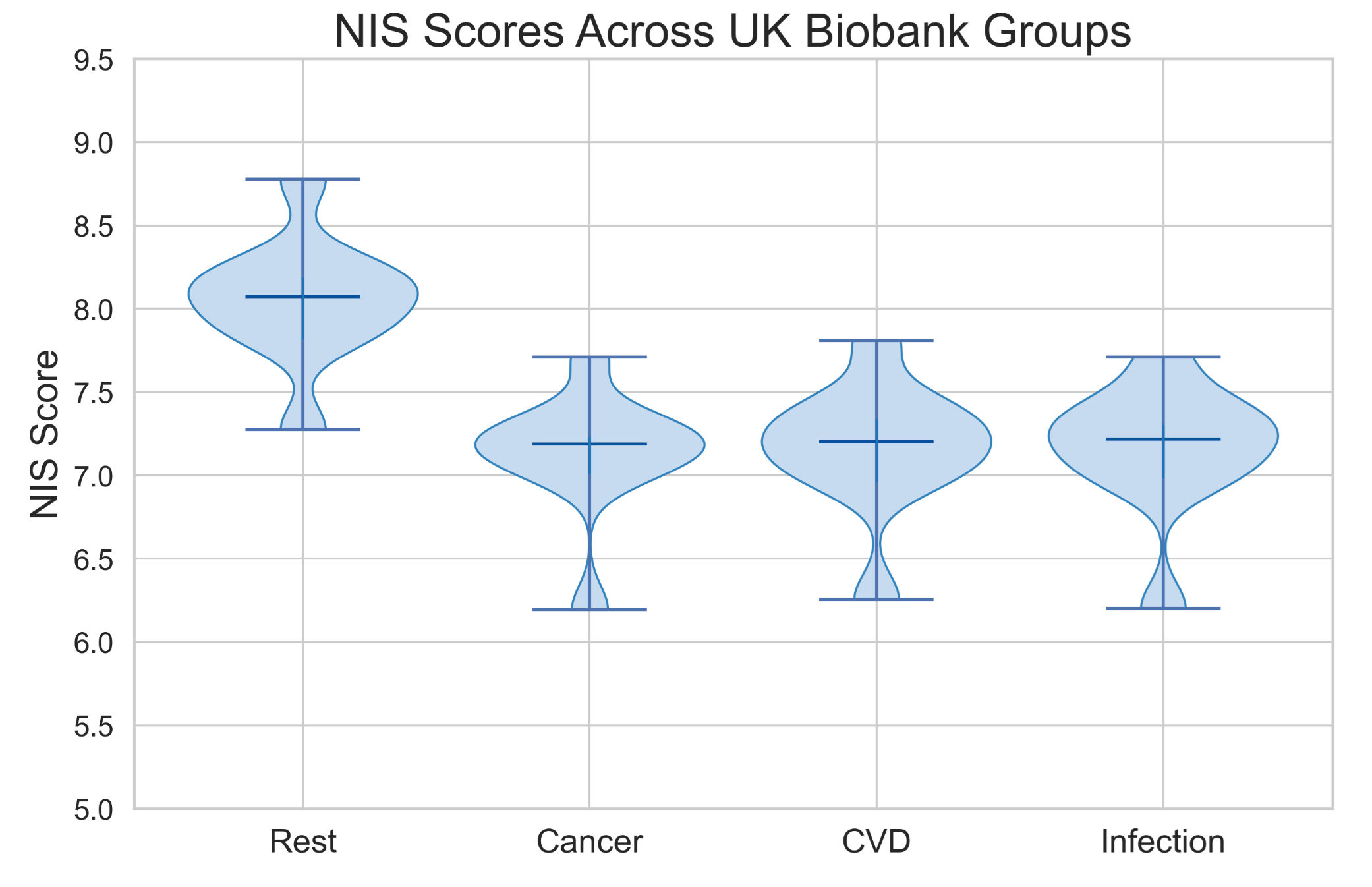}\\
    \includegraphics[width=0.85\textwidth]{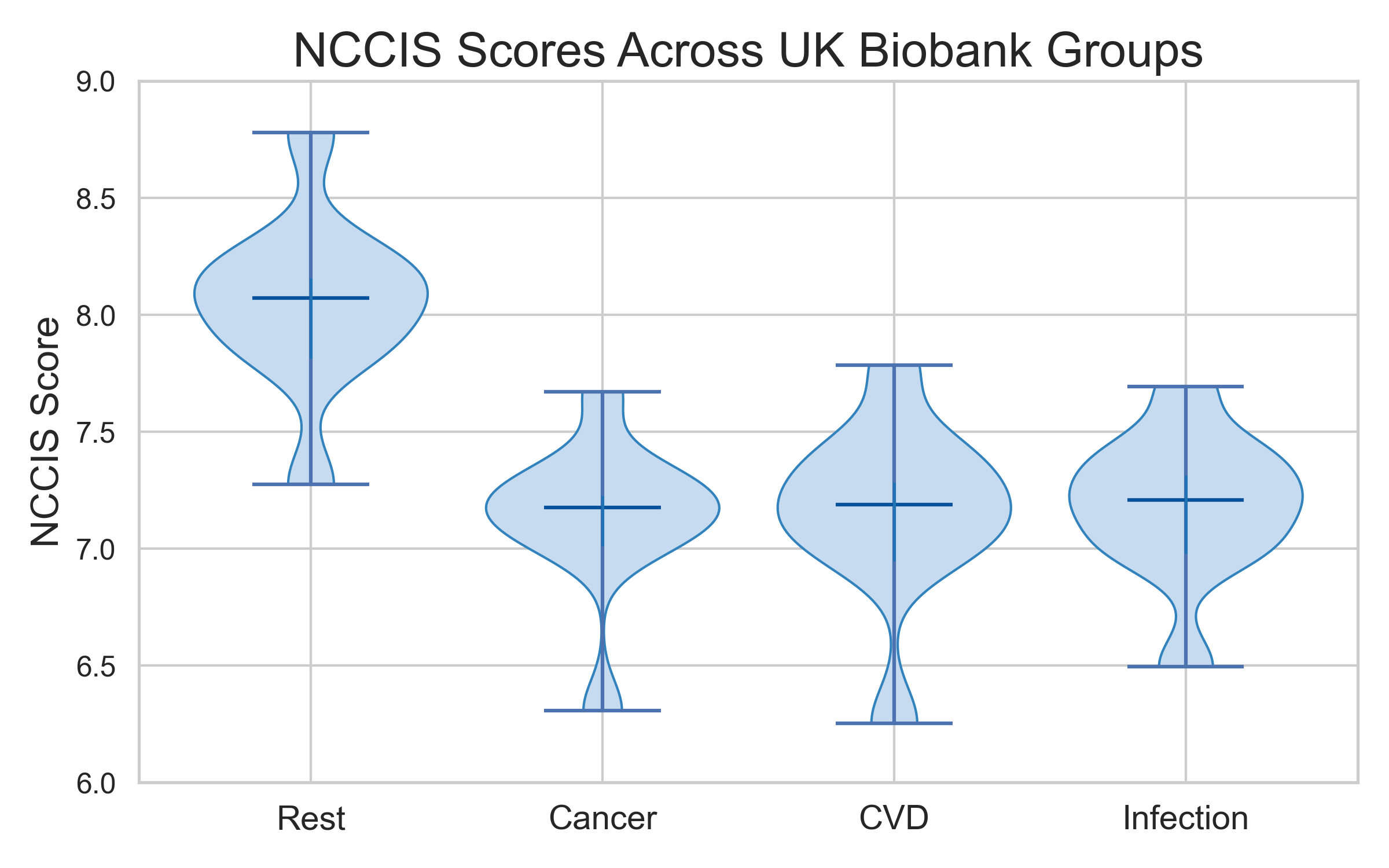}
    \caption{\textbf{Top:} Boxplots of composite NIS scores across UK Biobank groups validating the previous results with an independent cohort from the CDC NHANES database. The boxes represent the interquartile range (IQR) with the median marked as a horizontal line, and the whiskers indicate the minimum and maximum values within 1.5 IQR. \textbf{Bottom:}         Violin plot for NCCIS (Normalized Composite CBC Immune Score). The scores confirm a significant distinction between healthy and disease groups. The distributions are  consistent with the NIS score, indicating strong convergence between the two scoring systems. The p-values for NCCIS were comparable to those observed with the NIS index, showing strong statistical separation ($p<0.0001$) between all disease groups against the Rest cohort.
}
    \label{fig:UKBioNISBox}
\end{figure}

We first filtered the Rest (`healthy' control) cohort by removing all individuals whose third-time point CBC measurements deviated from established normal physiological ranges. This reduced the noise and the cohort size from $n=2312$ to $n=1741$.

For disease groups (Cancer, CVD, Infection), we kept only those individuals who exhibited at least four CBC values that fall outside the normal range. This filtering reduced the initial cohort sizes (Cancer: 666, CVD: 632, Infection: 396) to 438, 417, and 367 respectively.

Pairwise comparisons of composite NIS scores between these filtered groups are summarised in Table~\ref{tab:nis_comparisons}, and box plots of scores are presented in Fig.~\ref{fig:UKBioNISBox}.

The Shapiro–Wilk tests on the residuals yielded p-values $<$ 0.05 for all groups, thus we used the Kruskal–Wallis test followed by Bonferroni-corrected Mann–Whitney U tests for pairwise comparisons, as shown in Table~\ref{tab:nis_comparisons}.

As shown in Table~\ref{tab:nis_comparisons}, the high Cliff delta values ($\geq$ 0.987) for Rest versus each disease group indicate a strong effect of the NIS immune score on healthy individuals. The large Cohen’s d values, with narrow bootstrapped 95\% confidence intervals, further confirm robust and consistent effect sizes. Together with the highly significant p-values, these results underscore the low uncertainty and strong discriminatory power of the NIS between healthy and diseased groups. 

Furthermore, as revealed in Figs.~\ref{fig:UKBioNISBox}, a significant decrease in NIS (or NCCIS) scores is observed in the disease groups (Cancer, CVD, and Infection) compared to the healthy group (Rest). This validates our findings from the CDC NHANES dataset independent of the UK Biobank even under significant noise and two geographically different regions. Thus, we propose that immune scores could serve as a simple, interpretable clinical tool to stratify health outcomes and risk factors, supporting early intervention and public health surveillance in predictive medicine. 

\textcolor{red}{
As discussed, future work should also incorporate menopause status and hormonal descriptions available in UK Biobank to formally test whether the observed immune index inflection in females reflects endocrine transition rather than sampling heterogeneity. Such stratification would enable mechanistic refinement of the sex-specific immune ageing curve.
}

\textcolor{red}{
Lastly, while the NHANES cohort includes conditions with pronounced haematological disruption (e.g., leukemia, HIV), the UK Biobank validation extends the analysis to chronic cardiovascular disease and infection cohorts (Table 13). These represent systemic inflammatory and metabolic conditions rather than primary marrow-disruptive pathologies. The preservation of discriminatory performance in these cohorts suggests that the immune index captures broader physiological dysregulation of the blood–immune–inflammatory axis beyond overt haematological malignancy. Future studies may further evaluate the applicability of the immune index in developmental, neuropsychiatric and other systemic inflammatory and metabolic disorders, where immune perturbations are increasingly recognized as contributing factors.
}

\section{Conclusions}

In previous studies, biological age metrics have been based on biological tests that require specialised tests based on uncommon blood markers. Here, we have demonstrated the utility of common markers from a Complete Blood Count to discriminate and predict health status and age, showing that when the chronological and biologically derived age of these markers diverge, health status deteriorates and when it converges, better health is reported. We also validated the scores on a separate dataset (UK biobank) with statistical measures indicating strong discriminatory power of the scores to  as robust, generalisable tools across diverse health conditions.

With this information, we have introduced risk assessment scores for potential clinical use and studied their behaviour against typical cases of synthetic and empirical cases of various diseases and conditions showing sensitivity and specificity between health and disease with some intra- and inter-disease sensitivity to be further investigated. 

The indexes/scores introduced here are not intended to be used as a diagnostic tool alone, but are intended as a warning and flagging system capable of first triaging and general blood health screening. \textcolor{black}{We fully concur that abnormal blood counts naturally warrant medical evaluation. Our index is designed as an adjunctive computational triage tool that quantifies deviations in a standardized manner, complementing clinical decision-making rather than replacing expert clinical judgment.} We have shown that the index captures, in a single numerical value and a colour-coded scheme, the abnormal nature of blood results in terms of deviation from healthy reference values (absolute or adaptable). The index is sensitive to out-of-range values and increases its value and changes its colour as a function of how removed values are from lower and upper bounds of reference values according to the number of standard deviations from the medians. Although it has here been proven to be sensitive enough, it is not specific enough to quantify diseases or conditions.

Another aspect of this retrospective study is that it shows that while inter-person stability over longitudinal data points may be important~\cite{harvard,brodin1,brodin2}, useful single snapshot risk assessment scores are possible and of practical use when to their Cohen's d values are statistically significant like in this case. 

Taking this into account, we introduced a learning procedure to adapt normal versus abnormal reference values to personalised ranges from which we could show that single snapshots can still capture relevant information, as tested in the large CDC data set in 100\,000 records. We then validated the index on an independent cohort from the UK Biobank of very different nature, demonstrating that the numerical index still captures the non-healthy signals of the common blood markers for separating purposes.

We demonstrated the capabilities and digital applications of the scores for triaging purposes against two public databases, showing how the scores separate healthy from unhealthy cohorts, both as self-reported, with significant noise, and as determined by typical normal reference values.

The approach was shown to be informative for quick and entry-level patient health sorting, screening and triaging.

We further validated the NIS and NCCIS immune scores on a UK Biobank subset spanning diverse disease groups, including cancers, infections (bacterial, viral, fungal), and cardiovascular diseases (e.g., stroke, heart attack). As shown in Figure~\ref{fig:UKBioNISBox}, the scores demonstrated statistically significant discriminatory power in separating healthy individuals from each disease group.

Our findings demonstrate that interpretable scores can be combined with deep learning techniques such as feedforward ANN and Transformer models to leverage their predictive power. The  performance of these deep learning models can be attributed to their ability to capture complex, non-linear relationships within the training data. This highlights the potential of integrating advanced hybrid neuro-symbolic AI models guided by closed, explainable, and controllable algorithms, such as the risk assessment scores here introduced. 

\textcolor{black}{From an implementation standpoint, these indices could be embedded into electronic health record (EHR) systems as automated warning or longitudinal monitoring modules, providing clinicians with real-time quantitative feedback on patient trajectories.}

\textcolor{black}{Similar to widely used risk assessment scores, the proposed metric condenses deviations in blood test parameters into a single, symbolically interpretable, and directly comparable value for applications such as triage, where the alternative would be a first-come, first-served or random-order approach.}

Unlike statistical or traditional machine learning and black-box approaches, the hybrid approach introduced here helps guide the search, is more transparent to and controllable by health professionals or automated supervisors and by health consumers, facilitating rapid screening, regular and remote patient monitoring, and decision support.

The numerical index was also shown to be able to define an `immune age as a marker of health according to the classification test, with differences expected and reported for men and women, who are known to show greater fluctuations due to, for example, menopause, for which the recommendation is to take these life events into consideration for bias correction purposes. In the case of the estimation of immune age, the chronological age of healthy people was closely predicted by scores based on common haematological markers that are typically and massively collected from national health systems in CBC/FBC tests. For unhealthy people, whether self-reported or measured by typical abnormal reference values, the index diverged from chronological age, which was consistent with the theoretical expectation of a good predictor of health by divergence from chronological age when the subject is not in the healthy group.

Finally, the scores are not limited to CBC/FBC only and they can be personalised according to the analytes/indices of interest.




\section*{Declaration of interests Statement}

The authors declare themselves to be associated with Oxford Immune Algorithmics, a company associated with the Universities of Oxford, Cambridge, and King's College London. There are no other conflicts of interest.

\section*{Author Contributions}

H.Z. led the team and provided the general specifications and definitions of the scores, experiments to perform and data to analyse. H.Z. wrote the first draft of the paper and the main sections of the final version. H.Z., F.H.Q., S.H.O., A.U., and K.S.P. made contributions to the specifications. S.H.O. and A.U analyzed the data from the NHANES database and produced results and figures.

\section*{Acknowledgments and disclaimers}

\subsection*{Ethical Disclosure}
All methods were performed in accordance with the relevant guidelines and regulations as outlined by the institutional policies. As this research did not involve human subjects and used only publicly available anonymised data, it was exempted from consideration by an institutional review board and from ethical approvals. As the analysis was strictly computational, no further institutional approval was required.

\subsection*{Data Availability}

The National Health and Nutrition Examination Survey \cite{hanes} is a research program conducted by the National Center for Health Statistics (NCHS) of the United States of America. This program offers a public database that contains information related to the demographic, health, and nutritional status of adults and children in the United States. In particular, it offers individual laboratory results for the 13 indices, along with demographic information (such as age, biological sex, among others) for thousands of individuals collected over 20 years. All data related to this project are available through Zenodo at\\
\href{https://zenodo.org/records/10426175}{https://zenodo.org/records/10426175}.

\subsection*{Declaration of Competing Interest}

The authors declare the following competing financial interests that may be considered as potential competing interests: the authors are affiliated with Oxford Immune Algorithmics, which may commercially develop applications related to the index described in this work under its AI platform for precision healthcare called Algocyte and which they often refer to as Algocyte index.

\clearpage

\renewcommand{\thefigure}{S\arabic{figure}}
\renewcommand{\thetable}{S\arabic{table}}

\setcounter{figure}{0}  
\setcounter{table}{0}   

\section*{Supplementary Information}

\subsection{Machine Learning Methods}

 NHANES data on blood analytes (CBC markers) and NIS immune scores between patients were separated by health status for a binary classification problem: healthy vs. unhealthy. The data were pre-processed by imputing the missing NaN values using the mean values for each feature. The data set was then divided into training and test sets for classification, model building, and evaluation. Test sizes of 0.2 and 0.5 were used with a 10-fold cross-validation for analysis of performance metrics. Five statistical machine learning (ML) classifiers were used for initial assessment: Random Forest (RF), Support Vector Machine (SVM), K-Nearest Neighbours (KNN), KMeans clustering, and Gaussian Mixture Model. The latter two clustering methods delivered poorer performances in the data features, thus were removed from the screening. Each model was trained in the training set and evaluated in the discarded test set. Feature importance was analysed for the RF, SVM, and KNN models to identify the most predictive features. Pairwise analysis was also performed to inspect feature relationships. A correlation analysis was performed to identify any co-linear features before saving the results.

Two artificial neural network (ANN) architectures were explored as more flexible artificial intelligence (AI) tools. ANN models tested in binary health classification by analytes and immune scores included a feedforward neural network (FFNN) (also referred to as a multilayer perceptron MLP). The MLP consists of an input layer, one or more hidden layers, and an output layer. The model uses ReLU (Rectified Linear Unit) activation functions for the hidden layers and a sigmoid activation function for the output layer, which is typical for binary classification tasks.

Additionally, in order to make the ANN model more robust, we incorporated a more complex layered architecture with a Transformer network. The Keras and TensorFlow Python packages were used for the development of the neural network model. Statistical metrics such as ROC curves (area under the curve- AUC), sensitivity, specificity, mean square error (MSE) and classification accuracy were computed to evaluate the predictive performance of the ML algorithms. A description of the algorithms used is given in the Sup. Inf.

\begin{figure}[H]
\centering
\scalebox{.3}{\includegraphics{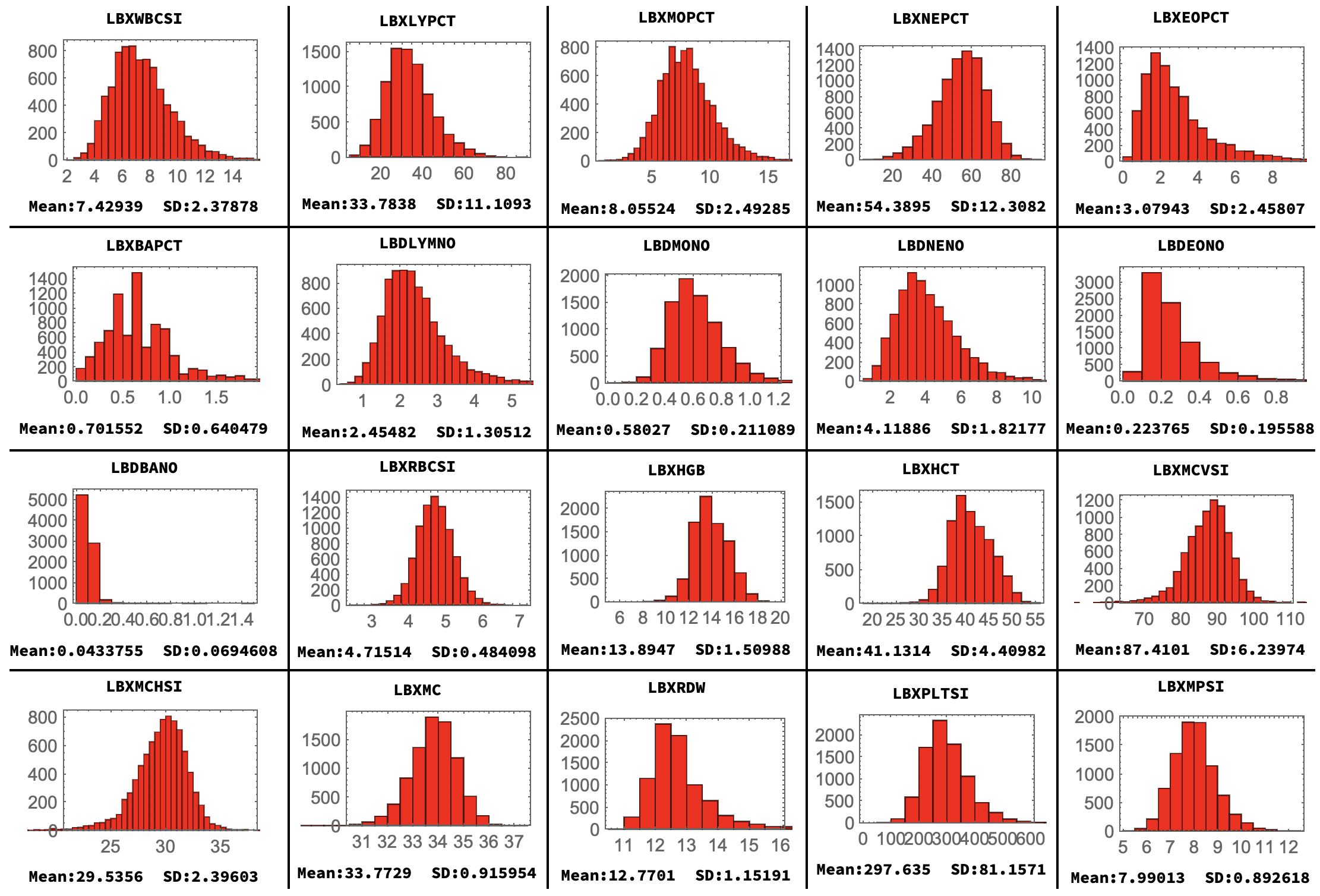}}
\caption{Population reference values and population average inadequacies for personal and precision healthcare: Distributions within `healthy' reference values are heavily skewed or display long tails in most cases, not following normal or uniform distributions. This and current clinical evidence ~\cite{brodin1,brodin2} calls for a personalised approach to reference values for abnormal results rather than one-size-fits-all current population approaches to blood test interpretation. The list of analyte labels is in the Supplementary Information.}
\label{nhanes}
\end{figure}

\begin{figure}[H]
	\centering
	\scalebox{.36}{\includegraphics{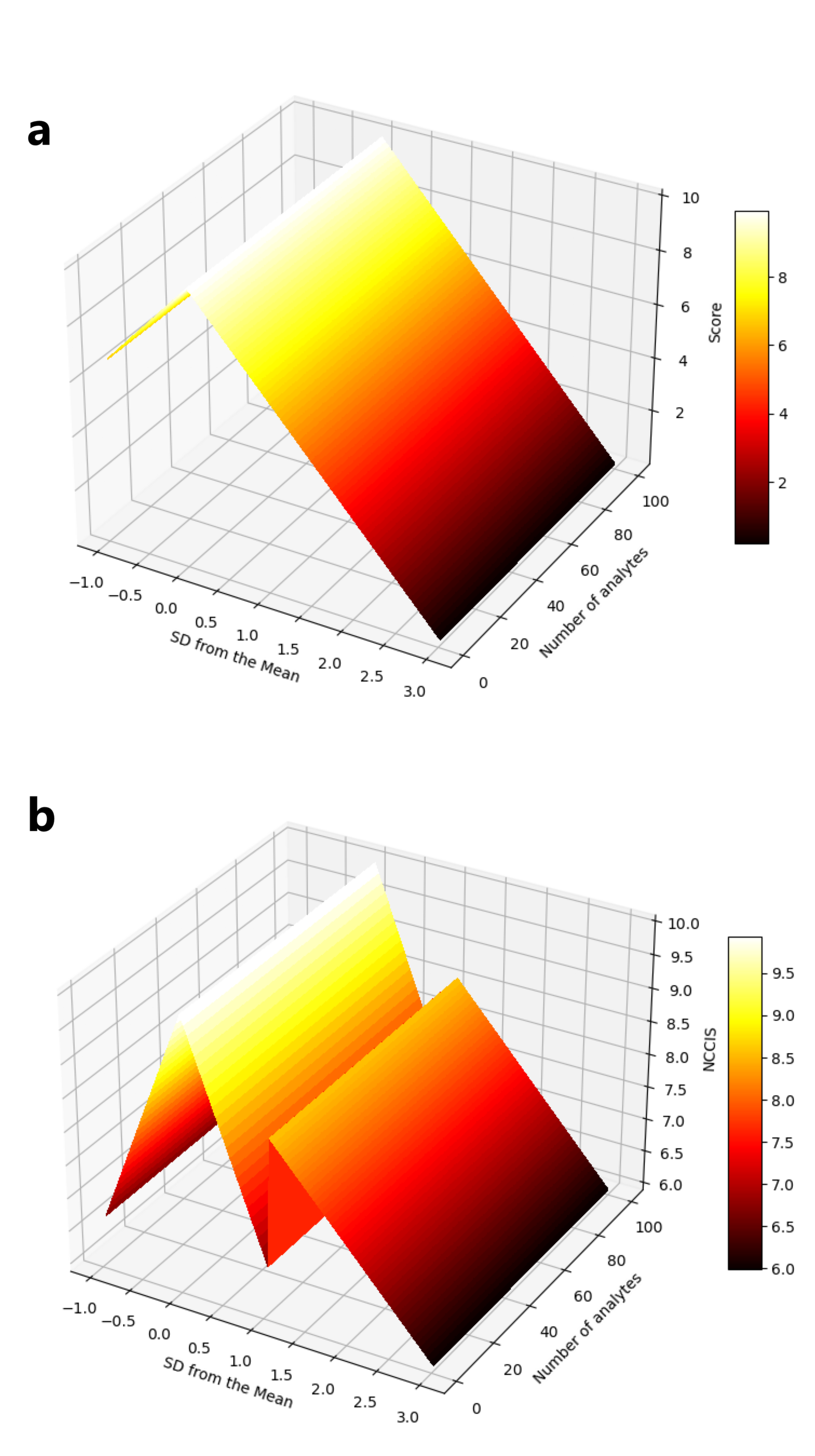}} 
	\caption{(a) Immune score space: The $x$ axis represents the deviation of all the analytes from the expected value, while the $y$ axis measures the total number of analytes. A \textit{negative} standard deviation indicates that the value is lower than expected and a positive one indicates a higher than expected value. The cut-off seen on the negative $x$ axis shows the fact that real-life values for analytes have a strict lower bound but, technically, have no upper bounds. For instance, it is not possible to have a negative white blood cell count. The behaviour of the numerical value is consistent through varying numbers of analytes and scales linearly with respect to deviation from the norm. This graph explores the behaviour of the immune score over synthetically generated patient values as a function of number of analytes and divergence from healthy reference values.(b) The behaviour of NCCIS over synthetically generated patients with differing values and a deviation from the norm with regard to the respective expected values. The $x$ axis measures the stated analyte, while the $y$ axis measures the total number of analytes. }
	\label{Score1and2:figure}
\end{figure}

 The score indicates how far the majority of all markers are from normal (healthy) reference values, the median, and an interval determined by published reference values for specific demographic or health conditions. The medians can be derived from empirical data (Fig.~\ref{nhanes}).

\begin{figure}[htb]
\begin{center}
\scalebox{.2}{\includegraphics{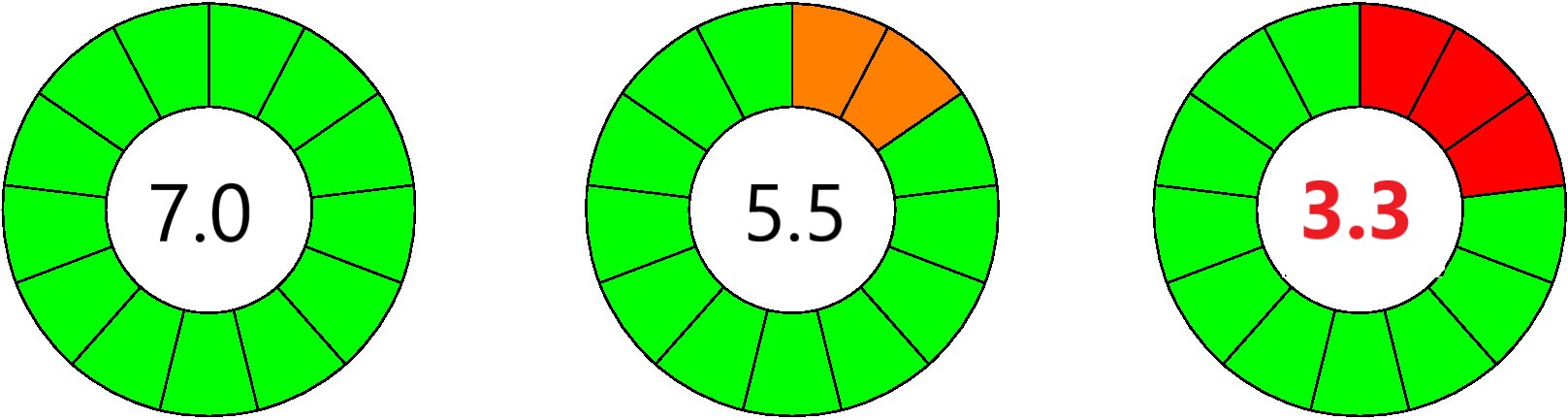}}
\end{center}
\caption{CCIS colour-based alerts. On the left, an individual with all values normal (healthy) and thus coded green. At the centre, an individual with two amber alerts. On the right-hand side, an individual with three red alerts, which produce a global red flag quantifying potential risk by colouring the numerical score depicted in the centre. In applications, colours following the guidelines of national health systems such as the NHS are recommended. For convenience, the numerical score has been inverted (with respect to $\text{NIS}^*$) to conform to the popular 0-10 grading scheme, where 10 is best and 0 is worst.}
\label{figure:ccis:green:amber:red}
\end{figure}
The information for the CCIS can be presented in a simple graphical way. We construct a doughnut graph with one slice for each analyte. Analytes assigned the same colour will be grouped together, which will produce doughnut graphs of at most 3 different colour-coded sections. Some examples for purposes of illustration are presented in Fig. \ref{figure:ccis:green:amber:red}
.

\subsection{Colour-based Ternary Score}

\noindent 
In addition to this numerical value, the score incorporates a second indicator that will be easy to read and interpret. The idea is to utilise colour codes as a means of flagging deviations from normal (healthy) reference values.
This is called the \emph{Colour-Coded Immune Score} or CCIS for short.

The CCIS uses the same analytes and their normal (healthy) ranges as before. There are thresholds for each of the analytes that trigger different types of alerts:

\begin{itemize}
 \item A value within normal healthy ranges: No alert, colour coded  \emph{green}.
 \item An abnormal value within one standard deviation above/below healthy reference values: \emph{amber} alert.
 \item An abnormal value beyond one standard deviation from the normal healthy values: \emph{red} alert. As in the normalised immune score, values are capped to a maximum of two standard deviations.
\end{itemize}

Keeping the range to within two standard deviations when assessing a score or metric has the following statistical implications: 1) About 95\% of observations fall within 2 standard deviations of the mean in a normal distribution. Thus, restricting the range cuts off the most extreme 5\% of outliers. This makes the statistics more robust by removing very unusual values (outliers) that could skew results. But it also reduces the variability seen in the data. However, it may also provide an overly narrow or skewed view of the true variability in the evaluated metrics. For the goals of our analysis, i.e., sample CBC analysis, assuming a normal distribution justifies this rationale and its implications.

The results for each analyte are also combined in a single global value, which will constitute the CCIS. The rules are as follows:

\begin{itemize}
 \item Green: all values within normal range (green).
 \item Amber: 1-3 individual amber alerts.
 \item Red: More than 3 amber alerts or one or more red alerts.
\end{itemize}

Clearly, the CCIS is not a numerical value but a colour-coded output, like the individual alerts for separate analytes. This clearly distinguishes the CCIS from  the normalised immune score.

\subsection{Numerical score Derived from the CCIS}

We explored an alternative score that would follow the colour-coded scheme, while also producing a numerical value. The intention was to combine the accessibility of the CCIS and the finer precision of the NIS in a single value. As with the latter, the value would range from 0 to 10.

We expected the following property: If $v_1, v_2$ are possible score values and $v_1 > v_2$, it should be the case that $v_2$ indicates greater cause for concern than $v_1$.

On the other hand, values should be closely related to the colour codes. That is, values $v_1$, $v_2$ and $v_3$ correspond to CCIS scores of green, amber, and red, respectively, and they should be ordered as follows:
$$v_1 > v_2 > v_3.$$

A more precise way of expressing the above ideas is the set of rules codified in Table~\ref{colours}.

In the following section, we present a procedure for translating the CCIS into a numerical value according to the above schema. We will call this value the NCCIS. The expected behaviour of the NCCIS concerning the number of analytes and its deviation from the norm can be observed in Fig.~\ref{Score1and2:figure},

\begin{table}[h!]
\caption{NHS reference values: Typical tabular presentation for illustration purposes of a Complete/Full Blood Count test. Minimum and maximum values show minimum and maximum between male/female values. Typical values used in the U.S. are available at \url{https://myhematology.com/red-blood-cells/full-blood-count-and-other-hematology-reference-ranges/}}
\centering

\begin{tabular}{l|r|r|r}
\toprule
\textbf{Parameter/Component} & Result & Low Ref Value & High Ref Value \\
\midrule
Haemoglobin (HGB) & 152.3 g/L & 115 g/L& 180 g/L\\
Total White Cell Count (WBC) & 9.3 g/L& 3.6 g/L& 11.00 g/L\\
Platelet count (PLT) & 286.5 $\times 10^9$/L& 140 $\times 10^9$/L& 400$\times 10^9$/L \\
Red cell count (RBC) & 5.5 $\times 10^{12}$/L & 3.8$\times 10^{12}$/L & 6.5$\times 10^{12}$/L\\
Mean Cell Volume (MCV) & 93.3 fL & 80 fL& 100 fL\\ 
Haematocrit (HCT)& 0.4 L/L & 0.37 L/L & 0.47 L/L \\
MC Haemoglobin (MCH)& 29 pg & 27 pg & 32 pg \\
MCH Concentration (MCHC)& 315 g/L & 310 g/L & 350 g/L \\
Neutrophils (NEUT)& 6.9 $\times 10^9$/L & 1.8 $\times 10^9$/L & 7.5$\times 10^9$/L \\
Lymphocytes (LYMPH)& 3.6$\times 10^9$/L & 1.0$\times 10^9$/L & 4.0$\times 10^9$/L \\
Monocytes (MONO)& 0.3$\times 10^9$/L & 0.2 $\times 10^9$/L& 0.8 $\times 10^9$/L\\
Eosinophils (EOS)& 0.2$\times 10^9$/L & 0.1$\times 10^9$/L & 0.4$\times 10^9$/L \\
Basophils (BAS)& 0.2$\times 10^9$/L & 0.02 $\times 10^9$/L& 0.1$\times 10^9$/L \\
[1ex] 
\bottomrule
\end{tabular}

\label{table:nhs}
\end{table}

\begin{table}[htb]
	\caption{Mean values and standard deviations found in the NHANES 2003-2016 database for the 13 analytes of the immune scores (same units as above).}
	\begin{center}
		\begin{tabular}{l|r|r|r|r}
			\toprule
			\textbf{Analyte}&Mean Male&Std Male&Mean Female&Std Female\\
			\midrule
			Haemoglobin (HGB)&150.72&11.51&133.33&11.62\\
		
			Total White Cells (WBC)&6.92&3.12&7.01&2.09\\
		
			Platelet count (PLT)&228&54.33&255.42&64.19\\
			
			Total Red Cells (RBC)&4.93&0.45&4.40&0.38\\
			
			MCV&89.96&5.13&89.22&5.72\\
		
			Haematocrit (HT)&0.44&0.03&0.39&0.03\\	
			MCH&26.96&10.33&26.55&10.47\\
		
			MCHC&336.55&18.41&335.02&20.46\\
			Neutrophils (NEUT)&4&1.85&4.16&1.66\\
		
			Lymphocytes (LYMPH)&2.11&2.21&2.11&0.79\\
			
			Monocytes (MONO)&0.57&0.21&0.52&0.18\\
			Eosinophils (EOS)&0.20&0.16&0.17&0.14\\
			
			Basophils (BAS)&0.04&0.06&0.04&0.05\\
			\bottomrule
		\end{tabular}
		\label{test}
	\end{center}
	\label{hanes:table:means:sd}
\end{table}


The behaviour of Fig.\ref{Score1and2:figure},illustrates how values at a remove from healthy ones are pushed towards higher values by design, as an alerting mechanism.


The computation of NCCIS is performed using the uninverted $\text{NIS}^*$. Some of the values used in the calculation of the $\text{NIS}^*$ include:

\begin{itemize}
    \item[] $w_a = \sqrt{100/13}$: the maximum normalised distance for each analyte in the normalised vector of a patient's readings.
    \item[] $e(g,i)$: the mean (or expected) value for analyte $i$ in group $g$.
    \item[]$\sigma(g,i)$: the standard deviation from the mean.
    \item[] $n(g,i) = (r_u(g,i)-r_l(g,i))/2$ : the normal (healthy reference) interval length from the mean value (assuming upper and lower limits are equidistant from the mean, an assumption to be revised in the future, both in the NIS and here in the NCCIS).
\end{itemize}

Now the raw (not normalised) borders between colours are

\begin{itemize}
    \item[]raw maximum value for green: $n(g,i)$.
    \item[]raw maximum value for amber: $n(g,i)+\sigma(g,i)$.
    \item[]raw maximum value for red: $n(g,i) + 2\sigma(g,i)$.
\end{itemize}

And the normalised version:

\begin{eqnarray*}
m_{gr}(g,i) &=& n(g,i) \times (w_a/(n(g,i) + 2\sigma(g,i)))\\
m_a(g,i) &=& (n(g,i) + \sigma(g,i)) \times (w_a/(n(g,i) + 2\sigma(g,i)))\\
m_r(g,i) &=& (n(g,i) + 2\sigma(g,i)) \times (w_a/(n(g,i) + 2\sigma(g,i))) = w_a
\end{eqnarray*}

Therefore the current $\text{NIS}^*$ intervals for the different colours for individual analytes are:

\begin{itemize}
    \item Green: $[0, m_{gr}(g,i)]$ (that is, from 0 distance from the mean value up to the normalised maximum value for green.)
   \item Amber: $(m_{gr}(g,i), m_a(g,i)]$.
   \item Red: $(m_r(g,i), w_a]$.
\end{itemize}

For global values, the calculation has to take into account both the definition of the CCIS and the fact that each analyte can have proportionally different normal ranges and standard deviations:

\begin{itemize}
    \item \emph{Green}: The minimum global green corresponds to 0 in every analyte. The maximum is $m_{gr}(g,i)$ in each analyte. This gives us the following interval:
$$[0, \sqrt{\sum^{13}_{i=1} m_{gr}(g,i)^2}]$$
    \item \emph{Amber}: The minimum global amber is 1 analyte value, just above green. The maximum is 3 maximum amber values and the rest (10) at the upper limit of green. Let analyte $j$ be such that $m_{gr}(g,j)$ is the minimum of the green upper limits, and let analytes $k$, $m$, $n$ be the three biggest of the amber upper limits. Then we have the following interval:
    
    \noindent $(m_{gr}(g,j),$
    
    $\quad\sqrt{(m_a(g,k)^2 + m_a(g,m)^2 + m_a(g,n)^2) + m_{gr}(g,i_1)^2 + \cdots m_{gr}(g, i_{10})^2}].$
    
    \item \emph{Red}. The minimum global red is 1 red analyte value or 3 amber analyte values. The maximum is, obviously, 10 (13 maximum individual analyte values). Let us suppose that the lowest minimum for red is $(m_r(g,k))$ and that there are no 3 upper limits for amber analytes whose sum is below this. Then the intervals are:
$$(m_r(g,k), 10].$$
\end{itemize}
We will designate the ends of these intervals $\mbox{min}_{green}$, $\mbox{max}_{green}$, $\mbox{min}_{amber}$, $\mbox{max}_{amber}$, $\mbox{min}_{red}$ and $\mbox{max}_{red}$.

The $\text{NCCIS}^*$ is calculated using the following function:
\textcolor{black}{
\begin{eqnarray*}
\mbox{NCCIS}^*(v) &=& \IF \mbox{CCIS}(v) = \mbox{green } \THEN a_{green}(\mbox{NIS}^*(v))\\
                && \ELSE \IF \mbox{CCIS}(v) = \mbox{amber } \THEN a_{amber}(\mbox{NIS}^*(v))\\
                && \qquad \ELSE a_{red}(\mbox{NIS}^*(v))
\end{eqnarray*}
}
where CCIS and \textcolor{black}{$\text{NIS}^*$} are functions calculating the respective scores and the functions $g_a$, $a_a$ and $r_a$ map the \textcolor{black}{$\text{NIS}^*$} to the intervals set in the table at the beginning of Section 4:
\begin{eqnarray*}
a_{green}(x) &=& w_{green} \times x\\
a_{amber}(x) &=& (10/3) + w_{amber}\times(x - \mbox{min}_{amber})\\
a_{red}(x) &=& (20/3) + w_{red}\times(x - \mbox{min}_{red})
\end{eqnarray*}
and the $\text{NIS}^*$ value is weighted according to a normalising weight:
\begin{eqnarray*}
w_{green} &=& 10/(3 \times \mbox{max}_{green})\\
w_{amber} &=& 10/(3 \times (\mbox{max}_{amber}-\mbox{min}_{amber}))\\
w_{red} &=& 10/(3 \times (10-\mbox{min}_{red}))
\end{eqnarray*}
\textcolor{black}{Finally, as in the case of NIS, we invert the scale to obtain the NCCIS: $$\text{NCCIS} = 10 - \text{NCCIS}^*.$$}

\begin{table}[htb]
\caption{Mapping of score segments and colours. 
\textcolor{black}{For convenience, the numerical score has been inverted to conform to the popular 0-10 grading scheme, where 10 is best and 0 is worst.}}
\begin{center}
\begin{tabular}{c|c}
\toprule
     \textbf{Colour from CCIS} & \textbf{Ideal interval for a numerical score}\\
     \midrule
     Green& \textcolor{black}{$(20/3,10]$}\\
     
     Amber&\textcolor{black}{$(10/3,20/3]$}\\
     
     Red&\textcolor{black}{$[1,10/3]$}\\
\bottomrule
\end{tabular}
\end{center}
\label{colours}
\end{table}

\subsection{Linearisation and Separability}


For example, infection can increase neutrophils and decrease lymphocytes. The parameters of the RBC can also be affected by complex environmental signals such as nutrient levels, iron deficiency, stress conditions, or disorders. We also assume that each analyte contributes equally to the immune index (this is likely to be only partially true, as some of the analytes are more or less medically informative than others, with this degree of informativeness itself varying with different conditions. Moreover, not all analytes are independent, and some analytes may be statistically dependent on others---suitable refinements will be introduced later). Some examples of analytes that are independent of each other are RBC and WBC count (each does not directly influence the other count). The counts of some analytes are dependent on each other. Examples include Neutrophils and WBC. In essence, neutrophils are a type of leukocyte (WBC), so the neutrophil count is included in and contributes to the total WBC. Another example involves Haemoglobin and Haematocrit levels. Haemoglobin measures the amount of this protein in red blood cells, while Haematocrit measures the percentage of blood volume taken up by red blood cells. Thus, Haematocrit is mathematically dependent on the haemoglobin concentration because one represents a subset or a ratio of the other. The full mathematical description is given in the Supplementary Information.

\textcolor{black}{While the calculation of the score can incorporate normal ranges adjusted for age, race, sex, pregnancy stage, geography and any other consideration warranted by the literature, the score is intended to go beyond this clinical utility by capturing medical knowledge.} A feature of the proposed (immune) score is the option to incorporate weights as multipliers in the form of scalars or piece-wise functions per analyte to modify its contribution relative to other analytes in a non-linear fashion, even under time sensitive
conditions \cite{day-time-var}. For example, in a blood differential test, conditions related to decreased white cell counts are milder than those associated with higher cell counts. However, the literature on conditions where there is a decrease of basophils, mast cells, monocytes and eosinophils in isolation is sparse, and reduced weights can be assigned to these possibly less-relevant markers.

Differences in cell shape and size, however, are usually more clinically significant and can be assigned greater weights. Typically, disorders affecting bone marrow function (i.e., blood cancers) result in the presence of abnormal (often immature) cells in peripheral blood; their presence in peripheral blood beyond this level would almost always be abnormal. A related point is whether the score can reflect subtle/minor changes in shape (variation in cell size, nuclear size, presence of other organelles) or only very crude and major deviations in size. For white blood cells, we do not yet understand what significance these have.

For example, the neutrophil-to-lymphocyte ratio relation is a key determinant of severity for sepsis, and involves 2 analytes. The synthetic analyte that can be added is the ratio itself, thus replacing a non-linear rule that would eventually make the score's description too convoluted to read with ease with 

We observed that the self-reported health status of the patients did not accurately predict their clinically assessed blood analytes. This was expected because self-reports may have reflected illness at various points in time in a person's life rather than ongoing conditions. In other words, with this experiment, we de-noised the data by disregarding self-reported status to test the score against cleansed data based exclusively on CBC results as a discriminant, according to medical guidance on normal population reference values, that is, the interpretation of their health status a person could have potentially received from a medical doctor based exclusively on their CBC. 

When healthy and unhealthy patients were separated using a subset of four CBC/FBC analyte measures, namely, MCH, HCT, neutrophil, and lymphocyte counts,  we obtained the second best $R^2$ value of 0.075, for both sexes. The $R^2$ values were 0.050 and 0.123, for the male and female groups, respectively. We found that neutrophil count alone, as a single clinical parameter, delivered a performance nearly equal to that of the composite measure of these four analyte parameters, with an $R^2$ of 0.072 when including both sexes. The inclusion of other white blood cell (WBC) parameters, such as monocyte count, eosinophil, and basophil, did not outperform these correlates. We observed that different subsets of a CBC/FBC by health status separated the data. All reported measures were statistically significant, with a two-way ANOVA P$<0.0001$. The weak correlation analysis suggests limited predictive power of CBC analytes alone for distinguishing healthy from unhealthy individuals, particularly when applied to heterogeneous populations with broadly defined health status using routine blood measures. Therefore, these findings highlight the rationale for proposing the immune score as a diagnostic tool, as it offers significantly stronger predictive power than individual analyte counts, repurposing routine, accessible blood tests into a more informative clinical measure.



This further suggests that the mutlivariate biopsychosocial determinants of health should also be integrated for a top-bottom analysis of the  disease predisposition and health dynamics of patients when assessing such self-reported measures. The irreducibility (multidimensionality) of these self-reported health measures may help explain why they are highly noisy and intersect with multiple other health factors, including multi-scale stressors such as the time of day, basic survival factors such as nutritional status/intake, sleep-related mood, psychological state, and social determinants of health, together with other health parameters.

\subsection{Adaptive Reference Values}

The reference ranges found in the medical literature are based on population-wide statistics that often fail to account for individual differences ~\cite{brodin1,brodin2}. To address this, we introduced an adaptive version of the immune score with the objective of providing personalised reference ranges to an individual's medical history data. 
Our rationale is that, in applying the immune score to the resulting adaptive ranges, we obtain an immune score that is more significant in the statistical and clinical sense. 

The main underlying assumption of this adaptive immune score definition is that if a significant event (in the clinical sense) can be measured via an analyte, such an event will present itself as a value that breaks from the trend defined by the behaviour of the analyte over time. A second assumption is that such events can occur within the \emph{normal range} as defined by a population reference range. A third assumption is that while a trend becomes more incontrovertible the more data points there are, past values are less significant than proximate events. The next assumption is that the trend can be non-monotonous. The final assumption is that standard reference ranges can detect such events with acute sensitivity, while low specificity can thus be improved upon by narrowing the ranges but not by expanding them.


The adaptive feature of the immune score starts with a mathematical function that models the behaviour of an analyte as a time series (observed values over time). For each given point, the function assigns a linear approximation to the current and past data over a fixed time span (called a \emph{time step}). The differential of this approximation is weighed against the differentials obtained for previous data via a linear combination of the current and previous derivatives. Finally, this linear approximation is evaluated over the mean time for the corresponding time-step. The resulting model is a function that carries the momentum of all previous values, and only previous values, yet remains adaptable for future long-term trends while being robust in the face of temporal disruptions to the trend. 
The speed of this adaptation can be controlled by hyper-parameters such as the size of the time step and the coefficients for the linear combination of differentials. 


The function described in the preceding paragraph represents the \emph{adaptive mean}, from which the \emph{adaptive ranges} are inferred. The adaptive ranges consist of two functions, the \emph{upper limit} and the \emph{lower limit}. The upper limit is defined as the adaptive mean plus the standard deviation of all the values that are above the mean, up to the point of time in question. Similarly, the lower limit is the adaptive mean minus the standard deviation of all the values that are below the adaptive mean up to the relevant point in time.  Next, we smooth the transition from the population ranges to the adaptive ranges by means of a linear combination of both ranges, such that the adaptive range becomes the dominant factor in the time function. 

Furthermore, this linear combination ensures that the adaptive ranges stay within the population ranges by defaulting to the population ranges whenever the resulting adaptive maximum is above the population maximum or the adaptive minimum is below the adaptive mean.






For ease of comparison between successive versions of the score, all values are normalised within the range $[0,10]$. Following user feedback, we then reverse the scale by subtracting the normalised value from 10. This means that each analyte will contribute with a maximum weight of $$w_a = \sqrt{\frac{100}{N}},$$ 
where $N$ is the number of analytes. From now on we will consider the example of $N=13$ typical for a CBC or FBC. Before the scale inversion, from a geometrical point of view, a vector of values in the multidimensional immune space will be mapped to a point in a $N$D-``quadrant'' going from 0 to $w_a$ that will be called the \emph{normalised immune space}. This mapping will allow us to observe the evolution of an individual's score as a trajectory in the normalised immune space, where a notion of distance will come in handy.

While the score proposed is agnostic, in the sense that it can incorporate, or not, any number of analytes, and requires no other input, it can be adapted to incorporate medically relevant information by assigning weights that serve as multipliers of an analyte's effect according to its criticality for different profiles. 

The adoption of a score within a maximum range of $[0,10]$ can be approached in different ways. One is to assume no theoretical maximum values for analytes and asymptotically approach the minimum score of 0, without ever reaching it in individual cases. We adopted an alternative approach, capping values to a pre-established ``reasonable'' maximum, beyond which all specific values mean the same. The pre-established maximum is two standard deviations beyond normal (healthy) ranges, which sets the statistically significant critical value alpha at 0.05 (i.e., 95\% confidence interval).

In an initial test, reference tables from \emph{Haematology Reference Ranges} from the NHS were used for normal (healthy) ranges for each analyte according to age, sex, and pregnancy status~\cite{nhsYork}. $G$ will denote the set of possible categories:

\begin{center}
\begin{tabular}{ccl}
     $G$&=&$\{$Adult male, non-pregnant adult female, pregnant adult female,\\
     &&newborn child, two-month-old child, six-month-old child,\\
     &&one-year old child, 2-6-year-old child, 6-12-year-old child, 12-18-year-old teenager$\}$\\
\end{tabular}
\end{center}

\noindent Let $g$ denote any value in $G$.

For each analyte, $r_u(g,i)$ is the upper limit in the normal range of the analyte $i$ for an individual belonging to group $g$. Similarly, $r_l(g,i)$ is the lower limit.

An efficient computer implementation of the score can pre-calculate the following values:

\begin{itemize}
  \item The \emph{expected vector}, which contains the mean value of each analyte for a given group:
$$\bar e(g) = (e(g,1), \dots, e(g,N)).$$
Mean values will derive from the analysis of suitable data to be obtained from reliable databases or directly collected by us. For ease of explanation, in the following we provisionally select the arithmetical average of the lower and upper limits in the NHS table:
$$e(g,i) = \frac{r_l(g,i)+r_l(g,i)}{2}.$$
  \item The standard deviation for an analyte $i$ in group $g$ is denoted by
  $$\sigma(g,i).$$

Again, these values will be empirically calculated. In the meantime, we will use the distance from the mean to either limit within the normal (healthy) range:
  $$\sigma(g,i) = \frac{(r_u(g,i)-r_l(g,i))}{2}.$$
  \item The \emph{maximum difference vector}, containing the maximum possible distance from the mean value of each analyte:
$$\bar m(g) = (m(g,1), \dots m(g,N)),$$
where $$m(g,i) = \frac{(r_u(g,i)-r_l(g,i))}{2} + 2\sigma(g,i).$$
(twice standard deviations from the limits of the healthy ranges).
  \item The \emph{weighted vectors} of each group (which will be used to normalise the possible values to a maximum immune score, before reversion, of 10):
$$\bar w(g) = (w(g,1), \dots, w(g,N)),$$
where $w(g,i) = w_a/m(g,i)$.
\end{itemize}

Let $\bar c = (c_1, \dots, c_{N})$ be a vector of analyte values for an individual of group $g$. Then the normalised vector is calculated using the following formula:
$$\bar n(\bar c) = (f(c_1,e(g,1),m(g,1),w(g,1)), \dots, f(c_{N},e(g,N),m(g,N),w(g,N))),$$
where
$$f(x,y,z,w) = \IF |x-y| \geq z \THEN w_a \ELSE |x-y| \times  w.$$

The value obtained by subtracting the normalised vector \(\bar{n}(\bar{c})\) from 10 is defined as the Normalised Immune Score, abbreviated as NIS:
\[\text{NIS} = 10 - \|\bar{n}(\bar{c})\|.\]
We will denote the uninverted score by \(\text{NIS}^*\), which is defined as \(\text{NIS}^* = \|\bar{n}(\bar{c})\|\). 

One novel aspect of NIS is the ability to incorporate medical knowledge into the score calculation using weights (scalar or piece-wise functions). This allows adding clinical expertise to the data-driven score in a transparent mathematical way, rather than using opaque statistical black-box approaches. Statistical black-box approaches refer to complex statistical modelling techniques that are not easily interpretable (e.g., neural networks). The weights make it possible to adjust analyte contributions in a nonlinear fashion if needed. Synthetic analytes can also be introduced to simulate nonlinear relationships in a clear mathematical form, rather than dealing with nonlinearities directly in a statistically obscure way. Unlike statistical or machine learning approaches, the distance-metric-based score is transparent, so that clinicians can fully understand how it works, and even reproduces the calculation by hand with a justification at every step without loss of generality.

\subsection*{ML Algorithms' Descriptions}

\textbf{K-Nearest Neighbors (KNN)}: The K-NN algorithm, implemented using the \texttt{sklearn.neighbors} package, is a non-parametric method used for classification. The primary hyperparameters include: \texttt{n\_neighbors} set to a default of 5. The \texttt{weights} parameter determines how the votes of the neighbours are weighted during the decision trees. It was kept at default settings of 'uniform'. The \texttt{algorithm} parameter was also kept in the default state of 'auto' option, as it attempts to choose the optimal algorithm based on the input data structure. Additionally, the \texttt{leaf\_size} parameter was set to a value of 30.

\textbf{Support Vector Machine (SVM)}: Available through the \texttt{sklearn.svm} package, this is a powerful supervised learning model. The key hyperparameters include \(C\), the regularisation parameter to regulate the trade-off between achieving a low error on the training data and minimising the model's complexity. \(C\) was set to a default value of 1.0. The \texttt{kernel} parameter was kept at 'rbf', being the default. The \texttt{gamma} parameter (the kernel coefficient for 'rbf') was kept at a default value 'scale', which uses \( \frac{1}{\text{n\_features} \times \text{X.var()}} \). All other parameters were kept at default.

\textbf{Random Forest (RF)}: Implemented using the \texttt{sklearn.ensemble} package, this is an ensemble learning method for classification that constructs multiple decision trees during training and outputs the mode of the classes. The algorithm detects implicit regularisation of patterns. The primary hyperparameters for RF include \texttt{n\_estimators} (the number of trees in the forest) at a default value of 100. The \texttt{criterion} parameter 
that defines the function to measure the quality of a split was kept at `gini'. The \texttt{max\_depth} was kept at default (None) and \texttt{min\_samples\_split} was kept at a value of 2.

\textbf{Artificial Neural Network (ANN - Feedforward Neural Network)}: A multilayer perceptron with one input layer and two hidden layers was used. The first hidden layer consisted of 64 neurons with ReLU activation, followed by a dropout layer with a dropout rate of 0.5 to prevent overfitting. The second hidden layer had 32 neurons with ReLU activation. The output layer consisted of 1 neuron with Sigmoid activation for binary classification. The primary hyperparameters included the Adam optimiser with a learning rate of 0.001, the loss function (binary cross-entropy), and accuracy metrics. Changing the first hidden layer to 128 neurons and the second hidden layer to 64 neurons did not make a difference as the initial model was already at optimal performance.

\textbf{Transformer-Based Neural Network (Transformer-ANN)}: Also implemented using the \texttt{keras} package from \texttt{tensorflow}. One input layer and one output layer (sigmoid activation function) with one neuron each. The input layer was reshaped into a sequence dimension before the attention mechanism. The attention mechanism was followed by dense layers with 64 neurons each and dropout layers. The dropout layer used a dropout rate of 0.5. The self-attention mechanism, using \texttt{keras.layers.MultiHeadAttention} consisted of two additional hyperparameters: The \texttt{num\_heads} (Number of attention heads) and the \texttt{key\_dim} (size of each attention head) were both set to 2. This was designed to handle sequential data for the detection of long-range dependencies. A flatten layer was applied to flatten the output of the attention layer prior to output (binary classification).

\subsection*{Notation}

\begin{table}[htbp]
\caption{List of blood-related label headers in the NHANES database}
\label{table:labels:nhanes}
\begin{center}
\begin{tabular}{ll}
\hline
\textbf{Label} & \textbf{Description} \\ \hline
LBXWBCSI & White blood cell count (1000 cells/uL) \\ 
LBXLYPCT & Lymphocyte percent (\%) \\ 
LBXMOPCT & Monocyte percent (\%) \\ 
LBXNEPCT & Segmented neutrophils percent (\%) \\ 
LBXEOPCT & Eosinophils percent (\%) \\ 
LBXBAPCT & Basophils percent (\%) \\ 
LBDLYMNO & Lymphocyte number \\ 
LBDMONO & Monocyte number \\ 
LBDNENO & Segmented neutrophils number \\ 
LBDEONO & Eosinophils number \\ 
LBDBANO & Basophils number \\ 
LBXRBCSI & Red blood cell count (million cells/uL) \\ 
LBXHGB & Haemoglobin (g/dL) \\ 
LBXHCT & Haematocrit (\%) \\ 
LBXMCVSI & Mean cell volume (fL) \\ 
LBXMCHSI & Mean cell haemoglobin (pg) \\ 
LBXMC & MCHC (g/dL) \\ 
LBXRDW & Red cell distribution width (\%) \\ 
LBXPLTSI & Platelet count SI (1000 cells/uL) \\ 
LBXMPSI & Mean platelet volume (fL) \\ \hline
\end{tabular}
\end{center}
\end{table}

\begin{itemize}
    \item 20 years of age or older as stated by \texttt{RIDAGEYR}.
    \item The \texttt{general health condition (all self-reported)} variable (\texttt{HSD010}) was stated as \texttt{Good}, \texttt{Very Good}, or \texttt{Excellent}.
    \item The values of the 13 analytes, along with general health condition, age and sex (\texttt{RIAGENDR}), were present in the data.
\end{itemize}


\subsection*{Supplementary Methods: Additional Clock Estimators}

Using our pooled CBC datasets across the NHANES cohort datasets, we benchmarked our immune score against CBC analytes-computable ``clock-like'' estimators constrained to the following features available across datasets: WBC, RBC, MCV, MCH, and MCHC. We computed (i) a Mahalanobis distance from the healthy reference distribution in this 5-feature CBC space (multivariate deviation from physiological norm), (ii) a z-distance (Euclidean distance from the healthy centroid in standardized CBC space), and (iii) a CBC Age Clock (delta age, or $\Delta$Age) trained on healthy subjects using ridge regression (best regularization achieved at $\alpha = 56.23$) to predict chronological age. This results in predicted CBC-age and age-acceleration (predicted minus chronological age). These CBC-only baselines provide a reasonable comparison to our immune score as there are currently no available CBC-only clocks. Further, existing biological ageing-based clocks such as methylation clocks (e.g., Horvath/Hannum) and composite biomarker clocks like PhenoAge or GrimAge, require DNA methylation or broader clinical chemistry panels rendering them incomputable from CBC counts alone. This limitation further highlights the clinical feasibility and accessibility of our routine blood test-derived score for patient care translation. By benchmarking against multivariate deviation metrics and immune-age proxies derivable strictly from hematologic counts, we ensure a modality-consistent comparison that reflects the maximal information extractable from routine CBC testing without incorporating multiomics or non-CBC biomarkers.

Further, we compared our immune score to real-world inflammatory immune ratios used in clinical context, namely: 1) NLR (Neutrophil-to-Lymphocyte Ratio) $=$ Neutrophils / Lymphocytes, 2) PLR (Platelet-to-Lymphocyte Ratio) $=$ Platelets / Lymphocytes, 3) SII (Systemic Immune-Inflammation Index) $=$ Platelets $\times$ Neutrophils / Lymphocytes, and 4) SIRI (Systemic Inflammation Response Index) $=$ Neutrophils $\times$ Monocytes / Lymphocytes. These are low-dimensional ratio-based markers and thus, reduce the immune complexity into 2--3 interacting cell types and are predicted to underperform compared to the multivariate models.

\begin{figure}[H]
\centering
\includegraphics[width=\linewidth]{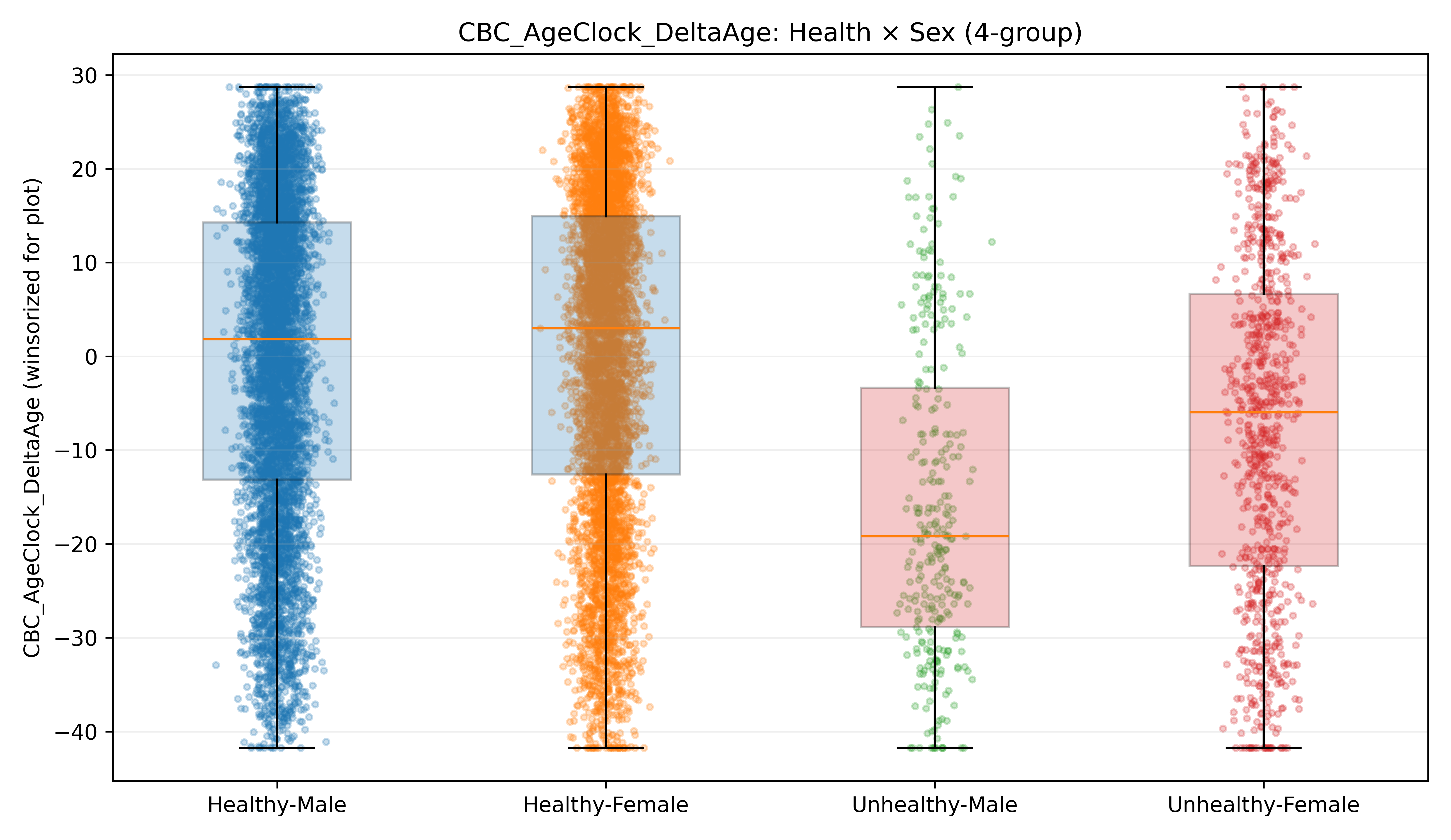}
\caption{\textbf{CBC Age Acceleration (DeltaAge) stratified by health and sex.} CBC\_AgeClock\_DeltaAge $=$ Predicted CBC-age $-$ chronological age, where Predicted CBC-age is generated by a ridge-regression CBC age clock ($\alpha \approx 56.23$) trained on healthy subjects using CBC-only features (WBC, RBC, MCV, MCH, MCHC). Unhealthy participants show significantly lower DeltaAge than healthy overall and within each sex. Among unhealthy individuals, males exhibit more negative DeltaAge than females.}
\label{fig:S4_deltaage}
\end{figure}

Figure~\ref{fig:S4_deltaage} shows the delta age from the Ridge Regression-based CBC clock simulation, where delta age is defined as Predicted CBC-age $-$ chronological age, wherein the Predicted CBC-age comes from a ridge-regression CBC age clock ($\alpha \approx 56.23$) trained only on healthy subjects using the shared CBC features (i.e., WBC, RBC, MCV, MCH, MCHC). Positive DeltaAge means the CBC profile appears older than expected for that person’s age (age acceleration), while negative DeltaAge means it appears younger. In the pooled comparison, unhealthy individuals show markedly lower DeltaAge than healthy (Healthy mean 0.00 vs Unhealthy mean $-9.79$; Cohen’s $d = 0.556$; Welch $p = 2.16\times 10^{-56}$). Stratifying by sex clarifies the pattern: within females, Healthy vs Unhealthy remains significant (mean 0.31 vs $-7.58$; $d = 0.443$; $p = 1.17\times 10^{-27}$), and within males the separation is even larger (mean $-0.30$ vs $-15.70$; $d = 0.890$; $p = 1.45\times 10^{-38}$). Within the unhealthy group, males are more negative than females (mean $-15.70$ vs $-7.58$; $d = -0.439$; $p = 4.76\times 10^{-11}$), while in healthy subject’s male--female differences are small/borderline ($p \approx 0.055$).

\begin{figure}[H]
\centering
\includegraphics[width=\linewidth]{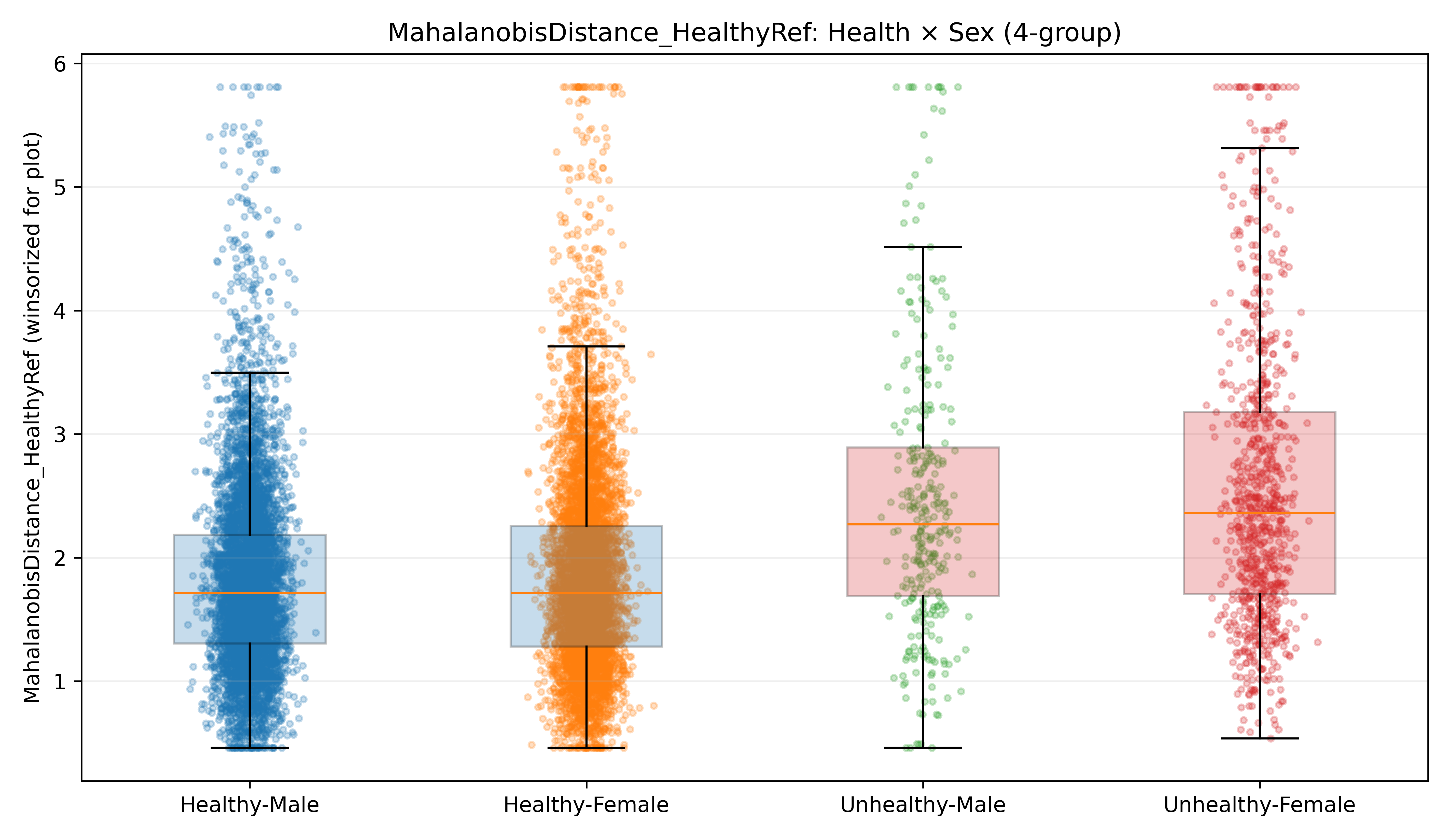}
\caption{\textbf{Multivariate CBC deviation from healthy reference stratified by health and sex.} Mahalanobis distance was computed in standardized CBC 5-feature space relative to the healthy training distribution. Unhealthy individuals exhibit increased multivariate hematologic deviation compared to healthy subjects, with minimal sex-dependent effects within strata.}
\label{fig:S5_mahal}
\end{figure}

Mahalanobis distance (Fig.~\ref{fig:S5_mahal}) shows clear separation between healthy and unhealthy individuals as a disease deviation hematologic manifold, with unhealthy subjects exhibiting substantially greater multivariate deviation from the normative hematologic centroid. Across the full cohort, the effect size for Healthy vs Unhealthy is moderate-to-large (Cohen’s $d \approx 0.68$; $p \ll 10^{-30}$), and the separation remains significant within both females ($d \approx 0.77$) and males ($d \approx 0.60$). In contrast, sex differences are negligible within both healthy and unhealthy strata ($p > 0.1$ in both cases). Thus, Mahalanobis distance, as a manifold learning algorithm, primarily captures health-related deviation rather than sex-driven differences (i.e., health-state discrimination), whereas the above defined delta age may be more insightful for capturing biological ageing dynamics.

\begin{figure}[H]
\centering
\includegraphics[width=\linewidth]{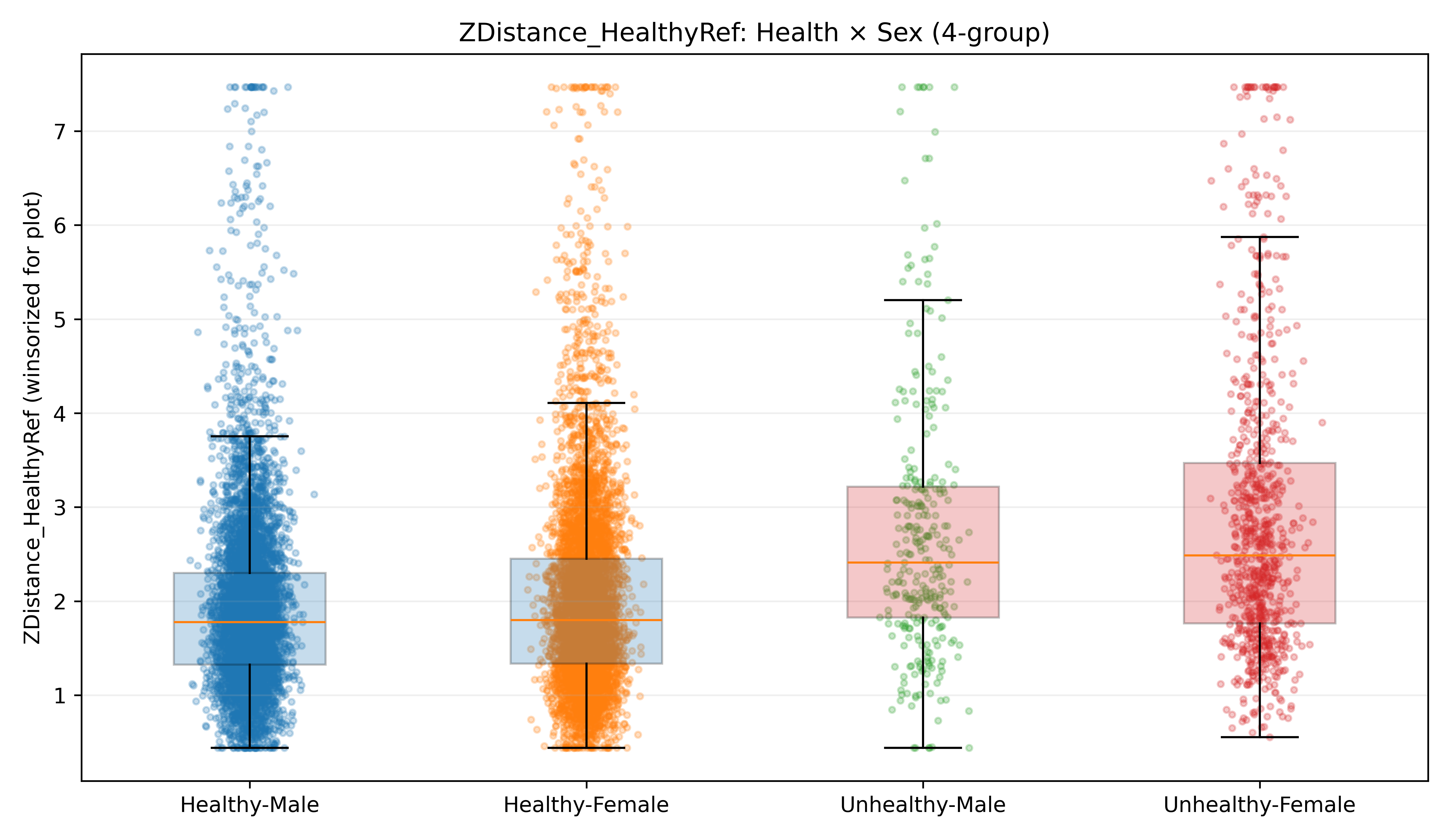}
\caption{\textbf{Z-distance from healthy CBC reference stratified by health and sex.} ZDistance quantifies Euclidean distance in standardized CBC feature space (WBC, RBC, MCV, MCH, MCHC) from the healthy reference centroid, thereby measuring global hematologic deviation from normative structure. Unhealthy individuals show substantially greater multivariate deviation than healthy subjects (mean 3.005 vs 2.024; median 2.487 vs 1.799), corresponding to a large effect size (Cohen’s $d \approx 0.76$) and highly significant separation (Welch $p \approx 2.1 \times 10^{-32}$). This discrimination remains robust within both males and females, while sex differences themselves are comparatively small, indicating that health status and not sex is the dominant driver of variation.}
\label{fig:S6_zdist}
\end{figure}

\begin{figure}[H]
\centering
\includegraphics[width=\linewidth]{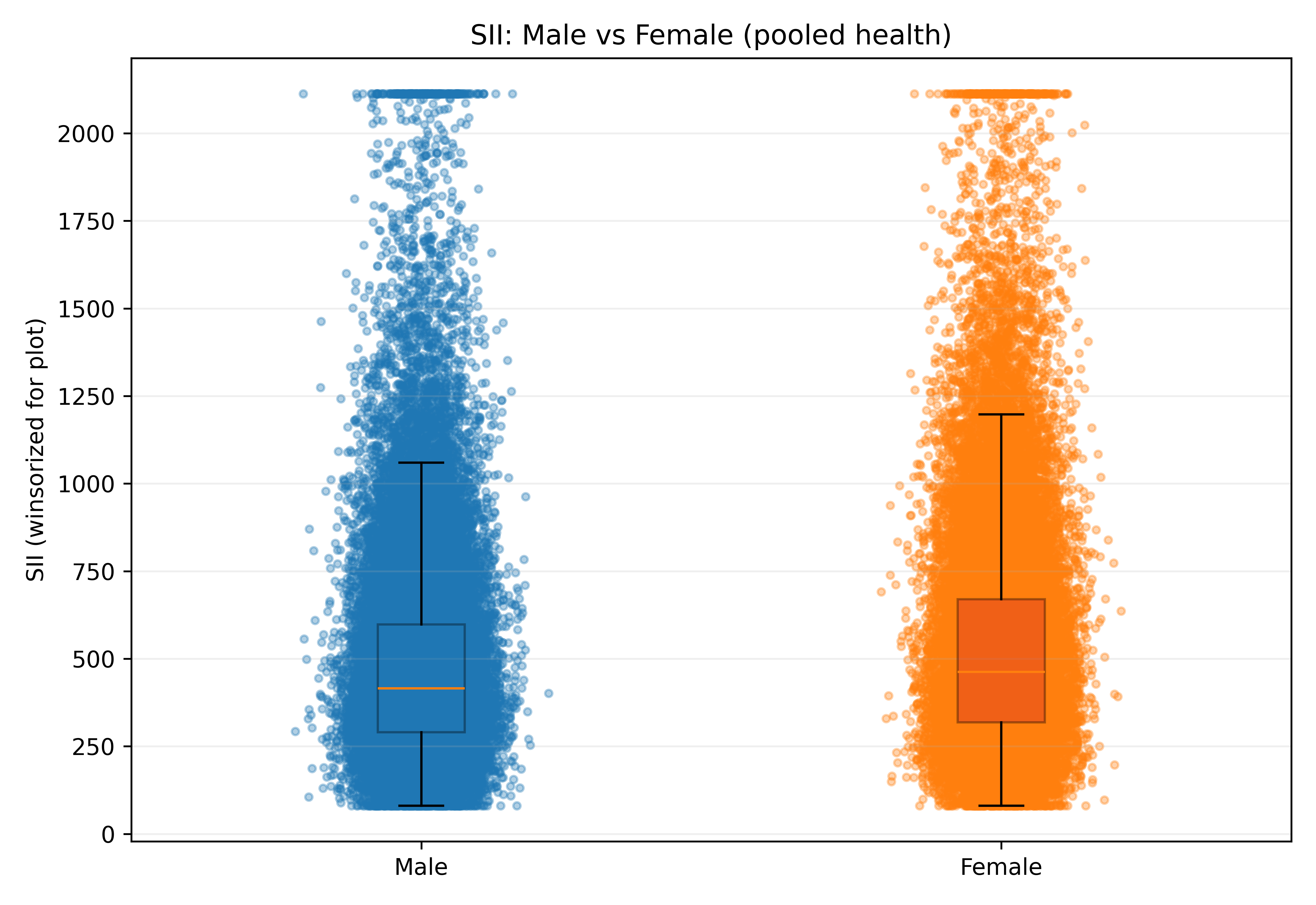}
\caption{\textbf{Systemic Immune-Inflammation Index (SII; platelets $\times$ neutrophils / lymphocytes) shows only a very small sex effect} (mean 490.2 vs 540.9; Cohen’s $d \approx 0.14$) despite extreme statistical significance from large sample size ($p \approx 4.8 \times 10^{-90}$), indicating weak biological separation compared to multivariate measures like Z-distance or Mahalanobis.}
\label{fig:S7_sii}
\end{figure}

\begin{figure}[H]
\centering
\includegraphics[width=\linewidth]{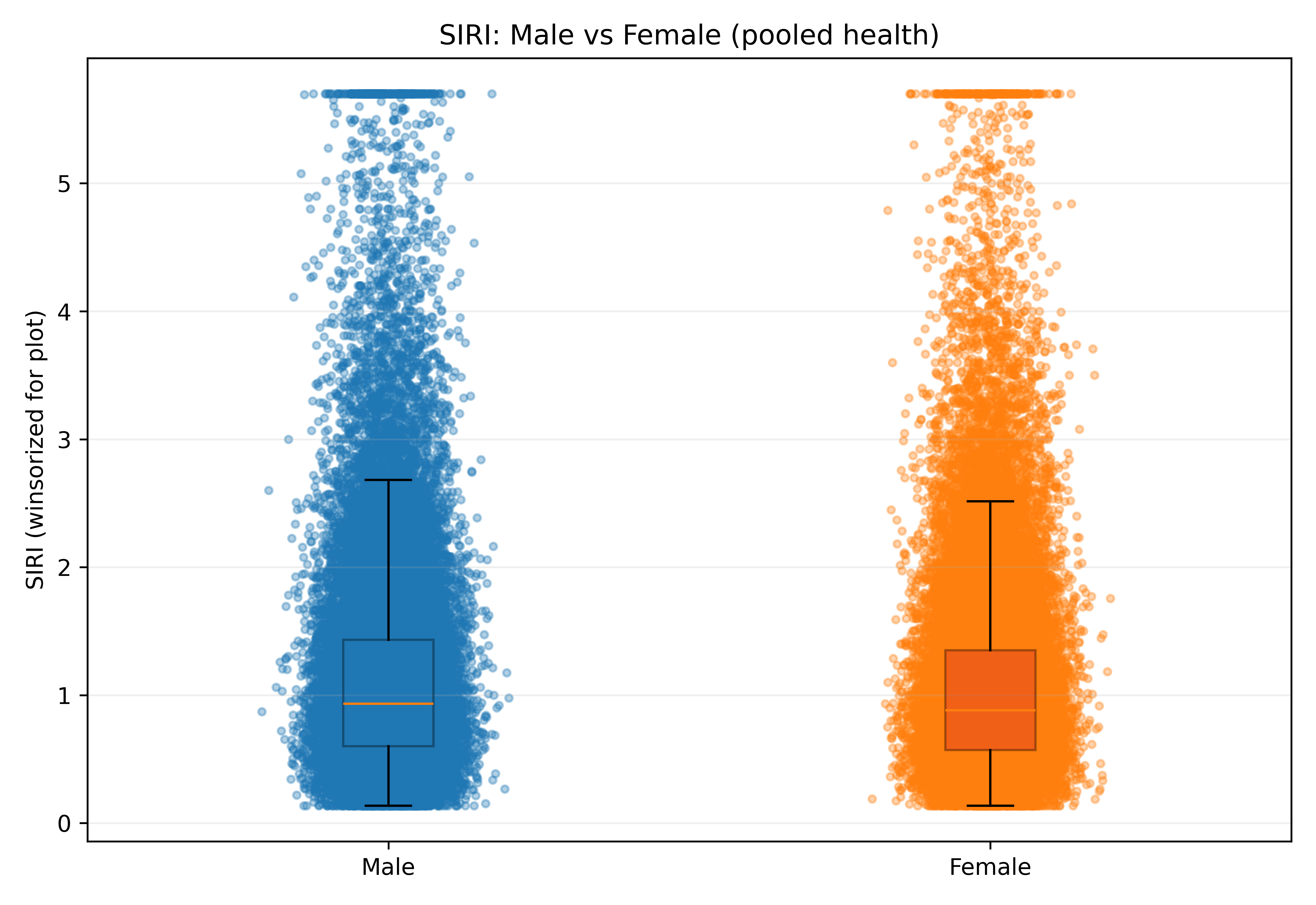}
\caption{\textbf{Systemic Inflammation Response Index (SIRI; neutrophils $\times$ monocytes / lymphocytes) demonstrates an even smaller sex effect} (mean 1.17 vs 1.09; Cohen’s $d \approx 0.08$) despite strong statistical significance from large sample size ($p \approx 1.4 \times 10^{-29}$), indicating biologically minimal separation and substantially weaker discriminative value than multivariate measures such as Z-distance or Mahalanobis.}
\label{fig:S8_siri}
\end{figure}

\begin{figure}[H]
\centering
\includegraphics[width=\linewidth]{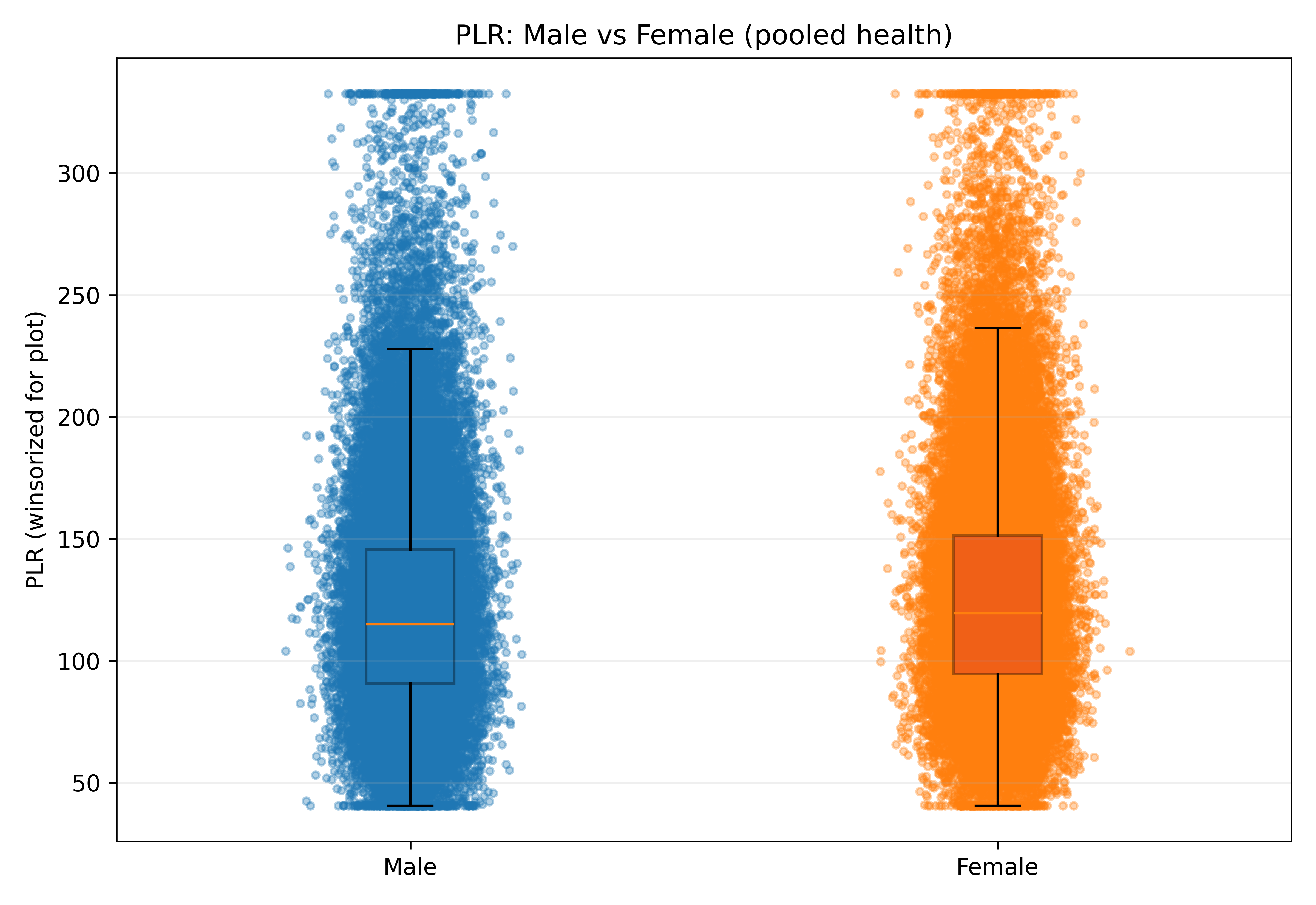}
\caption{\textbf{Platelet-to-Lymphocyte Ratio (PLR; platelets / lymphocytes) shows only a small sex effect} (Cohen’s $d \approx 0.095$) despite strong statistical significance driven by large sample size, indicating minimal biological separation between males and females. Its discriminative performance is modest---slightly stronger than SIRI and similar to SII---but substantially weaker than multivariate measures such as Mahalanobis distance or Z-distance, which capture coordinated hematologic structure rather than single-ratio shifts.}
\label{fig:S9_plr}
\end{figure}

\begin{figure}[H]
\centering
\includegraphics[width=\linewidth]{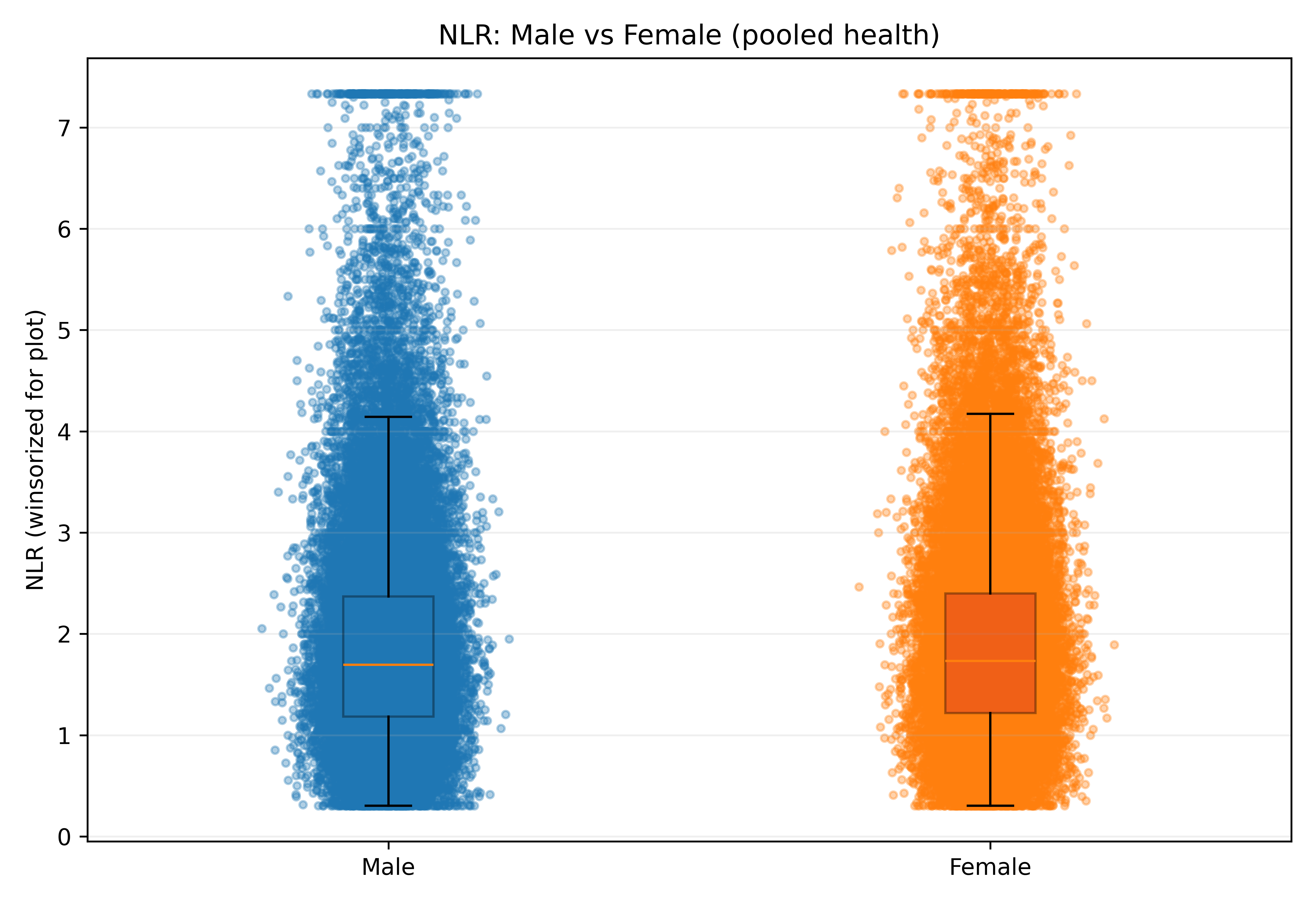}
\caption{\textbf{Neutrophil-to-Lymphocyte Ratio (NLR; neutrophils / lymphocytes) is the weakest of the evaluated indices,} with an essentially negligible effect size (Cohen’s $d \approx 0.016$), indicating biologically trivial male--female separation despite statistical significance driven by large sample size. It performs substantially worse than PLR, SII, and SIRI, and far below multivariate approaches such as Mahalanobis distance or Z-distance.}
\label{fig:S10_nlr}
\end{figure}

\subsection*{Conclusion}

Across models, multivariate approaches consistently outperformed immune ratio-based inflammatory indices. Among the distance-based methods, Z-distance demonstrated the largest effect sizes for Healthy versus Unhealthy discrimination (Fig.~\ref{fig:S6_zdist}), with Mahalanobis distance showing a similar separation due to covariance weighting (Fig.~\ref{fig:S5_mahal}). The ridge regression-derived CBC age clock ($\Delta$Age) captured a distinct, age-structured deviation and was sub-optimal in capturing health status discrimination (Fig.~\ref{fig:S4_deltaage}). In comparison, our immune score achieved discrimination comparable to or exceeding these distance-based metrics while maintaining greater clinical translatability and interpretability, as it models complex systems multivariate hematologic patterns rather than purely geometric deviation from a healthy feature space centroid. Our findings indicate that while Z-distance provides maximal raw separation, our immune score offers a more physiologically meaningful complex systems representation of system-level immune perturbation.

Further, our immune score outperforms these traditional ratio-based immune markers because it captures global hematologic structure rather than a single inflammatory axis (Figs.~\ref{fig:S7_sii}--\ref{fig:S10_nlr}). SII, SIRI, PLR, and NLR are low-dimensional, highly correlated composites derived from two or three cell types, limiting their capacity to detect coordinated immune--hematologic perturbations. By integrating multiple orthogonal CBC features and modeling their joint covariance structure, our score identifies system-level dysregulation rather than isolated ratio shifts, resulting in substantially larger, biologically meaningful effect sizes and clearer discrimination of health status.


\begin{thebibliography}{99}

\bibitem{naturescore} Bortz, J., Guariglia, A., Klaric, L. et al. (2023). "Biological age estimation using circulating blood biomarkers". Commun Biol \textbf{6}, 1089. doi:\url{https://doi.org/10.1038/s42003-023-05456-z}

\bibitem{levine2018epigenetic}
Levine, M. E., Lu, A. T., Quach, A., Chen, B. H., Assimes, T. L., Bandinelli, S., Hou, L., Baccarelli, A. A., Stewart, J. D., Li, Y., Whitsel, E. A., Wilson, J. G., Reiner, A. P., Aviv, A., Lohman, K., Liu, Y., Ferrucci, L., \& Horvath, S. (2018). “An epigenetic biomarker of aging for lifespan and healthspan.” \textit{Aging (Albany NY)}, \textbf{10}(4): 573–591. doi: \href{https://doi.org/10.18632/aging.101414}{10.18632/aging.101414}. URL: \url{https://www.aging-us.com/article/101414}.

\bibitem{mitrabio2023}
Mitra Bio (2023). “Which Biological Clock to Use for Skin's Anti-Ageing Interventions?” URL: \url{https://mitrabio.tech/resources/which-biological-clock-to-use-for-skins-anti-ageing-interventions/}. Accessed: 2025-05-07.

\bibitem{lehallier2019undulating}
Lehallier, B., Gate, D., Schaum, N., Nanasi, T., Lee, S. E., Yousef, H., Moran Losada, P., Berdnik, D., Keller, A., Verghese, J., Sathyan, S., Franceschi, C., Milman, S., Barzilai, N., \& Wyss-Coray, T. (2019). “Undulating changes in human plasma proteome profiles across the lifespan.” \textit{Nature Medicine}, \textbf{25}(12): 1843–1850. doi: \href{https://doi.org/10.1038/s41591-019-0673-2}{10.1038/s41591-019-0673-2}. URL: \url{https://www.nature.com/articles/s41591-019-0673-2}.

\bibitem{jaiswal_clonal_CHIP} Jaiswal, S. \& Ebert, B.L. (2019). "Clonal hematopoiesis in human aging and disease." \textit{Science} \textbf{366}(6472), eaan4673. doi:\url{https://doi.org/10.1126/science.aan4673}.

\bibitem{abegunde_tet2_tnfa_2018} Abegunde, S.O., Buckstein, R., Wells, R.A., \& Rauh, M.J. (2018). "An inflammatory environment containing TNFa favors Tet2-mutant clonal hematopoiesis." \textit{Experimental Hematology} \textbf{59}, 60–65. doi:\url{https://doi.org/10.1016/j.exphem.2017.11.002}.

\bibitem{bruserud_hematopoiesis_inflammaging} Bruserud, Ø., Vo, A.K., \& Rekvam, H. (2022). "Hematopoiesis, inflammation and aging — the biological background and clinical impact of anemia and increased C-reactive protein levels on elderly individuals." \textit{Journal of Clinical Medicine} \textbf{11}(3), 706. doi:\url{https://doi.org/10.3390/jcm11030706}.

\bibitem{geiger_inflammation_epigenetic_2020} Geiger, S.S., \& Essers, M.A.G. (2020). "Inflammation's epigenetic footprint in hematopoietic stem cells." \textit{Cell Stem Cell} \textbf{26}(5), 611–612. doi:\url{https://doi.org/10.1016/j.stem.2020.04.015}.

\bibitem{zhu_inflammation_epigenetics_aging_2021} Zhu, X., Chen, Z., Shen, W., Huang, G., Sedivy, J.M., Wang, H., \& Ju, Z. (2021). "Inflammation, epigenetics, and metabolism converge to cell senescence and ageing: the regulation and intervention." \textit{Signal Transduction and Targeted Therapy} \textbf{6}, 245. doi:\url{https://doi.org/10.1038/s41392-021-00646-9}.


\bibitem{barrett_stress_immune} Barrett, T.J., et al. (2021). "Chronic stress primes innate immune responses in mice and humans." \textit{Cell Reports} \textbf{36}(10), 109595. doi:\url{https://doi.org/10.1016/j.celrep.2021.109595}.

\bibitem{oriordan_gut_immune_brain_2025} O'Riordan, K.J., Moloney, G.M., Keane, L., Clarke, G., \& Cryan, J.F. (2025). "The gut microbiota-immune-brain axis: Therapeutic implications." \textit{Cell Reports Medicine} \textbf{6}(3), 101982. doi:\url{https://doi.org/10.1016/j.xcrm.2025.101982}.

\bibitem{1}
Lawrence Liao, David F. Kong, Linda K. Shaw, Michael H. Sketch Jr., Carmelo A. Milano, Kerry L. Lee, \& Daniel B. Mark (2005). “A new anatomic score for prognosis after cardiac catheterization in patients with previous bypass surgery.” \textit{J Am Coll Cardiol.}, \textbf{46}(9): 1684–1692. doi: \href{https://doi.org/10.1016/j.jacc.2005.06.074}{10.1016/j.jacc.2005.06.074}.

\bibitem{2}
I. Ringqvist, L. D. Fisher, M. Mock, K. B. Davis, H. Wedel, B. R. Chaitman, E. Passamani, R. O. Russell Jr., E. L. Alderman, N. T. Kouchoukas, G. C. Kaiser, T. J. Ryan, T. Killip, \& D. Fray (1983). “Prognostic value of angiographic indices of coronary artery disease from the Coronary Artery Surgery Study (CASS).” \textit{J Clin Invest.}, \textbf{71}(6): 1854–1866. doi: \href{https://doi.org/10.1172/JCI110941}{10.1172/jci110941}.

\bibitem{3}
F. H. Edwards, F. L. Grover, A. L. Shroyer, M. Schwartz, \& J. Bero (1997). “The Society of Thoracic Surgeons National Cardiac Surgery Database: current risk assessment.” \textit{Ann Thorac Surg.}, \textbf{63}(3): 903–908. doi: \href{https://doi.org/10.1016/S0003-4975(97)00017-9}{10.1016/S0003-4975(97)00017-9}.

\bibitem{4}
Jack E. Zimmerman, Andrew A. Kramer, Douglas S. McNair, \& Fern M. Malila (2006). “Acute Physiology and Chronic Health Evaluation (APACHE) IV: hospital mortality assessment for today’s critically ill patients.” \textit{Crit Care Med.}, \textbf{34}(5): 1297–1310. doi: \href{https://doi.org/10.1097/01.CCM.0000215112.84523.F0}{10.1097/01.CCM.0000215112.84523.F0}.

\bibitem{5}
Quinn R. Pack, Aruna Priya, Tara Lagu, Penelope S. Pekow, Richard Engelman, David M. Kent, \& Peter K. Lindenauer (2016). “Development and Validation of a Predictive Model for Short- and Medium-Term Hospital Readmission Following Heart Valve Surgery.” \textit{J Am Heart Assoc.}, \textbf{5}(9). doi: \href{https://doi.org/10.1161/JAHA.116.003544}{10.1161/JAHA.116.003544}.

\bibitem{6}
P. W. Wilson, R. B. D’Agostino, D. Levy, A. M. Belanger, H. Silbershatz, \& W. B. Kannel (1998). “Prediction of coronary heart disease using risk factor categories.” \textit{Circulation}, \textbf{97}(18): 1837–1847. doi: \href{https://doi.org/10.1161/01.CIR.97.18.1837}{10.1161/01.CIR.97.18.1837}.

\bibitem{13}
Mohammad Madjid \& Omid Fatemi (2013). “Components of the Complete Blood Count (CBC) as risk predictors for coronary heart disease: in-depth review and update.” \textit{Tex Heart Inst J.}, \textbf{40}(1): 17–29.

\bibitem{intermountain}
Benjamin D. Horne, Heidi T. May, Abdallah G. Kfoury, Dale G. Renlund, Joseph B. Muhlestein, Donald L. Lappé, Kismet D. Rasmusson, T. Jared Bunch, John F. Carlquist, Tami L. Bair, Kurt R. Jensen, Brianna S. Ronnow, \& Jeffrey L. Anderson (2010). “The Intermountain Risk Score (including the red cell distribution width) predicts heart failure and other morbidity endpoints.” \textit{Eur J Heart Fail}, \textbf{12}(11): 1203–1213.

\bibitem{aldrighi}
Aldrighi, J. M., Oliveira, R. L. S., D’Amico, E., Rocha, T. R. F., Gebara, O. E., Rosano, G. M. C., \& Ramires, J. A. F. (2005). “Platelet activation status decreases after menopause.” \textit{Gynecological Endocrinology}, \textbf{20}(5): 249–257. doi: \href{https://doi.org/10.1080/09513590500097549}{10.1080/09513590500097549}.

\bibitem{alberro2021inflammaging}
Alberro, A., Iribarren-Lopez, A., Sáenz-Cuesta, M., Matheu, A., Vergara, I., \& Otaegui, D. (2021). “Inflammaging markers characteristic of advanced age show similar levels with frailty and dependency.” \textit{Scientific Reports}, \textbf{11}(1): 4358. doi: \href{https://doi.org/10.1038/s41598-021-83991-7}{10.1038/s41598-021-83991-7}.

\bibitem{fulop2019biomarkers}
Fulop, T., Cohen, A., Wong, G., Witkowski, J. M., \& Larbi, A. (2019). “Are There Reliable Biomarkers for Immunosenescence and Inflammaging?” In \textit{Biomarkers of Human Aging} (Vol. 10, pp. 231–251). Springer International Publishing, Cham.

\bibitem{fulop2021immunology}
Fulop, T., Larbi, A., Pawelec, G., Khalil, A., Cohen, A. A., Hirokawa, K., et al. (2021). “Immunology of aging: the birth of inflammaging.” \textit{Clinical Reviews in Allergy \& Immunology}. doi: \href{https://doi.org/10.1007/s12016-020-08827-8}{10.1007/s12016-020-08827-8}.

\bibitem{hanes}
Centers for Disease Control and Prevention, National Center for Health Statistics (2016). “National Health and Nutrition Examination Survey.” URL: \url{https://wwwn.cdc.gov/nchs/nhanes/Default.aspx}.

\bibitem{alpert2019immuneage}
Alpert, A., Pickman, Y., Leipold, M., Rosenberg-Hasson, Y., Ji, X., Gaujoux, R., et al. (2019). “A clinically meaningful metric of immune age derived from high-dimensional longitudinal profiling.” \textit{Nature Medicine}, \textbf{25}: 487–495. doi: \href{https://doi.org/10.1038/s41591-019-0381-y}{10.1038/s41591-019-0381-y}.

\bibitem{nancy}
Aneke, J., Ibeh, N., Okocha, C., Nkwazema, K., \& Manafa, P. (2016). “Changes in Haematological Indices of Women at Different Fertility Periods in Nnewi, South-East, Nigeria.” \textit{Journal of Medical Research}, \textbf{2}: 166–169. doi: \href{https://doi.org/10.31254/jmr.2016.2610}{10.31254/jmr.2016.2610}.

\bibitem{cbc}
Anderson, J. L., Ronnow, B. S., Horne, B. D., Carlquist, J. F., May, H. T., Bair, T. L., Jensen, K. R., \& Muhlestein, J. B. (2007). “Usefulness of a Complete Blood Count-derived risk score to predict incident mortality in patients with suspected cardiovascular disease.” \textit{American Journal of Cardiology}, \textbf{99}(2): 169–174. doi: \href{https://doi.org/10.1016/j.amjcard.}{10.1016/j.amjcard.}.

\bibitem{biobank1}
Bahcall, O. G. (2018). “UK Biobank — a New Era in Genomic Medicine.” \textit{Nature Reviews Genetics}, \textbf{19}(12): 737. doi: \href{https://doi.org/10.1038/s41576-018-0065-3}{10.1038/s41576-018-0065-3}.

\bibitem{upmc}
University of Pittsburgh Department of Pathology (2022). “Case Index by Patient History.” URL: \url{https://path.upmc.edu/cases/}.

\bibitem{biobank2}
Watts, G. (2012). “UK Biobank Opens Its Data Vaults to Researchers.” \textit{BMJ}, \textbf{344}: e2459. doi: \href{https://doi.org/10.1136/bmj.e2459}{10.1136/bmj.e2459}.

\bibitem{cruickshank}
Cruickshank, J. M. (1970). “Some Variations in the Normal Haemoglobin Concentration.” \textit{British Journal of Haematology}, \textbf{18}(5): 523–530. doi: \href{https://doi.org/10.1111/j.1365-2141.1970.tb00773.x}{10.1111/j.1365-2141.1970.tb00773.x}. URL: \url{https://onlinelibrary.wiley.com/doi/abs/10.1111/j.1365-2141.1970.tb00773.x}.

\bibitem{day-time-var}
Erdemir, I. (2013). “The comparison of blood parameters between morning and evening exercise.” \textit{European Journal of Experimental Biology}, \textbf{3}(1): 559–563.

\bibitem{nhsYork}
York Teaching Hospital NHS Foundation Trust (n.d.). “Case Index by Patient History.” URL: \url{https://www.yorkhospitals.nhs.uk/seecmsfile/?id=2396}.

\bibitem{brodin1}
Brodin, P., Jojic, V., Gao, T., Bhattacharya, S., Lopez Angel, C. J., Furman, D., Shen-Orr, S., Dekker, C. L., Swan, G. E., Butte, A. J., Maecker, H. T., \& Davis, M. M. (2015). “Variation in the human immune system is largely driven by non-heritable influences.” \textit{Cell}, \textbf{160}(1–2): 37–47. doi: \href{https://doi.org/10.1016/j.cell.2014.12.020}{10.1016/j.cell.2014.12.020}.

\bibitem{brodin2}
Brodin, P., \& Davis, M. M. (2017). “Human immune system variation.” \textit{Nature Reviews Immunology}, \textbf{17}(1): 21–29. doi: \href{https://doi.org/10.1038/nri.2016.125}{10.1038/nri.2016.125}.


\bibitem{edwards1997}
Edwards, F. H., Grover, F. L., Shroyer, A. L., Schwartz, M., \& Bero, J. (1997). “The Society of Thoracic Surgeons National Cardiac Surgery Database: current risk assessment.” \textit{The Annals of Thoracic Surgery}, \textbf{63}(3): 903–908. doi: \href{https://doi.org/10.1016/S0003-4975(97)00017-9}{10.1016/S0003-4975(97)00017-9}.

\bibitem{erdemir2013}
Erdemir, I. (2013). “The comparison of blood parameters between morning and evening exercise.” \textit{European Journal of Experimental Biology}, \textbf{3}(1): 559–563.

\bibitem{factorial}
Zou, M.-X., Pan, Y., Huang, W., Zhang, T.-L., Escobar, D., Wang, X.-B., Jiang, Y., She, X.-L., \& Lv, G.-H. (2020). “A four-factor immune risk score signature predicts the clinical outcome of patients with spinal chordoma.” \textit{Clinical and Translational Medicine}, \textbf{10}(1): 224–237. doi: \href{https://doi.org/10.1002/ctm2.4}{10.1002/ctm2.4}.

\bibitem{ferrucci2018inflammageing}
Ferrucci, L., \& Fabbri, E. (2018). “Inflammageing: chronic inflammation in ageing, cardiovascular disease, and frailty.” \textit{Nature Reviews Cardiology}, \textbf{15}(9): 505–522. doi: \href{https://doi.org/10.1038/s41569-018-0064-2}{10.1038/s41569-018-0064-2}.

\bibitem{harvard}
Foy, B. H., Petherbridge, R., Roth, M. T., Zhang, C., De Souza, D. C., Mow, C., Patel, H. R., Patel, C. H., Ho, S. N., Lam, E., Powe, C. E., Hasserjian, R. P., Karczewski, K. J., Tozzo, V., \& Higgins, J. M. (2024). “Haematological setpoints are a stable and patient-specific deep phenotype.” \textit{Nature}. doi: \href{https://doi.org/10.1038/s41586-024-08264-5}{10.1038/s41586-024-08264-5}.

\bibitem{Ref2}
Franceschi, C., Bonafè, M., Valensin, S., Olivieri, F., De Luca, M., Ottaviani, E., \& De Benedictis, G. (2000). “Inflamm-aging. An evolutionary perspective on immunosenescence.” \textit{Annals of the New York Academy of Sciences}, \textbf{908}: 244–254. doi: \href{https://doi.org/10.1111/j.1749-6632.2000.tb06651.x}{10.1111/j.1749-6632.2000.tb06651.x}.

\bibitem{Ref3}
Franceschi, C., Capri, M., Monti, D., Giunta, S., Olivieri, F., Sevini, F., Panourgia, M. P., Invidia, L., Celani, L., Scurti, M., Cevenini, E., Castellani, G. C., \& Salvioli, S. (2006). “Inflammaging and anti-inflammaging: a systemic perspective on aging and longevity emerged from studies in humans.” \textit{Mechanisms of Ageing and Development}, \textbf{128}(1): 92–105. doi: \href{https://doi.org/10.1016/j.mad.2006.11.016}{10.1016/j.mad.2006.11.016}.

\bibitem{Ref4}
Franceschi, C., \& Campisi, J. (2014). “Chronic inflammation (inflammaging) and its potential contribution to age-associated diseases.” \textit{The Journals of Gerontology Series A: Biological Sciences and Medical Sciences}, \textbf{69}(1): 4–9. doi: \href{https://doi.org/10.1093/gerona/glu057}{10.1093/gerona/glu057}.

\bibitem{Ref5}
Franceschi, C., Garagnani, P., Vitale, G., Capri, M., \& Salvioli, S. (2017). “Inflammaging and ‘Garb-aging’.” \textit{Trends in Endocrinology and Metabolism}, \textbf{28}(3): 199–212. doi: \href{https://doi.org/10.1016/j.tem.2016.09.005}{10.1016/j.tem.2016.09.005}.

\bibitem{Ref6}
Furman, D., Chang, J., Lartigue, L., Bolen, C. R., Haddad, F., Gaudilliere, B., Ganio, E. A., Fragiadakis, G. K., Spitzer, M. H., Douchet, I., Daburon, S., Moreau, J.-F., Nolan, G. P., Blanco, P., Déchanet-Merville, J., Dekker, C. L., Jojic, V., Kuo, C. J., Davis, M. M., \& Faustin, B. (2017). “Expression of specific inflammasome gene modules stratifies older individuals into two extreme clinical and immunological states.” \textit{Nature Medicine}, \textbf{23}(2): 174–184. doi: \href{https://doi.org/10.1038/nm.4267}{10.1038/nm.4267}.

\bibitem{Ref8}
Furman, D., Campisi, J., Verdin, E., Carrera-Bastos, P., Targ, S., Franceschi, C., Ferrucci, L., Gilroy, D. W., Fasano, A., Miller, G. W., Miller, A. H., Mantovani, A., Weyand, C. M., Barzilai, N., Goronzy, J. J., Rando, T. A., Effros, R. B., Lucia, A., Kleinstreuer, N., \& Slavich, G. M. (2019). “Chronic inflammation in the etiology of disease across the life span.” \textit{Nature Medicine}, \textbf{25}(12): 1822–1832. doi: \href{https://doi.org/10.1038/s41591-019-0675-0}{10.1038/s41591-019-0675-0}.


\bibitem{jupiter}
Horne, B. D., Anderson, J. L., Muhlestein, J. B., Ridker, P. M., \& Paynter, N. P. (2014). “Complete Blood Count risk score and its components, including RDW, are associated with mortality in the JUPITER trial.” \textit{European Journal of Preventive Cardiology}, \textbf{22}(4): 519–526. doi: \href{https://doi.org/10.1177/2047487313519347}{10.1177/2047487313519347}.


\bibitem{kovanen}
Kovanen, V., Aukee, P., Kokko, K., Finni, T., Tarkka, I. M., Tammelin, T., Kujala, U. M., Sipilä, S., \& Laakkonen, E. K. (2018). “Design and protocol of Estrogenic Regulation of Muscle Apoptosis (ERMA) study with 47 to 55-year-old women’s cohort: novel results show menopause-related differences in blood count.” \textit{Menopause}, \textbf{25}(9): 1020–1032. doi: \href{https://doi.org/10.1097/GME.0000000000001117}{10.1097/GME.0000000000001117}.


\bibitem{Ref1}
López-Otín, C., Blasco, M. A., Partridge, L., Serrano, M., \& Kroemer, G. (2013). “The Hallmarks of Aging.” \textit{Cell}, \textbf{153}(6): 1194–1217. doi: \href{https://doi.org/10.1016/j.cell.2013.05.039}{10.1016/j.cell.2013.05.039}.

\bibitem{lord2024covid}
Lord, J. M., Veenith, T., Sullivan, J., et al. (2024). “Accelerated immune ageing is associated with COVID-19 disease severity.” \textit{Immun Ageing}, \textbf{21}: 6. doi: \href{https://doi.org/10.1186/s12979-024-00322-7}{10.1186/s12979-024-00322-7}.




\bibitem{nakada}
Nakada, D., Oguro, H., Levi, B. P., Ryan, S. J., Kitano, A., Saitoh, Y., Takeichi, M., Wendt, G. R., \& Morrison, S. J. (2014). “Estrogen increases haematopoietic stem-cell self-renewal in females and during pregnancy.” \textit{Nature}, \textbf{505}(7484): 555–558. doi: \href{https://doi.org/10.1038/nature12932}{10.1038/nature12932}.

\bibitem{cbc2}
Niu, X., Liu, G., Huo, L., Zhang, J., Bai, M., Peng, Y., \& Zhang, Z. (2018). “Risk stratification based on components of the Complete Blood Count in patients with acute coronary syndrome: A classification and regression tree analysis.” \textit{Scientific Reports}, \textbf{8}(1): 2838. doi: \href{https://doi.org/10.1038/s41598-018-21139-w}{10.1038/s41598-018-21139-w}.



\bibitem{routine}
Kristensen, M., Iversen, A. K. S., Gerds, T. A., Østervig, R., Linnet, J. D., Barfod, C., Lange, K. H. W., Sölétormos, G., Forberg, J. L., Eugen-Olsen, J., Rasmussen, L. S., Schou, M., Køber, L., \& Iversen, K. (2017). “Routine blood tests are associated with short term mortality and can improve emergency department triage: a cohort study of $>$12,000 patients.” \textit{Scandinavian Journal of Trauma, Resuscitation and Emergency Medicine}, \textbf{25}(1): 115. doi: \href{https://doi.org/10.1186/s13049-017-0458-x}{10.1186/s13049-017-0458-x}.







\end{thebibliography}
\end{document}